\documentclass[structabstract]{aa}  
\usepackage{graphicx} 
\usepackage{txfonts}
\usepackage{natbib}
\usepackage{hyperref}
\usepackage{siunitx}  
\usepackage{xcolor}   
\newcommand{\HII}{\mbox{H\,{\sc ii}}}
\newcommand{\HeI}[1]{\mbox{He\,{\sc i}~$\lambda${#1}}}
\newcommand{\HeII}[1]{\mbox{He\,{\sc ii}~$\lambda${#1}}}

\newcommand{\LiI}[1]{\mbox{Li\,{\sc i}~$\lambda${#1}}}
\newcommand{\CII}[1]{\mbox{C\,{\sc ii}~$\lambda${#1}}}

\newcommand{\CIIIt}[1]{\mbox{C\,{\sc iii}~$\lambda\lambda\lambda${#1}}}

\newcommand{\NIIIt}[1]{\mbox{N\,{\sc iii}~$\lambda\lambda\lambda${#1}}}
\newcommand{\NIV}[1]{\mbox{N\,{\sc iv}~$\lambda${#1}}}

\newcommand{\NVd}[1]{\mbox{N\,{\sc v}~$\lambda\lambda${#1}}}
\newcommand{\OII}[1]{\mbox{O\,{\sc ii}~$\lambda${#1}}}
\newcommand{\OIIId}[1]{\mbox{[O\,{\sc iii}]~$\lambda\lambda${#1}}}
\newcommand{\SiIII}[1]{\mbox{Si\,{\sc iii}~$\lambda${#1}}}

\newcommand{\GG}{\mbox{$G$}}
\newcommand{\GBP}{\mbox{$G_{\rm BP}$}}
\newcommand{\GRP}{\mbox{$G_{\rm RP}$}}
\newcommand{\GBPmGRP}{\mbox{$G_{\rm BP}-G_{\rm RP}$}}
\newcommand{\GGc}{\mbox{$G^\prime$}}
\newcommand{\GGcmax}{\mbox{$G^\prime_{\rm max}$}}
\newcommand{\GBPc}{\mbox{$G_{\rm BP}^{\prime}$}}
\newcommand{\GRPc}{\mbox{$G_{\rm RP}^{\prime}$}}
\newcommand{\GBPmGRPc}{\mbox{$G_{\rm BP}^\prime-G_{\rm RP}^\prime$}}

\newcommand{\microas}{\mbox{$\mu$as}}
\newcommand{\mci}[1]{\multicolumn{1}{c}{#1}}
\newcommand{\mcii}[1]{\multicolumn{2}{c}{#1}}
\newcommand{\dCC}{\mbox{$d_{\rm CC}$}}
\newcommand{\Cstar}{\mbox{$C^*$}}
\newcommand{\pmra}{\mbox{$\mu_{\alpha *}$}}
\newcommand{\pmdec}{\mbox{$\mu_{\delta}$}}
\newcommand{\pmrag}{\mbox{$\mu_{\alpha *,{\rm g}}$ }}
\newcommand{\pmdecg}{\mbox{$\mu_{\delta,{\rm g}}$}}
\newcommand{\pic}{\mbox{$\varpi_{\rm c}$}}
\newcommand{\Teff}{\mbox{$T_{\rm eff}$}}
\newcommand{\sigmas}{\mbox{$\sigma_{\rm s}$}}
\newcommand{\sigmai}{\mbox{$\sigma_{\rm int}$}}
\newcommand{\sigmae}{\mbox{$\sigma_{\rm ext}$}}
\newcommand{\pig}{\mbox{$\varpi_{\rm g}$}}
\newcommand{\spig}{\mbox{$\sigma_{\varpi_{\rm g}}$}}
\newcommand{\Nf}{\mbox{$N_{\rm f}$}}

\newcommand{\RV}{\mbox{$R_{5495}$}}

\newcommand{\VO}[1]{Villafranca~O-{#1}}

\newcommand{\lili}{LiLiMaRlin}

\begin{document}

   \title{The Villafranca catalog of Galactic OB groups \linebreak
          II. From \textit{Gaia} DR2 to EDR3 and ten new systems with O stars}
   \titlerunning{Villafranca OB groups: II. From \textit{Gaia} DR2 to EDR3 and ten new systems}

   \author{J. Ma\'{\i}z Apell\'aniz \inst{1}
           \and
           R. H. Barb\'a\inst{2}
           \and
           R. Fern\'andez Aranda\inst{1,3,4}
           \and
           M. Pantaleoni Gonz\'alez\inst{1,3}
           \and
           P. Crespo Bellido\inst{1,3}
           \and
           \\
%           S. Sim\'on-D{\'\i}az\inst{4,5}
           A. Sota\inst{5}
           \and
           E. J. Alfaro\inst{5}
%           J. A. Caballero\inst{1}
           }
   \authorrunning{J. Ma\'{\i}z Apell\'aniz et al.}

   \institute{Centro de Astrobiolog\'{\i}a, CSIC-INTA. Campus ESAC. 
              C. bajo del castillo s/n. 
              E-\num{28692} Villanueva de la Ca\~nada, Madrid, Spain. \linebreak
              \email{jmaiz@cab.inta-csic.es} 
              \and
              Departamento de Astronom{\'\i}a, Universidad de La Serena.
              Av. Cisternas 1200 Norte.
              La Serena, Chile 
              \and
              Departamento de Astrof{\'\i}sica y F{\'\i}sica de la Atm\'osfera, Universidad Complutense de Madrid. 
              E-\num{28040} Madrid, Spain. 
              \and
              Institute of Astrophysics, University of Crete.
              \num{70013} Heraklion, Greece.
%              \and
%              Instituto de Astrof\'{\i}sica de Canarias. 
%              E-38\,200 La Laguna, Tenerife, Spain. 
%              \and
%              Departamento de Astrof\'{\i}sica. Universidad de La Laguna.
%              E-38\,205 La Laguna, Tenerife, Spain. 
              \and
              Instituto de Astrof\'{\i}sica de Andaluc\'{\i}a-CSIC. 
              Glorieta de la Astronom\'{\i}a s/n. 
              E-\num{18008} Granada, Spain.
             }

   \date{Received 4 October 2021 / Accepted XX XXX 2021}

% \abstract{}{}{}{}{} 
% 5 {} token are mandatory
 
  \abstract
   {This is the second paper of a series on Galactic OB groups that uses astrometric and photometric data from \textit{Gaia} and 
    spectral classifications from the Galactic O-Star Spectroscopic Survey (GOSSS) and the Library of Libraries of Massive-star high-Resolution 
    spectra (\lili). The previous paper was based on the second \textit{Gaia} data release (DR2). Since then, the early third \textit{Gaia} data 
    release (EDR3) has appeared with new astrometry and photometry.}
   {The two aims of this paper are to revise the results for the sample from paper I using \textit{Gaia} EDR3 data and to expand the sample of 
    analyzed stellar groups to 26, from \VO{001} to \VO{026}.}
   {We used GOSSS to select Galactic stellar groups with O stars and an updated version of the method in
    paper 0 of this series, combining \textit{Gaia} EDR3 G + \GBP\ + \GRP\ photometry, positions, proper motions, and parallaxes to assign 
    memberships and measure distances. We present 99 spectra from GOSSS and 32 from \lili\ for stars in the analyzed groups or in their foreground.}
   {We derived distances to the 26 stellar groups with unprecedented precision and accuracy, with total (random plus systematic) uncertainties lower than
    1\% for distances within 1~kpc and of $\sim$3\% around 3~kpc, which are values almost four times better than for \textit{Gaia}~DR2. 
    We provide homogeneous spectral types for 110 stars and correct a number of errors in the literature, especially for objects in \VO{023} 
    (Orion nebula cluster).  For each group, we discuss its membership and present possible runaway and walkaway stars.
    At least two of the studied groups, \VO{O12}~S in NGC~2467 and \VO{014}~NW in the North America nebula, are orphan clusters in which the most massive 
    stars have been ejected by dynamical interactions, leaving objects with a capped mass function. The existence of such clusters has important 
    consequences for the study of the initial mass function (IMF), the distribution of supernova explosions across the Galaxy, and the population and dynamics 
    of isolated compact objects. We fit pre-main-sequence (PMS) isochrones to the color-magnitude diagrams (CMDs) of four clusters to derive ages of 
    2.0$\pm$0.5~Ma for \VO{026} ($\sigma$~Orionis cluster), 4$\pm$2~Ma for \VO{016} 
    (NGC~2264), 5.0$\pm$0.5~Ma for \VO{021} (NGC~2362), and 8$\pm$2~Ma for \VO{024} ($\gamma$~Velorum cluster).
   }
   {}
   \keywords{astrometry --- catalogs --- Galaxy: structure --- open clusters and associations: general ---
             stars: kinematics and dynamics --- stars: early-type}
   \maketitle
%
%________________________________________________________________

\section{Introduction}

$\,\!$\indent We are building the Villafranca catalog of Galactic OB stellar groups based on {\it Gaia} astrometry and photometry \citep{Prusetal16} and spectroscopic 
information from the Galactic O-Star Spectroscopic Survey (GOSSS, \citealt{Maizetal11}) and the Library of Libraries of Massive-Star High-Resolution Spectra (\lili,
\citealt{Maizetal19a}). In \citet{Maiz19}, from now on Villafranca~0, we present the method we used to identify members and measure distances to stellar groups and
we applied it to the first two examples. \citet{Maizetal20b}, from now on Villafranca~I, describes the project aims and analyzes 16 stellar groups using the second 
{\it Gaia} data release (DR2).

As described in Villafranca~I, OB stars are formed in stellar groups that can be either bound (clusters) or unbound (associations or subassociations), with the difference 
between the two categories being difficult to ascertain in some circumstances due to the limitations of the available observational data and to the circumstances of each
specific case. Some clusters have double cores, some associations have bound cores at their centers, and the limits between neighbor associations may be poorly defined.
This requires each case to be analyzed in detail if we are to understand how massive stars form and how to determine the nature and form of the initial mass function
(IMF).

\begin{table*}[h!]
%\fontsize{8}{9}\selectfont
\caption{Sample of new O-type stellar groups in this paper.} % The other IDs may refer to the cluster or the associated H\,{\sc ii} region.}
\centerline{
\begin{tabular}{lllll}
ID    & Other IDs            & O/WR/LBV stars                    & Spectral type                & Ref. \\
\hline
O-017 & Heart nebula         & HD~\num{15570}                    & O4 If                        & S11  \\
      & IC 1805              & HD~\num{15558}~A$^\dagger$        & O4.5 III(fc)p + O8:          & M19  \\
      & Sh 2-190             & HD~\num{15629}                    & O4.5 V((fc))                 & S11  \\
      & W4                   & KM~Cas                            & O9.2 IVn                     & TW   \\
      &                      & At least five others              &                              &      \\
O-018 & Lagoon nebula        & 9~Sgr~A,B                         & O3.5 V((f*)) + O5-5.5 V((f)) & R12  \\
      & M8                   & HD~\num{165052}                   & O6.5 V((f))z + O6.5 V((f))z  & TW   \\
      & NGC 6523+6530        & Herschel~36                       & O7.5 V + O9 V + B0.5 V       & A10  \\
      & Sh 2-25              & HD~\num{164816}                   & O9.5 V + O9.7 V              & TW   \\
      & W 29                 & Possibly two others               &                              &      \\
O-019 & Eagle nebula         & HD~\num{168076}~A,B$^\dagger$     & O4 IV(f)                     & M16  \\
      & M16                  & BD~$-$13~4923$^\dagger$           & O4 V((f)) + O7.5 V           & M19  \\
      & NGC 6611             & ALS~\num{15360}                   & O6.5 V((f))z                 & TW   \\
      & Sh 2-49              & HD~\num{168075}                   & O6.5 V((f)) + B0-1 V         & S09  \\
      & RCW 165              & At least eight others             &                              &      \\
      & W 37                 &                                   &                              &      \\
O-020 & Rosette nebula       & HD~\num{46223}                    & O4 V((f))                    & S11  \\
      & NGC 2244             & HD~\num{46150}                    & O5 V((f))z                   & S11  \\
      & Sh 2-275             & At least four others              &                              &      \\
      & W 16                 &                                   &                              &      \\
O-021 & NGC 2362             & $\tau$~CMa~Aa$^\dagger$           & O9.2 Iab:                    & M20  \\
      & $\tau$ CMa cluster   & $\tau$~CMa~Ab$^{\dagger\dagger}$  & O9 III                       & M20  \\
O-022 & NGC 6231             & WR~79                             & WC7 + O4 III-I(f)            & TW   \\
      & Collinder 315        & HD~{152233}~Aa,Ab                 & O6 II(f)                     & S14  \\
      &                      & HD~{152248}~Aa,Ab                 & O7 Iabf + O7 Ib(f)           & S14  \\
      &                      & At least eleven others            &                              &      \\
O-023 & Orion nebula         & $\theta^1$~Ori~Ca,Cb$^\dagger$    & O7 f?p var                   & M19  \\
      & M42+M43              & $\theta^2$~Ori~A$^\dagger$        & O9.2 V + B0.5: V(n)          & M19  \\
      & NGC 1976+1982        &                                   &                              &      \\
O-024 & $\gamma$ Vel cluster & $\gamma^2$~Vel$^{\dagger\dagger}$ & WC8 + O7/8.5 III-II          & TW   \\
      & Pozzo 1              &                                   &                              &      \\
      & Brandt 1             &                                   &                              &      \\
O-025 & Trumpler 16 E        & $\eta$~Car$^\dagger$              & LBV + O:                     & S06  \\
      &                      & HDE~\num{303308}~A,B              & O4.5 V((fc))                 & S14  \\
      &                      & CPD~$-$59~2641                    & O6~V((fc))                   & S14  \\
      &                      & At least ten others               &                              &      \\
O-026 & $\sigma$ Ori cluster & $\sigma$~Ori~Aa,Ab,B$^\dagger$    & O9.5 V + B0.2 V + B0.2 V(n)  & M21  \\
%O-027 & Westerlund 1         &                                   &                              &      \\
\hline
Notes. & \multicolumn{4}{l}{$^\dagger$: Bad RUWE or $\sigmae > 0.1$~mas. $^{\dagger\dagger}$: Not in {\it Gaia} EDR3.} \\
Ref.:  & \multicolumn{4}{l}{A10: \citet{Ariaetal10},  M16: \citet{Maizetal16},  M19: \citet{Maizetal19b},} \\
       & \multicolumn{4}{l}{M20: \citet{MaizBarb20},  M21: \citet{Maizetal21b}, R12: \citet{Rauwetal12a},} \\
       & \multicolumn{4}{l}{S06: \citet{Smit06b},     S09: \citet{Sanaetal09},  S11: \citet{Sotaetal11a}, S14: \citet{Sotaetal14}, TW:  This work.} \\
\end{tabular}
}
\label{sample}                   
\end{table*}

Since Villafranca~I was published, the early third {\it Gaia} data release (EDR3) has become available \citep{Browetal21}. {\it Gaia}~EDR3 provides astrometry with
reduced uncertainties \citep{Lindetal21a} that require the application of a correction to eliminate the parallax bias as a function of position, magnitude, and color 
\citep{Lindetal21b}. Once the correction is applied, the remaining systematic uncertainty is of only 10.3~\microas\ and the angular covariance as a function of
separation is well characterized \citep{Maizetal21c}, leading to the possibility of significantly improving the distances to stellar groups derived from {\it Gaia}~DR2
data, such as the ones in Villafranca~I. Furthermore, \citet{Maizetal21c} measured that the \citet{Lindetal21b} correction worked very well except for the brightest
stars, which could be recalibrated using open clusters with bright stars such as the ones in nearby OB groups. This leads to the possibility not only of determining the
distances to stellar groups with better accuracies but also of using the same data to improve the \citet{Lindetal21b} correction with stellar clusters, a possibility
suggested in that paper.

In this paper we reanalyze the sample of 16 stellar groups of Villafranca~I using {\it Gaia}~EDR3 data and include ten additional objects. In a companion paper 
\citep{Maiz21}
we combine the results from those 26 stellar groups with the results for the six globular clusters of \citet{Maizetal21c}
and for the LMC and SMC of \citet{Lurietal21} to derive an improved zero point and a characterization of the external uncertainties for {\it Gaia}~EDR3 parallaxes. 
In the next section we present our data and methods, we then describe our results for the individual stellar groups, and we finish our paper with an analysis and an
exposition of the next steps in the Villafranca project.

\section{Data and methods}

\subsection{OB groups sample selection}

$\,\!$\indent The long-term goal of the Villafranca project is to include all optically accessible Galactic OB groups with O stars identified as such from GOSSS and \lili\ spectroscopy. We 
plan to achieve it in subsequent papers increasing the sample in steps that add 30-100 O~stars each. Villafranca~I concentrated on OB groups with very-early O-type stars (O2-O3.5). In this 
paper, we add ten OB groups that contain O stars, are of low-extinction (hence, nearby: all are within 2.5~kpc), and have a rich stellar population accessible to {\it Gaia}. Those criteria 
yield groups that are at the same time scientifically interesting (being some of the better studied OB groups) and useful for the goals of the companion paper of deriving an improved zero 
point and characterizing the uncertainties of {\it Gaia}~EDR3 parallaxes.

\begin{table*}
%\fontsize{8}{9}\selectfont
\caption{Notation used in this paper.} % The first two blocks refer to photometric and astrometric quantities of individual stars, respectively. The last two blocks list the group input (Table~\ref{filters}) and output (Table~\ref{results}) values, respectively.}
\centerline{
\begin{tabular}{lll}
Symbol               & Name                                                             & Description/Reference                                 \\
\hline
\GG                  & Very broad band photometric magnitude                            & Obtained directly from {\it Gaia}~EDR3                \\
\GBPmGRP             & Photometric color                                                & Obtained directly from {\it Gaia}~EDR3                \\
\GGc                 & Corrected \GG                                                    & Weiler et al. (in preparation)                        \\
\GBPc                & Corrected \GBP                                                   & Weiler et al. (in preparation)                        \\ 
\GRPc                & Corrected \GRP                                                   & Weiler et al. (in preparation)                        \\ 
\Cstar               & Corrected color excess factor                                    & Eqn.~6 in \citet{Rieletal21}                          \\
\hline
$\varpi$             & Individual parallax                                              & Obtained directly from {\it Gaia}~EDR3                \\
RUWE                 & Renormalized Unit Weight Error                                   & Obtained directly from {\it Gaia}~EDR3                \\
$Z_{\rm EDR3}$       & Parallax zero point                                              & \citet{Lindetal21b} and companion paper               \\
\pic                 & Individual corrected parallax                                    & $\varpi-Z_{\rm EDR3}$, \citet{Maizetal21c}            \\
\sigmai              & Individual parallax/proper motion internal uncertainty           & Obtained directly from {\it Gaia}~EDR3                \\
\sigmas              & Individual parallax/proper motion systematic uncertainty         & 10.3~\microas\ for parallaxes, \citet{Maizetal21c}    \\
                     &                                                                  & 23~\microas/a for proper motions, \citet{Lindetal21a} \\
$k$                  & Multiplicative constant for parallax/proper motion uncertainties & 1.1-2.7 depending on \GG, \citet{Maizetal21c}         \\
                     &                                                                  & $\;\;$ and companion paper                            \\
\sigmae              & Individual parallax/proper motion external uncertainty           & $\sqrt{k^2\sigmai^2+\sigmas^2}$, \citet{Maizetal21c}  \\
\hline
\Nf                  & Number of {\it Gaia}~EDR3 sources in the field                   & Table~\ref{filters}, Villafranca~I                    \\
$\alpha,\delta$      & Filtering center J2000 Equatorial coordinates                    & Table~\ref{filters}, Villafranca~I                    \\
$r$                  & Filtering separation from the center in the plane of the sky     & Table~\ref{filters}, Villafranca~I                    \\
\pmra,\pmdec         & Filtering central proper motion                                  & Table~\ref{filters}, Villafranca~I                    \\
$r_\mu$              & Filtering separation from the center in proper motion            & Table~\ref{filters}, Villafranca~I                    \\
$\Delta(\GBPmGRPc)$  & Filtering color displacement with respect to isochrone           & Table~\ref{filters}, Villafranca~I                    \\
\GGcmax              & Filtering maximum corrected \GG\ magnitude                       & Table~\ref{filters}                                   \\
\hline
$N_{*,0}$            & Number of stars in group before normalized parallax filtering    & Table~\ref{results}, Villafranca~I                    \\
$N_*$                & Number of stars in group after normalized parallax filtering     & Table~\ref{results}, Villafranca~I                    \\
$t_\varpi$           & Normalized $\chi^2$ test for the group parallax                  & Table~\ref{results}, Villafranca~0                    \\
$t_{\mu_{\alpha *}}$ & Normalized $\chi^2$ test for the group proper motion in $\alpha$ & Table~\ref{results}, Villafranca~0                    \\
$t_{\mu_{\delta}}$   & Normalized $\chi^2$ test for the group proper motion in $\delta$ & Table~\ref{results}, Villafranca~0                    \\
\pig                 & Group parallax                                                   & Table~\ref{results}, Eqn.~5 in \citet{Maizetal21c}    \\
\spig                & Group parallax uncertainty                                       & Table~\ref{results}, Eqn.~7 in \citet{Maizetal21c}    \\
\pmrag               & Group proper motion in $\alpha$                                  & Table~\ref{results}, Villafranca~1                    \\
\pmdecg              & Group proper motion in $\delta$                                  & Table~\ref{results}, Villafranca~1                    \\
$d$                  & Group distance                                                   & Table~\ref{results}, \citet{Maiz01a,Maiz05c},         \\
                     &                                                                  & $\;\;$ \citet{Maizetal08a}                            \\
\hline
\end{tabular}
}
\label{notation}
\end{table*}

Table~\ref{sample} provides the basic information for the new OB groups in this paper. See Table~1 in Villafranca~I for the equivalent information for the original sixteen OB groups. We 
continue the notation of the previous paper by assigning an ID of the type \VO{XXX}, where the O refers to the existence of at least an O star in the group and XXX is a sequential number, with
001 to 016 corresponding to the groups present in Villafranca~I and 017 to 026 corresponding to the new ones. For each group we list their common names in the literature, their O, WR, and
LBV stars, and the spectral classifications of those. A given name can refer to the stellar group (usually a cluster), a part of it, the associated nebulosity, both the group and the nebulosity, or
even be unclear in the literature. For that reason, we include in the text brief descriptions for the possibly confusing cases.

Of the new ten groups, nine are real clusters. The exception is \VO{025} (Trumpler~16~E), which is a subassociation that is part of Trumpler~16 (\VO{003} is Trumpler~16~W), which in
turn is part of the Carina nebula association or Car~OB1. Car~OB1 is the richest OB association within 3 kpc and also includes \VO{002} (Trumpler~14) and several other clusters and
non-bound OB groups and we divide it into pieces due to the criterion established in Villafranca~I of defining groups in a ``cluster scale'' (sizes of a few~pc), independently of their 
true nature as clusters, instead of in an ``association scale'' (sizes of 10~pc or more). Besides the reasons for this described in the previous paper, this condition helps us maintain a
small dispersion in real physical distances along the line of sight, which is useful for the purposes of the companion paper on the {\it Gaia}~EDR3 parallax zero point.

\subsection{Distances to and membership of OB groups}

$\,\!$\indent The supervised method used to determine the distances to and membership of stellar groups has already been described in the previous papers: \citet{Campetal19}, 
Villafranca~0, and Villafranca~I. For the convenience of the reader, we list in Table~\ref{notation} the notation developed in those references, \citet{Maizetal21a}, and the 
companion paper. The method uses a series of filters (defined in Table~\ref{notation} and listed for each group in Table~\ref{filters}) to establish an initial membership of 
$N_{*,0}$ stars in each group, which is further culled by a 3-sigma cut in normalized parallax (the corrected parallax minus the group parallax divided by the external uncertainty) 
that yields $N_*$ stars. As explained in Villafranca~I, we use restrictive filters
as our goal is not to derive a complete sample of group members but rather one in which there are none or very few false positives. That makes the derived distances very robust,
as shown by Monte Carlo simulations in Villafranca~0. 

Here we describe the differences in the procedure introduced since Villafranca~I. The most important (and obvious) one is that we use {\it Gaia}~EDR3 parallaxes instead of DR2 ones,
which are of significantly better quality \citep{Lindetal21a}. The second most important difference is that we apply
a magnitude, color, and Galactic latitude dependent parallax zero point instead of a constant one for all stars (see below for details). A first consequence of this is that the 
individual (or stellar) parallaxes are corrected before calculating the group parallax, so there is no need to distinguish between uncorrected and corrected group parallaxes, as was done
in Villafranca~I. A second consequence is that there is no need to add an extra amount in quadrature to the group parallax uncertainty. A third difference regarding the astrometry is that
for bright stars we apply the proper motion corrections of \citet{CanGBran21}.

When using {\it Gaia}~DR2 photometry, the \GG\ photometry has to be corrected and one has to distinguish between a bright and a faint magnitude range for \GBP\ \citep{MaizWeil18}. Those
effects are small but it is important to specify whether one uses the catalog magnitudes (\GG) or the corrected ones (\GGc). For {\it Gaia}~EDR3 photometry we have developed similar
corrections for \GG, \GBP, and \GRP\ (Weiler et al. in preparation). We have also substituted the DR2 \dCC\ parameter by the new \Cstar\ \citep{Rieletal21}. For one group we introduced a 
maximum corrected \GG\ magnitude, \GGcmax, to exclude the numerous population of faint field stars in the region, and for two groups we introduced a maximum $\Delta(\GBPmGRPc)$ in addition to the
minimum one.
$\Delta(\GBPmGRPc)$ is the value used as a displacement in color with respect to a reference isochrone used to filter out field stars, see the bottom left plots in Appendix~A of
Villafranca~I where both the reference and displaced isochrones are plotted.

The last filtering step of the process is the 3-sigma cut in normalized parallax, which is dependent on the parallax zero point $Z_{\rm EDR3}$ and on the multiplicative constant $k$ used to
convert from internal to external parallax uncertainties (\citealt{Maizetal21c} and Table~\ref{notation}). The first effect should be small, as for most sources the variations in $Z_{\rm EDR3}$
are small compared to \sigmai. The second effect should also be small unless $k$ is off by $\sim 50\%$, the amount necessary to shift 2-sigma deviations to 3~sigmas. 
To address these issues we followed a two-step process.
First, we used the zero point of \citet{Lindetal21b} and the $k$ value of \citet{Maizetal21c} to perform an initial selection for each group. As described in the companion paper, in that selection 
there were eleven stars in the 26 Villafranca OB groups that were highly likely to be members but had been excluded based on the normalized parallax criterion. Those eleven stars were manually added 
to the sample in the companion paper and the whole sample (from the OB groups plus six globular clusters, the SMC, and the LMC) was used to derive a new $Z_{\rm EDR3}$ and a new $k$. Those two were
used in the second step to re-derive the membership of the 26 Villafranca groups and those are the results presented here. We note that the eleven stars that had to be manually added in the companion
paper satisfy the normalized parallax criterion in the second step. Also, the $k$ calculated in the companion paper applies strictly only to parallax uncertainties
(note that the high-RUWE and 6-parameter solutions cases have a special treatment) but for lack of an alternative we also apply it to proper motion uncertainties, as both are calculated as part of a joint 
astrometric solution so, to first degree, one would expect $k$ to be similar for both.

Table~\ref{filters} lists the field sizes and filters used for the sample and is the equivalent to Table~2 in Villafranca~I. For \VO{001} to \VO{016} most of the values are similar to those in the 
previous paper but some tweaks have been carried out to include a few additional known sources and to uniformize criteria. \VO{014} is a special case that is described below.

In summary, the distances to the stellar groups are based on a careful membership selection supervised with an interactive tool for each individual case using {\it Gaia}~EDR3 data. The final
individual parallaxes use the zero point derived in the companion paper. The distances use the Galactic disk prior of \citet{Maiz01a,Maiz05c} with the \citet{Maizetal08a} parameters but they are
quite insensitive to the choice of prior \citep{Pantetal21}.

%\subsection{Spectral classifications and AstraLux data}
\subsection{Spectral classifications}

$\,\!$\indent As we did in Villafranca~I, we use GOSSS and \lili\ spectral classifications whenever possible and literature ones otherwise to spectroscopically describe the most relevant
stars in the OB groups. The reader is referred to the GOSSS \citep{Sotaetal11a,Sotaetal14,Maizetal16} and \lili\ \citep{Maizetal19a,Maizetal19b} papers for details on how they are obtained 
and processed. In some cases, we present new GOSSS and \lili\ spectra and spectral classifications. \lili\ spectra are degraded to a spectral resolution of 2500 in order to compare them with 
the GOSSS standards. The new GOSSS spectral classifications are given in Tables~\ref{GOSSS_spclas_01}~and~\ref{GOSSS_spclas_02} and the \lili\ classifications in Table~\ref{LiLi_spclas}.
The new GOSSS spectrograms are shown in Figs.~\ref{GOSSS_spectra_01}~to~\ref{GOSSS_spectra_17} and the \lili\ spectrograms in Figs.~\ref{LiLi_spectra_01}~to~\ref{LiLi_spectra_06}.

\subsection{Literature distances}

$\,\!$\indent In Villafranca~I we did an extensive literature search for distances to \VO{001} through \VO{016} with the goal of comparing our values with those of other methods and
with other {\it Gaia}~DR2 distances. Having established such comparison, in this paper we concentrate on determining the changes in precision between the new {\it Gaia}~EDR3 distances and
the previous {\it Gaia}~DR2 ones. Nevertheless, for the new stellar groups (\VO{017} through \VO{026}) we also make comparisons with some selected literature distances.

\section{Results}

$\,\!$\indent In this section we describe our results for the 26 stellar groups in the sample. The calculated parameters are given in Table~\ref{results}. 
We first briefly cover the objects in the Villafranca~I sample. As those groups were 
extensively discussed in that paper, we simply indicate the most significative changes between {\it Gaia}~DR2 and EDR3 and present some new GOSSS and \lili\ results. For the 10 new 
objects we present a more extensive discussion of each.

\subsection{Sample from Villafranca I}

\subsubsection{The distant region of the Carina-Sagittarius arm} 

$\,\!$\indent NGC~3603 (\VO{001}) and Gum~35 (\VO{006}) were the two most distant groups in Villafranca~I. It was mentioned there that they could be physically related if 
they were at similar distances, as they are close in the sky and have similar extinctions. The uncertainties are much lower now and the results are still consistent with a 
very similar distance for the two around 7~kpc. Therefore, it is possible that the two are part of a previously unrecognized distant OB association.

For \VO{001} we present a \lili\ spectrum of the B1.5~Ia supergiant NGC~3603~HST-5 (a.k.a. Sher~25), see \citet{Tayletal14} for the original data. For \VO{006} the only new 
spectrum is for ALS~\num{18552}, which is just outside the search radius for the cluster but is most likely a cluster member. 
It is a SB2 system with spectral types O6.5~IV((f))~+~B0:~V. We have detected a typo in the \VO{006} subsection in Villafranca~I: ALS~2063 was correctly identified in Table~1 
but was misnamed as ALS~2067 in the text. 

\subsubsection{The Carina nebula association} 

$\,\!$\indent With the {\it Gaia}~EDR3 results, Trumpler~14 (\VO{002}) and Trumpler~16~W (\VO{003}) have parallaxes still consistent with being at the same 
distance but now the uncertainties are much lower, $\sim60$~pc. 
The distance is also consistent with the geometric value for $\eta$~Car of 2350$\pm$50~pc of \citet{Smit06a}. Below we analyze the case of the Trumpler~16~E (\VO{025})
subassociation. 

We present in Fig.~\ref{GOSSS_spectra_01} a new GOSSS spectrum for HD~\num{93129}~B with a spectral classification of O3.5~V((fc))z, where the c suffix 
\citep{Walbetal10a} has been added thanks to a cleaner spatial separation with the A component. In Fig~\ref{LiLi_spectra_01} we show a clean velocity separation of HD~\num{93161}~A into 
two components, O7~V((f)) and O9~IV, using \lili\ data. 
In Figs.~\ref{GOSSS_spectra_01}~and~\ref{GOSSS_spectra_02} we present the spectra for eight early-B stars in \VO{002} and one in \VO{003}. The most 
interesting one is ALS~\num{15203} (or CPD~$-$58~2608~B, note it is not selected as a \VO{002} member by the algorithm because it has a RUWE of 1.69), 
a previously unrecognized B+B+B SB3 system with a visual companion, ALS~\num{15204} (or CPD~$-$58~2608~A), which is an O+O 
SB2 system \citep{Maizetal16}, a good example of the high-order multiplicity achieved by some massive stars. Another B+B SB2 system is ALS~\num{15219}.
One object in \VO{002}, ALS~\num{15863}, is one of the eleven stars included in the second step of the membership selection process (see above).

%\paragraph{\VO{004}.} Westerlund~2.

\subsubsection{The nearby region of the Carina-Sagittarius arm} 

$\,\!$\indent Two Villafranca~I groups are located at similar distances in the closest region of the Carina-Sagittarius arm: Pismis~24 (\VO{005}) and M17 (\VO{009}). With {\it Gaia}~EDR3
data we derive similar distances for both despite being $\sim22^{\rm o}$ apart in the sky.

%Pismis~24, GOSSS spectrum for WR~93 (RUWE=1.41) not shown because missing lines for WC star.
Figure~\ref{GOSSS_spectra_02} includes a spectrogram of the interesting star ALS~\num{19692}, located in the halo of \VO{005} (i.e. just outside the search radius) and with a borderline RUWE of
1.40. The star has a variable spectrum with \NIV{4058} in emission and \NVd{4604,4640} in absorption that indicates the presence of a very early-type star and that is incompatible with the observed
\HeI{4471}/\HeII{4542} ratio if it were a single star (in \citet{Maizetal16} we classified it as O5.5~IV(f)). The spectrum must be composite despite the non-detection of double lines and we assign 
it an (uncertain) O3:~III(f)~+~O7:~III((f)) classification. If confirmed, the primary would be a new member of the rare O3 subtype. 

Figure~\ref{GOSSS_spectra_03} includes four O stars in the \VO{009} field of view that had no prior GOSSS classifications but with different degrees of certainty regarding the group membership.
BD~$-$16~4818 is classified as O9.5~V and had been identified as O8.5~V by \citet{Crametal78} but it has a very large RUWE and other indications of poor astrometry in {\it Gaia}~EDR3, so the
adscription to M17 is unclear. ALS~\num{19612} is a clear member and is classified here as O6.5~V((f))z with GOSSS data (it had previously received an O6:~V classification by \citealt{Crametal78}).
ALS~\num{19611} also has poor {\it Gaia}~EDR3 astrometry which does not allow us to confirm its \VO{009} membership and a relatively noisy GOSSS spectrogram yields an O7:~V(n) spectral classification
(\citealt{Hoffetal08} gave O6~V). Finally, ALS~\num{19608} is a borderline exclusion by the membership algorithm (normalized parallax of 3.27) with a GOSSS spectrogram that yields O9.5~V
(\citealt{Hoffetal08} gave O9~V). 

\subsubsection{Cygnus OB2} 

$\,\!$\indent In Villafranca~I we analyzed the two clusters in Cyg~OB2, Bica~1 (\VO{007}) and Bica~2 (\VO{008}). With the {\it Gaia}~EDR3 parallaxes, they have moved slightly closer and their 
measurements are consistent with the two being at same distance. 

Figure~\ref{GOSSS_spectra_02} includes three stars in Cyg~OB2 without previous GOSSS spectral classifications but analyzed by \citet{Kimietal07}. One of them, ALS~\num{15128}, is in 
\VO{007} and was classified as O8~V by \citet{Kimietal07}. Here we give it a slightly earlier classification of O7.5~Vz. The other two, ALS~\num{15123} and Cyg~OB2-23, are in \VO{008} and their
previous classifications were O9~V in both cases. The GOSSS classifications are slightly later: O9.5~V(n) and O9.5~V, respectively.

\subsubsection{\VO{010} = NGC 6193} 

$\,\!$\indent We present spectral types for three early-B stars in \VO{010} in Fig.~\ref{GOSSS_spectra_03}. HD~\num{150041} is a group member relatively far from the core, so it could have been born in an 
earlier stellar generation. We classify it as B0~II, the same spectral type as in \citet{Feasetal55}. The other two stars, CPD~$-$48~8705~and~CPD~$-$48~8704, are B dwarfs.

\subsubsection{\VO{011} = Berkeley 90} 

$\,\!$\indent In this group it is worth mentioning three stars. 2MASS~J20350798+4649321 was classified as O8~V by \citet{MarcNegu17}, who noted that its IPHAS photometry indicated H$\alpha$ emission. In our 
blue violet spectrum (Fig.~\ref{GOSSS_spectra_04}) there is also incipient H$\beta$ and \NIIIt{4634,4641,4642} emission. We tentatively assign it the spectral type O8~IV((f))e and suggest 
future spectroscopic observations to detect if \CIIIt{4647,4650,4651} is variable and can appear in emission. If that were the case, it would make this object a member of the rare magnetic Of?p class, of which 
there are only six known Galactic examples (the last addition being $\theta^1$~Ori~Ca,Cb, \citealt{Maizetal19b}). However, 2MASS~J20350798+4649321 has a RUWE of 6.73 in {\it Gaia}~EDR3, so its adscription to the
cluster (which has another two O-type systems, \citealt{Maizetal15a}) is likely but tentative at this stage. The second star is ALS~\num{11448}~B (or LS~III~$+$46~11~B), that was classified as B4~V in \citet{Maizetal15b}
but not formally included in GOSSS at that time. For the third source, IRAS~20342$+$4645, we do not have a spectrum but its parallax, location (just outside the search radius), and proper motion suggest that it was
ejected from the cluster $\sim 130$~ka ago. The source is relatively faint in the optical but very red and becomes brighter than any cluster member in the IR, making it a possible massive pre-main-sequence (PMS) runaway 
star. This hypothesis requires confirmation, as with the data at hand it could also be a late-type star unrelated to \VO{011}. 

\subsubsection{\VO{012} = NGC 2467} 

$\,\!$\indent For the two components of this double cluster (\VO{012}~S or Haffner~18 and \VO{012}~N or Haffner~19) we obtain {\it Gaia}~EDR3 distances with much lower uncertainties than in Villafranca~I and confirm that
they are physically associated, with a difference of less than one sigma. In Villafranca~I we indicated that the two earliest O-type systems, HD~\num{64568} and HD~\num{64315}~A,B appear to have been ejected from \VO{012}~S 
in nearly opposite directions $\sim400$~ka ago, a hypothesis that is still consistent with the {\it Gaia}~EDR3 astrometry. We increased the search radius of \VO{012}~S to include the cluster halo located to the south. 
Doing so, a new O~star is added to the census, CPD~$-$26~2711 (Fig.~\ref{GOSSS_spectra_04}). We obtain a spectral type of O7~Vz and \citet{Lode66} had previously classified it as O7. Another star in \VO{012}~S, ALS~832, is
an early-type B with a RUWE of 1.78 that is classified here as B0.5~V(n). 

%\paragraph{\VO{013}.} Sh~2-158.

\subsubsection{\VO{014} = North America nebula = Bermuda and Gulf of Mexico clusters} 

$\,\!$\indent The most relevant result in Villafranca~I regarding \VO{014} is that there is no significant stellar group around the Bajamar star, the main source of ionizing photons of the North America and Pelican nebulae.
However, while the paper was being reviewed, another analysis by \citet{Kuhnetal20} proposed that the Bajamar star was ejected as a walkaway star \citep{Renzetal19b} from the most significant cluster in the region, their 
group D located in the North Atlantic Ocean and Pelican regions, to the NW of the star (which can be dubbed the Bermuda cluster, following the geographical nomenclature of the North America nebula). {\it Gaia}~DR2 astrometry 
placed the other known O star in the field, HD~\num{199579} (which can be dubbed the Toronto star, following the geographical nomenclature of the North America nebula), beyond the North America nebula. The Toronto star
has a GOSSS spectral classification of O6.5~V((f))z but is a spectroscopic binary with a weak companion, likely of early-B type \citep{Willetal01}.

In our analysis here we define two regions in the North America nebula, \VO{014}~NW and \VO{014}~SE, which correspond to groups D and E, respectively, in \citet{Kuhnetal20}, the latter being a smaller cluster in the Gulf 
of Mexico region of the nebula. We derive group parallaxes of $1.254\pm0.010$~mas and $1.344\pm0.016$~mas for the two regions, respectively, values that are slightly larger than the uncorrected {\it Gaia}~DR2 ones of 
\citet{Kuhnetal20} but compatible after applying a DR2 zero point of $40\pm 10$~\microas\ (and similar to the more recent {\it Gaia}~EDR3 value of \citealt{KuhnHill20}). The derived distances to \VO{014}~NW and \VO{014}~SE 
are $798\pm 6$~pc and $744\pm 9$~pc, respectively, confirming the finding by \citet{Kuhnetal20} that the young stars in the Gulf of Mexico are located closer to us than the ones in the Bermuda cluster.

In a separate paper 
\citep{Maizetal21f}
we analyze the relationship between the Bajamar and Toronto stars with the Bermuda cluster in the light of the {\it Gaia}~EDR3 data. There, we propose that
the two O-type SB2 systems have been ejected as walkaways from \VO{014}~NW along with other stars ejected as walkaways or runaways, leaving this group as an ``orphan cluster'' without its more massive stars due to the dynamical 
interactions among them \citep{Ohetal15}. We also detect three objects, 2MASS~J20531694$+$4424298, 2MASS~J20522503$+$4437562, and Tyc~3179-00756-1 which could be O or B0 stars (the first two are cluster members and the last one 
a walkaway star) but for which we do not have spectroscopic confirmation at this point.

%Tyc 3179-01133-1: Brightest member of NW at B3 V (to verify) but RUWE=2.87
%Bajamar star: $\pic = 1.3301\pm0.0401$~mas, $d=754^{+24}_{-22}$~pc.
%HD~\num{199579}: $\pic = 1.2207\pm0.0948$~mas, $d=840^{+74}_{-63}$~pc.

We present in Figs.~\ref{GOSSS_spectra_04}~and~\ref{GOSSS_spectra_05} the GOSSS spectra of four stars in this region. 
57~Cyg is the brightest object in the field and is a foreground twin SB2 B5~V~+~B5~V. Tyc~3179-01209 is located in Georgia (i.e. just to the north of
Florida) and moving quickly away from \VO{014}~NW, at whose position was located $\sim 300$~ka ago. Nevertheless, it is somewhat in front of the cluster and its GOSSS spectral type is A0~III so it does not appear to be a young 
star ejected from \VO{014}~NW. On the other hand, the last two objects, HD~\num{195965} and HD~\num{201795}, are confirmed early-B-type runaway stars that were ejected 1.61~Ma and 1.91~Ma ago, respectively, from the Bermuda cluster
\citep{Maizetal21f}.

%OLD%Possible ejection from 313.10, 44.57. 
%Possible ejection from 313.07, 44.58. 
%Distances: 
%OLD%56\arcmin\ for Bajamar star
%OLD%48\arcmin\ for HD~\num{199579} 
%53\arcmin\ for Bajamar star
%44\arcmin\ for HD~\num{199579} 
%Position differences with respect to possible ejection (fixed by proper motions):
%OLD%(+37\arcmin,-42\arcmin) for Bajamar star
%OLD%(+44\arcmin,+20\arcmin) for HD~\num{199579}
%(+32\arcmin,-42\arcmin) for Bajamar star
%(+39\arcmin,+20\arcmin) for HD~\num{199579}
%Proper motion differences with respect to cluster:
%OLD%(-0.205,-4.640)-(-1.629,-3.056) = (1.424,-1.584) for Bajamar star    (2.130)
%OLD%(+0.464,-2.129)-(-1.629,-3.056) = (2.093,+0.927) for HD~\num{199579} (2.289)
%(-0.205,-4.640)-(-1.365,-3.065) = (1.164,-1.575) for Bajamar star    (1.958)
%(+0.464,-2.129)-(-1.365,-3.065) = (1.829,+0.936) for HD~\num{199579} (2.054)
%Check:
%OLD% 37./-42. = -0.88, 1.424/-1.584 = -0.90
%OLD% 44./ 20. =  2.20, 2.093/ 0.927 =  2.26
% 32./-42. = -0.76, 1.164/-1.575 = -0.74
% 39./ 20. =  1.95, 2.093/ 0.927 =  1.95
%Travel times
%OLD% 56\arcmin/2.130 mas/a = 1.58 Ma for Bajamar star
%OLD% 48\arcmin/2.289 mas/a = 1.26 Ma for HD~\num{199579}
% 53\arcmin/1.958 mas/a = 1.62 Ma for Bajamar star
% 44\arcmin/2.054 mas/a = 1.29 Ma for HD~\num{199579}
% If we add 1 sigma in dec cluster proper motion (0.33 mas/a), agreement is much better -> possible original proper motion of CM

%\paragraph{\VO{015}.} Collinder~419, one new B star.

\subsubsection{\VO{016} = NGC 2264} 

$\,\!$\indent For the two components of this double cluster we measure a distance difference of just 4~pc but with a combined uncertainty of 8~pc. Their separation in the plane of the sky (22\arcmin) also
corresponds to 4~pc, so at this point we are on the verge of being able to peek into the 3-D structure of \VO{016}, something that may be possible with {\it Gaia}~DR4. However, we caution that 22\arcmin\ is
dangerously close to the half-amplitude of the checkered pattern in the {\it Gaia}~EDR3 parallaxes \citep{Maizetal21c}, so systematic effects will have to be improved in future {\it Gaia} data releases to
accomplish that goal. In Figs.~\ref{GOSSS_spectra_05}~and~\ref{GOSSS_spectra_06} we present GOSSS spectra for twelve~B and one A~star in \VO{016}, including two B+B SB2 systems, HD~\num{47732} and 
HD~\num{47755}, and one very fast rotator, HDE~\num{261903}. HD~\num{47755} is one of the eleven stars included in the second step of the membership selection process (see above). 
See Fig.~\ref{CMDs} for a {\it Gaia}~EDR3 CMD of \VO{016} that includes 15~Mon~Aa,Ab and B 
(the latter with an estimated \GBPmGRP\ from the photometry of \citealt{Maizetal19b} and the spectral types of \citealt{Maizetal18a}) 
and the two potential runaways from Villafranca~I.

\subsection{New groups}

$\,\!$\indent

\subsubsection{\VO{017} = Heart~nebula = IC~1805 = Sh~2-190 = W~4}

$\,\!$\indent The region formed by the Heart (IC~1805 or W~4), Soul (IC~1848 or W~5), and Fishhead (IC~1795 or W~3) nebulae is the most prominent optical \HII\ complex in the second Galactic quadrant and together they
form the Cas~OB6 association \citep{GarmSten92,Carpetal00,Megeetal08}. Here we analyze only the stellar group in IC~1805 and leave the other two and their relationship for future Villafranca papers, noting only that
IC~1795 is the youngest of the three. The Heart nebula itself is a cavity created by the stellar group at its center. There are nine known
O stars in \VO{O17}, with the earliest ones being HD~\num{15570} (one of the O4~If spectral classification standards, \citealt{Maizetal16}), HD~\num{15558}~A (an SB2 that is part of a complex system with thirteen
components in the WDS catalog, \citealt{Masoetal01}, some of them likely chance alignments from the cluster, \citealt{Maizetal19b}), and HD~\num{15629} (O4.5~V((fc)), another spectral classification standard, 
\citealt{Maizetal16}).

In the GOSSS papers we have published the spectra and spectral types of eight of the nine known O-type systems in \VO{017}. The missing one is KM Cas, an O9.2~IVn star (Fig.~\ref{GOSSS_spectra_07}) that was previously 
classified as O9.5~V((f)) by \citet{Massetal95a} and that is relatively far from the cluster center and close to the western edge of the cavity drawn by the Heart. In 
Figs.~\ref{GOSSS_spectra_07},~\ref{GOSSS_spectra_08},~and~\ref{LiLi_spectra_01} we also show GOSSS and \lili\ spectra for seven stars in \VO{017}, including one supergiant (BD~$+$60~493), two B+B SB2 systems (BD~$+$60~496
and ALS~7320), and one fast rotator (ALS~\num{15320}).

\VO{017} has a core surrounded by a rich halo. The core itself is not very prominent and the halo is just the central region of the Cas~OB6 association, with IC~1795 towards the NW and the more distant IC~1848 towards the
SE. The group population is not well differentiated in proper motion from the background disk population around 3-4~kpc that dominates for fainter stars and this manifests in the relatively large number of stars rejected 
by the normalized parallax criterion. Indeed, it is possible that there is a slight remaining contamination in the final sample that would artificially push the measured distance out but, if present, we estimate that to be 
a small effect ($\sim$20~pc). Of the nine O-type systems, the membership algorithm selects eight, with the missing one (HD~\num{15558}~A) rejected due to its large RUWE and \sigmae, and the additional five not mentioned 
above being BD~$+$60~497, BD~$+$60~498, BD~$+$60~499, BD~$+$60~501, and BD~$+$60~513.

The previous distance measurements to IC~1805 span a large range of values from 1.45~kpc \citep{Zucketal20} to 3.79~kpc \citep{FostMacW06}, to mention only results in the last two decades. Our value of 
$2075^{+44}_{-42}$~pc is in between those two and also between the distances of two analyses from isochrone fitting of optical photometry from the last decade: 1.7~kpc of \citet{Kharetal12} and 2.4~kpc of \citet{Sungetal17}. 
It is also within two sigmas of the two precise maser distances to the nearby IC~1795 \citep{Megeetal08}.

From the positions, parallaxes, and proper motions we detect six potential runaway stars from \VO{017}: ALS~7310, 2MASS~J02332978$+$6135238, 2MASS~J02344503$+$6125106, 2MASS~J02334807$+$6142036, 2MASS~J02334886$+$6141045, 
and 2MASS~J02331163$+$6116585. 
The first two have published spectral types of B1~V \citep{Ishi70} and G7~Ib (\citealt{ShiHu99}, possibly a PMS object), respectively. 

\subsubsection{\VO{018} = Lagoon~nebula = M8 = NGC~6523+6530 = Sh~2-25 = W~29}

$\,\!$\indent The Lagoon nebula (M8), one of the most famous \HII\ regions in the sky, contains at least two entries in the NGC catalog, NGC~6523 and NGC~6530, that we will use to refer to two subclusters (the first towards the W 
and the second towards the E but see \citealt{Tothetal08} for the confusing NGC nomenclature regarding the region, which includes two additional entries) that will be analyzed together here. The earliest type system in M8 is 
9~Sgr~A,B, which was classified as O4~V((f)) in \citet{Maizetal16} but is a low-velocity-separation eccentric SB2 \citep{Fabretal21b} composed of an O3.5~V((f*)) primary and an O5-5.5~V((f)) secondary according to 
\citet{Rauwetal12a}\footnote{The O4 classification is the reason why \VO{018} was not included in Villafranca~I.}. 9~Sgr~A,B is located in NGC~6523, the western and brightest part of M8, and is its main source of ionization.
Close to it we find Herschel~36, a high-order O-type multiple system \citep{Ariaetal10,SanBetal14,Campetal19}, that is the origin of the Hourglass nebula \citep{Maizetal15d} and is still partially embedded in its primordial cloud,
hence having a higher extinction and with a large value of \RV\ \citep{MaizBarb18}. Contrary to the hollowed-out appearance of the Heart nebula, M8 is still filled with nebulosity and actively forming stars \citep{BarbAria07}.
Another two O-type SB2 systems are present in \VO{018}, HD~\num{164816} and HD~\num{165052}, and discussed below. A fifth O star, HD~\num{164536}, is outside the search radius we use for \VO{018} and also has a poor RUWE and a large 
\sigmae. A sixth O star, HD~\num{165246} is also outside the search radius.

In Fig.~\ref{LiLi_spectra_01} we present new \lili\ spectra for HD~\num{164816} and HD~\num{165052}. In the first case the new spectral classification is O9.5~V~+~O9.7~V, which makes the secondary slightly earlier than in 
\citet{Sotaetal14}, where we classified it as B0~V. In the second case the new classification is O6~V((f))z~+~O7~V((f))z, where the primary remains as O6 as in \citet{Maizetal16} and the difference between the two components is
of one spectral subtype as in \citet{Ariaetal02} and \citet{Ferretal13}. The slight variations in spectral classifications arise in this case because of the introduction of a new classification criterion (the z suffix in this case, 
see \citealt{Ariaetal16}), the intrinsic difficulties in disentangling in velocity systems with maximum separations of just $\sim$200~km/s, and the existence of a Struve-Sahade effect \citep{Stic97,Ariaetal02,Lindetal07}. 
In addition to the above, in Fig.~\ref{GOSSS_spectra_08} we present GOSSS spectra for three B-type stars in \VO{018}.

\VO{018} has a peculiar structure. The largest concentration of stars (what may be termed the main core) is the eastern subcluster, NGC~6530, but three of the four O-type systems are in the western subcluster, NGC~6523, the fourth
one, HD~\num{165052}, is located further east in the halo and if we include HD~\num{164536} as a fifth O-type member it is even more distant. Nevertheless, it is a very rich OB group (with the second highest $N_*$ in 
Table~\ref{results}) and a population that is partially distinguished in proper motion with respect to the dominant background population in {\it Gaia}~EDR3, located at distances of 3-4~kpc. The distinction is partial because there
is some overlap which translates into a relatively large difference between $N_{*,0}$ and $N_*$. The other factor that helps differentiating the \VO{018} population from the background is its configuration, similar to that of
\VO{016} (NGC~2264), in which the extinction associated with the cluster is behind it (i.e. \VO{018} is on the front side of its natal cloud) and the global {\it Gaia}~EDR3 source density increases with the separation from the 
center instead of the other way around. That is the reason for introducing a maximum $\Delta(\GBPmGRPc)$ in Table~\ref{filters} for this group. The membership algorithm selects three of the four O stars, with 9~Sgr~A,B excluded 
only because of its large \sigmae.

Most recent distance determinations to \VO{018} place it at a distance around 1.3-1.4~kpc, see Table~6 in \citet{Aideetal18} where the authors own measurement is a clear outlier with a value of 2245$\pm$215~pc. Another recent
distance determination is that of \citet{Wrigetal19} at $1326^{+77}_{-69}$~pc. Our value here of 1234$\pm$16~pc places it somewhat closer and is in good agreement with our previous determinations of 1.25~kpc \citep{Ariaetal06} using 
photometry processed with CHORIZOS \citep{Maiz04c} and of $1169^{+67}_{-60}$~pc \citep{Campetal19} using {\it Gaia}~DR2 parallaxes. There is one potential runaway star in the {\it Gaia}~EDR3 data for \VO{018}: CPD~$-$24~6114. Also,
the O-type HD~\num{165246} may be a walkaway star.

\subsubsection{\VO{019} = Eagle~nebula = M16 = NGC~6611 = Sh~2-49 = RCW~165 = W~37}

$\,\!$\indent The Eagle nebula (M16) is another famous \HII\ region, pictured in the iconic ``pillars of creation'' HST image, which are the consequence of the different rates of photoevaporation of the molecular gas in the natal 
cloud as a function of density \citep{Hestetal96}. Similarly to \VO{016} (NGC~2264) and \VO{0018} (M8), the \HII\ region is a concave hole in the front wall of a molecular cloud carved by the action of the O stars, leading to 
differential (in amount and type) extinction effects as a function of sightline (Fig.~7 in \citealt{MaizBarb18}). \VO{019} hosts a rich population of massive and PMS stars \citep{Oliv08} with at least 12 O-type systems. The two
earliest ones are HD~\num{168076}~A,B (combined spectral type of O4~IV(f), the visual companion has no resolved spectral type yet, \citealt{Maizetal16}) and BD~$-$13~4923 (an SB2 system with spectral types O4~V((f))~+~ O7.5~V,
\citealt{Maizetal19b}). 

In Figs.~\ref{GOSSS_spectra_08}~and~\ref{GOSSS_spectra_09} we present GOSSS spectra for six O-type systems and three B~stars in \VO{019}. In Figs.~\ref{LiLi_spectra_01}~and~\ref{LiLi_spectra_02} we present \lili\ spectra for 
three of the same six O-type systems. Five of the O-type systems had not appeared in GOSSS or \lili\ before but all had prior classifications as O-type: BD~$-$13~4930 (O9.7~V here, O9.5~V in \citet{HiltIria55}), BD~$-$13~4929 
(an interesting SB3 system with a classification of O8~V~+~B0.5:~V~+~B0.5:~V previously identified and similarly classified by \citealt{Sanaetal09}), BD~$-$13~4928 (O9.7~IVnn here, O9.5~Vn in \citet{Hilletal93b}), ALS~4903
(O9~V here, O8.5~V in \citet{Hilletal93b}), and ALS~\num{15352} (O8~IV here, O8.5~V in \citealt{Hilletal93b}). The sixth O star is ALS~\num{15360}, classified as O7~V((f))z in \citet{Maizetal16} and here (with a better S/N
spectrogram) as O6.5~V((f))z. The three B~stars are ALS~\num{15369}, BD~$-$13~4921, and ALS~\num{15363}, the latter a peculiar object with narrow and deep lines that could be enriched in He.

%ALS~\num{15360}, new O spectrum here (currently listed as M20d). Also, ALS~\num{15352}, ALS~\num{4903}, BD~$-$13~4928, BD~$-$13~4930.
% Possible new B stars: ALS 15 369, BD -13 4921, ALS 15 363 (peculiarly narrow lines, He rich)

\VO{019} is a rich cluster with a well-defined core surrounded by a halo that is easily detected in the {\it Gaia}~EDR3 data despite the fact that the dominant population in the field has similar proper motions, even more so
than for the (relatively) nearby in the plane of the sky \VO{018}. The reason is that there are very few {\it Gaia}~EDR3 sources located beyond the cluster, as the associated molecular gas appears to be thick enough to create an
extinction wall that eliminates them from the sample. The OB stars in the sample have a broad range of colors, from \GBPmGRPc\ around 0.5 for the less extinguished one to values larger than 2.0 for the most extinguished ones
identified by spectral type. As noted by \citet{MaizBarb18}, extinction increases as we move from the core towards the NE and it is possible that some additional O stars remain undetected there. Of the known twelve O-type stars, 
the algorithm selects eight members: HD~\num{168075}, HD~\num{168137}~Aa,Ab, BD~$-$13~4927, BD~$-$13~4930, BD~$-$13~4928, ALS~4903, ALS~\num{15352}, and ALS~\num{15360}. The remaining four (HD~\num{168076}~A,B, QR~Ser, 
BD~$-$13~4929, and BD~$-$13~4923) have RUWE larger than 1.4. 
One object in \VO{019}, ALS~\num{15369}, is one of the eleven stars included in the second step of the membership selection process (see above).

The older distance estimations to \VO{019} had a wide range of values from 1.7~kpc to 3.4~kpc \citep{Oliv08} but values in the last two decades cluster in the narrower 1.6-1.9~kpc, a trend that follows that of the analysis for the
16 stellar groups in Villafranca~I that we presented in Table~5 there. Our value of 1697$^{+31}_{-30}$~pc is close to the middle of the latter range. The value is within one sigma of that of the nearby \VO{009} ($2.4^{\rm o}$ or
70~pc), so M16 and M17 are likely to be physically related. \VO{019} also has a rich stellar population but the relative orientation of the stellar group with respect to its natal molecular cloud (sideways rather than in front)
significantly increases the internal extinction and makes it harder to study. We detect two possible walkaways/runaways in \VO{019}: 2MASS~J18185057$-$1347338 (G5~III, \citealt{Evanetal05}), and 2MASS~J18182988$-$1336248 
(B9~III, \citealt{Evanetal05}).

\subsubsection{\VO{020} = Rosette~nebula = NGC~2244 = Sh~2-275 = W~16}
%   2 - HD 46 150               - Ma  - O5    V((f))z                     - OK
%   3 - HD 46 223               - Ma  - O4    V((f))                      - OK
%   6 - HD 46 149               - Ma  - O8.5  V                           - OK, K1+K2 too small for SB2 separation at R=2500
%   9 - HD 46 106               - Ma  - O9.7  V          + B2:   Vn       - OK, it is M21c in GOSC (LiLiMaRlin), new SB2, not SB2 in GOSSS
%  10 - HD 46 202               - Ma  - O9.2  V                           - RUWE=4.98,spi=0.376
%  12 - HD 46 056               - Ma  - O8    Vn                          - OK
%  14 - HDE 259 135             - S07 - B0    V          + B1:   V        - OK, B0 V + B1: V in LiLiMaRlin
%  24 - HDE 259 105             - S10 - B1    V                           - OK
%  29 - HDE 259 481             - S07 - B1.5: V(n)e                       - RUWE=1.66+dispm
%  31 - HDE 258 691             - S05 - O9.5  V                           - RUWE=6.97,disr=1517,spi=0.247,runaway. O9 V, Turner (1976), GOSSS+LiLiMaRlin
%  36 - ALS 8988                - S10 - B1    III                         - OK
%  82 - HD 46 056 B             - S07 - B2    V                           - original npi=3.38, fixed with new k

$\,\!$\indent The Rosette nebula (NGC~2244) is the prototype \HII\ region that has been hollowed out by a central cluster with O stars, a stage that shares with \VO{017} (IC~1805) and that takes place after the one in which 
the \HII\ region is still filled (e.g. \VO{018} and \VO{019}). Such a temporal evolution is also seen in more massive clusters \citep{MacKetal00,Maizetal04a} but what makes the Rosette nebula special is its excellent circular symmetry 
in the optical emission lines, which is nevertheless broken in CO and the IR dust emission with an extension towards the SE \citep{RomaLada08}. \VO{020} contains seven O stars: HD~\num{46223} (the earliest one at O4~V((f)),
\citealt{Sotaetal11a}), HD~\num{46150}, HD~\num{46149}, HD~\num{46106}, HD~\num{46202}, HD~\num{46056}, and HDE~\num{258691}.

We present GOSSS and \lili\ spectrograms for two O stars and three B stars in \VO{020} in Figs.~\ref{GOSSS_spectra_11}~and~\ref{LiLi_spectra_02}. HDE~\num{258691} had not appeared in previous GOSSS 
papers and is here classified as O9.5~V, having been previously identified as O9~V by \citet{Turn76}. HD~\num{46106} was classified as O9.7~III(n) in \citet{Sotaetal14}. We have analyzed ten \lili\ epochs to determine that 
the lines (of the primary) move with a peak-to-peak amplitude of $\sim$150~km/s and we have used the epochs near the extremes to detect the weak (and broad) lines of the secondary, whose peak-to peak amplitude appears to be 
$\sim$400~km/s, identifying this system as an SB2 for the first time. We obtain a spectral classification of O9.7~V~+~B2:~Vn, making this another example of how the combined spectral type of a late-O dwarf and an early-B dwarf
appears to be a late-O star of higher luminosity class and of how in such systems the secondary tends to be a fast rotator \citep{Walbetal14,Maizetal18a,Maizetal21b}. One of the B stars, HDE~\num{259135} is an eclipsing SB2
binary \citep{Hensetal00} for which we derive a spectral classification of B0~V~+~B1:~V and for which we confirm its membership to \VO{020}.
% Possible new B stars: HDE 259 135, HDE 259 481, HD 46 056 B

\VO{020} is relatively easy to detect in the {\it Gaia}~EDR3 proper motions with respect to the dominant background population, located at distances of 3-5~kpc. The selected members form a cluster core that extends towards the
SE, similarly to what is seen in CO and the IR dust emission, with another minor extension towards the SW and a halo around the core. Five of the seven O stars are selected as members by the algorithm, with the exceptions being 
HD~\num{46202} and HDE~\num{258691}. The first case can be explained by its RUWE of 4.98 and its \sigmae\ of 0.376~mas. HDE~\num{258691} is more interesting: it also has a large RUWE but it is also excluded for being far from the 
core (indeed, it is outside our search radius). However, its proper motion points away from the cluster core, suggesting it was ejected as a runaway from there 1.0-1.5~Ma ago.
One object in \VO{020}, HD~\num{46056}~B, is one of the eleven stars included in the second step of the membership selection process (see above).

Previous distance measurements to \VO{020} place the cluster at values of 1.4-1.7~kpc \citep{RomaLada08}, with {\it Gaia}~DR2 values concentrating at 1.55-1.59~kpc \citep{CanGetal18,Kuhnetal19,Muzietal19}. Our 
{\it Gaia}~EDR3 value of 1421$^{+21}_{-20}$~pc is significantly lower that the latter ones but we note that if we process the {\it Gaia}~DR2 data for \VO{020} in a manner similar as that in Villafranca~I we obtain 
1465$^{+70}_{-64}$~pc, so at least part of the difference must involve the sample selection and/or the treatment of systematic effects. Nevertheless, the usually reliable distance method of eclipsing binaries yields 
1.39$\pm$0.10~kpc for HDE~\num{259135} \citep{Hensetal00} which, as we pointed out above, is confirmed as a cluster member. That distance measurement is in excellent agreement with both of our {\it Gaia}~DR2~and~EDR3 values.

\subsubsection{\VO{021} = NGC~2362 = $\tau$~CMa~cluster}

\begin{figure*}
 \centerline{\includegraphics[width=0.33\linewidth]{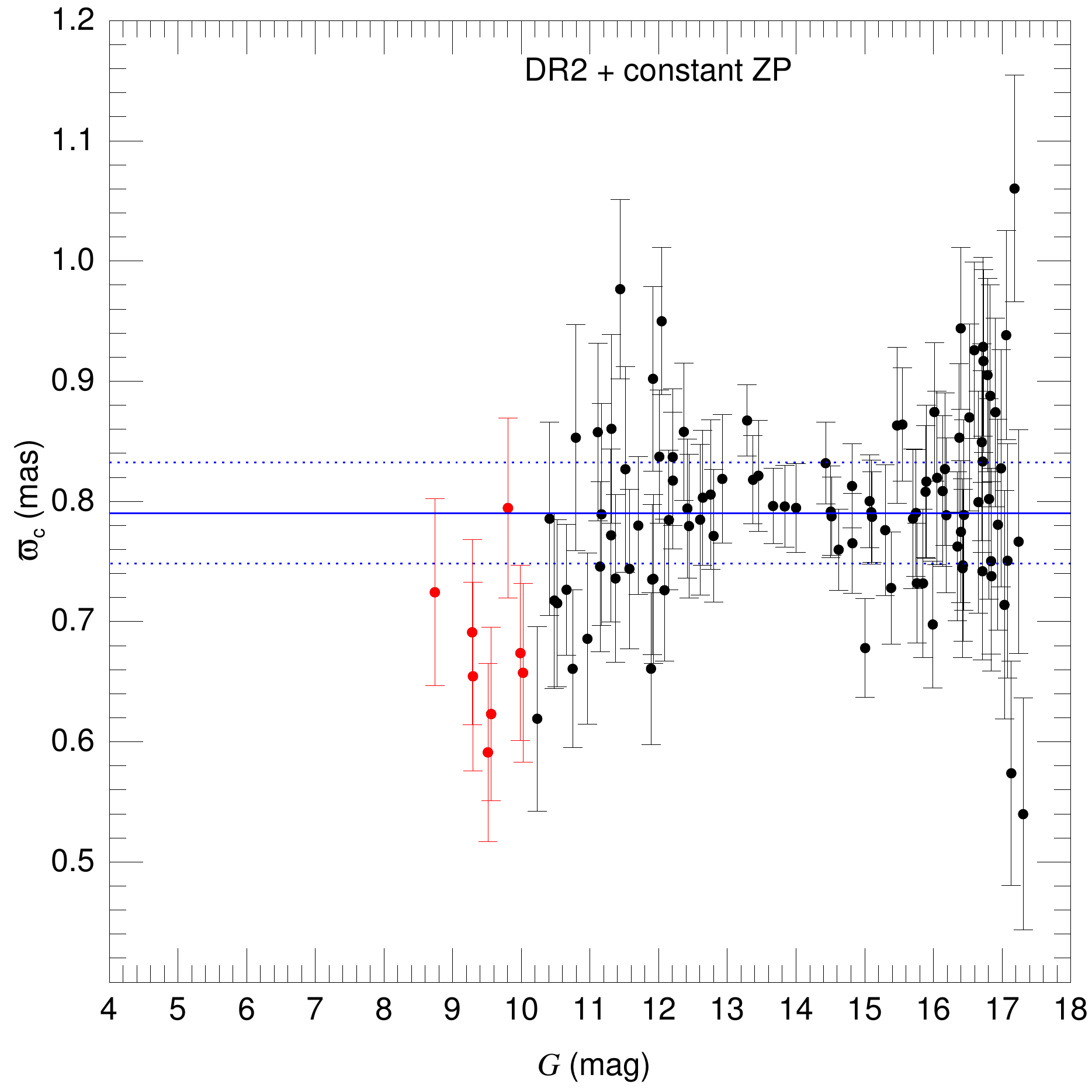} \
             \includegraphics[width=0.33\linewidth]{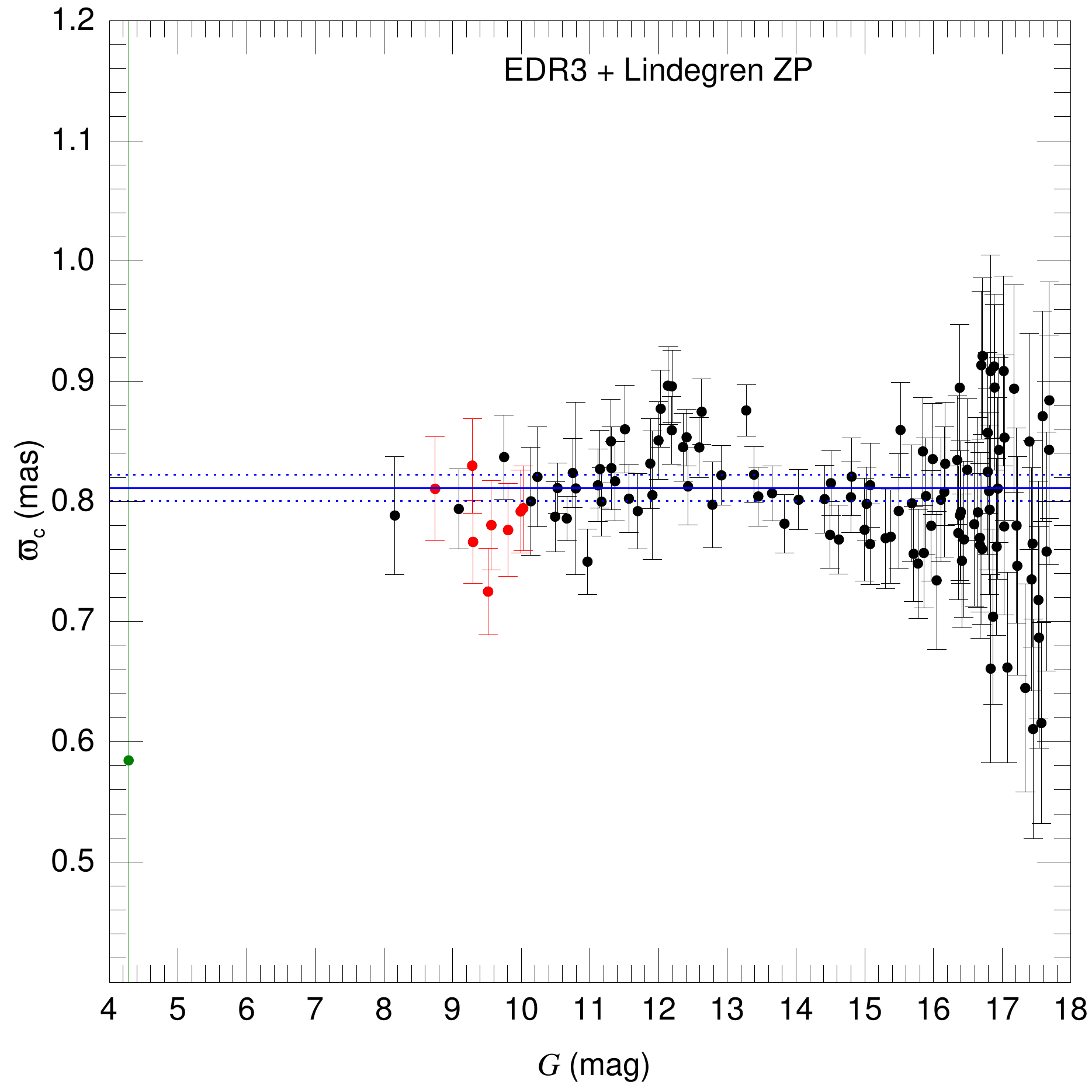} \
             \includegraphics[width=0.33\linewidth]{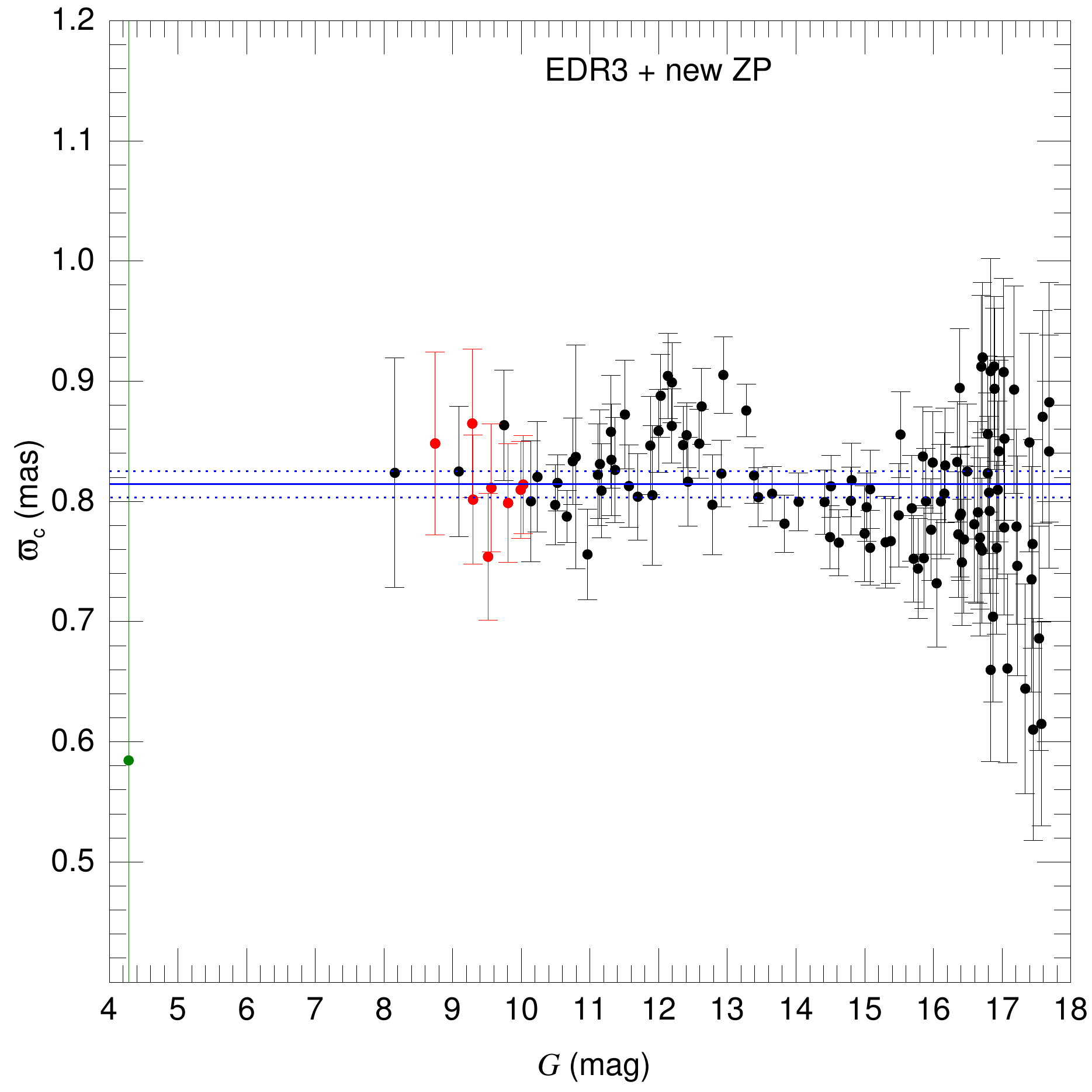}}
 \caption{Corrected individual parallaxes as a function of \GG\ for the \VO{021} (NGC~2362) stars selected by the membership algorithm using parallaxes from (left) {\it Gaia}~DR2 with $Z_{\rm DR2} = -40$~\microas, 
          (middle) {\it Gaia}~EDR3 with the Lindegren $Z_{\rm EDR3}$ and the $k$ from \citet{Maizetal21c}, and (right) {\it Gaia}~EDR3 with the new $Z_{\rm EDR3}$ and the $k$ from the companion paper. Red symbols are used for 
          the eight stars used to calculate the distance to the cluster in \citet{MaizBarb20}. The green point with the large error bar 
          near the left edge
          in the {\it Gaia}~EDR3 plots corresponds to (the not selected) $\tau$~CMa~Aa,Ab 
          (the star is not present in {\it Gaia}~DR2). The horizontal blue lines show the (corrected) group parallax (with the one sigma range plotted with dotted lines) calculated with each dataset.}
 \label{NGC_2362}
\end{figure*}

$\,\!$\indent NGC~2362 is different from the previous four stellar groups in that it has already dispersed the gas around it. Therefore, it is at an even later morphological evolutionary stage than 
\VO{017} (IC~1805) or \VO{020} (NGC~2244). Its older age of $\sim$5~Ma (see below) 
also manifests in most of its PMS population having already lost their circumstellar disks \citep{Dahm08b}. \VO{021} is dominated by its central multiple system, $\tau$~CMa. $\tau$~CMa~Aa,Ab is a 
visual binary with spectral types O9.2~Iab: and O9~III \citep{MaizBarb20} but the Aa,Ab pair hides two additional components, as one of the O stars (likely Aa) is an SB1 \citep{Sticetal98} and there are eclipses with a 
different period \citep{vanLvanG97}. Those two are the only O stars in the cluster (which has a lower mass than the previous four) and, barring any undetected compact objects, the only two stars to have significantly evolved 
from the main sequence. The earliest B-type star in \VO{021} is $\tau$~CMa~D, with a classification of B0.7~V \citep{MaizBarb20}. We do not present new GOSSS and \lili\ spectra for \VO{021}, as we recently did it in that paper.

The population of \VO{021} is easily distinguished from the rest of the stars in the {\it Gaia}~EDR3 field, as it is a rich and centrally concentrated cluster with a proper motion very different from the dominant background 
population at a distance of $\sim$5~kpc ({\it Gaia} peers into the distant outer Galactic disk along this sightline) and with well defined main- and pre-main-sequences in the CMD (Fig.~\ref{CMDs}). $\tau$~CMa~Aa,Ab (a single source 
in {\it Gaia}~EDR3) is not selected by the membership algorithm due to its RUWE (and other issues with its astrometry) but the early-type B stars $\tau$~CMa~D, CPD~$-$24~2213, and $\tau$~CMa~B are. HD~\num{57192} appears to be a 
foreground object while $\tau$~CMa~C (an extremely fast rotator classified as B5~Vnnn in \citealt{MaizBarb20}) appears to be a background star. We detect two possible walkaways/runawys from this cluster: CPD~$-$24~2223 and 
CPD~$-$24~2235.

\citet{Dahm08b} lists prior distances to NGC~2362 that range from 904~pc \citep{Hump78} to 2090~pc \citep{JohnMorg53}. In \citet{MaizBarb20} we used eight members to derive a {\it Gaia}~DR2 distance of $1.50^{+0.12}_{-0.10}$~kpc, 
which is significantly longer than the 1227$^{+17}_{-16}$~pc here and with much worse uncertainties. Analyzing the difference between the two results is instructive for the purposes of understanding results with different 
{\it Gaia} datasets and methods (Fig.~\ref{NGC_2362}). The lower uncertainty in this paper is easy to understand: {\it Gaia}~EDR3 has lower random and (uncorrected) systematic uncertainties and the new sample is significantly larger
(116 vs. 8), allowing us to reach the minimum uncertainty allowed by the angular covariance \citep{Maizetal21c}. The significant reduction in distance can also be explained: bright, blue stars are the ones most affected by the 
(non-applied) differential zero point in {\it Gaia}~DR2 (Fig.~4 in Villafranca~I) and all of the eight stars in \citet{MaizBarb20} happened to be in that category. If instead of using only those eight stars we apply the same 
selection criteria to the {\it Gaia}~DR2 data as we have done in this paper, we obtain 108 stars and derive a distance of 1267$^{+73}_{-65}$~pc, which is in good agreement with the new {\it Gaia}~EDR3 value. The lesson here is that 
systematic effects can be important in {\it Gaia} data and need to be accounted for. We have also included a middle panel in Fig.~\ref{NGC_2362} with the result of applying the Lindegren zero point instead of the new one to show 
how the new solution reduces the uncorrected parallax zero point for the eight bright stars in \citet{MaizBarb20}.

\subsubsection{\VO{022} = NGC~6231 = Collinder~315} 

$\,\!$\indent NGC~6231 is the richest (in OB stars) cluster among the ten new groups added to the Villafranca sample in this paper, especially if we consider it as the nucleus of the surrounding Sco~OB1 association (but would be
surpassed by Car~OB1 or Cyg~OB2 if we considered associations rather than clusters within 2.5~kpc). The cluster is significantly evolved, with the earliest isolated O-type system being HD~\num{152233}~Aa,Ab (O6~II(f), 
\citealt{Sotaetal14}) and with a Wolf-Rayet present (WR~79), which has an O star as a companion (see below). From a morphological point of view, it is in an evolutionary state intermediate 
between that of \VO{020} (NGC~2244) and that of \VO{021} (NGC~2362): 
the cluster has created a cavity so large around it that the remaining H$\alpha$ nebulosity is at a long distance and only partially surrounds it \citep{Reip08d}. Counting WR~79 (a short period WC+O binary system) , there
are fourteen O-type systems in \VO{022}, most of them late-type O dwarfs and giants. There are diverging age measurements for the cluster (and association), likely due to the presence of different populations and of a real age 
spread \citep{Reip08d}, but the absence of O~dwarfs with subtypes earlier than O8 indicates that massive-star formation stopped 4-5~Ma ago.

% Possible new B stars: HD 152 234, HDE 326 333, CPD -41 7724, HD 152 235, HDE 326 330, V945 Sco, HD 152 076
%                       CPD -41 7721 B, CPD -41 7723. HDE 326 320, ALS 14 761, 2MASS J16540272-4150286
%WR~79, new WR spectrum.
%GOSSS: 1 WR + 12 B
%\lili: 1 WR + 4 O + 1 B

We present GOSSS and \lili\ spectrograms for one WR, four O-type and twelve B-type stars in \VO{022} in 
Figs.~\ref{GOSSS_spectra_10},~\ref{GOSSS_spectra_11},~\ref{GOSSS_spectra_12},~\ref{GOSSS_spectra_13},~\ref{LiLi_spectra_02},~\ref{LiLi_spectra_03},~and~\ref{LiLi_spectra_04}. 
The WR Catalogue\footnote{\url{http://pacrowther.staff.shef.ac.uk/WRcat/}.} \citep{RossCrow15} lists a spectral classification of WC7~+~O5-8 for WR~79. 
We show a GOSSS spectrogram of the star in Fig.~\ref{GOSSS_spectra_10} that confirms the WC nature of the primary but extends only to
5500~\AA, which is not enough to assign it a spectral subtype. The equivalent \lili\ spectrogram in Fig.~\ref{LiLi_spectra_03}, on the other hand, extends to 5930~\AA\ and allows us to access the C lines in that region and
confirm the WC7 subtype \citep{Crowetal98}. We use our spectral classification software MGB \citep{Maizetal12} to fit the diluted He absorption components in the GOSSS and \lili\ spectrograms (with \HeII{4542} significantly more 
intense than \HeI{4471}) to establish a classification for the secondary of O4~III-I(f). This is earlier than the previous spectral type and is a likely sign that the secondary has been rejuvenated by the mass transfer between the
two components. 

As for the O stars with new spectra in \VO{022}, the \lili\ data for HD~\num{152248}~Aa,Ab gives a spectral classification of O7~Ib(f)~+~O7~Ib(f), which is the same as the GOSSS one in \citet{Sotaetal14} but downgrading the
luminosity class of the primary from Iab to Ib. Another small change takes place for HD~\num{152218}, going from O9~IV~+~B0:~V: in \citet{Sotaetal14} to O8.5~IV~+~B0~IV here using a \lili\ spectrogram. Similarly, 
V1034~Sco goes from O9.2~IV~+~B1:~V in \citet{Maizetal16} to O9.5~IV~+~B1:~V here using \lili\ data. For CPD~$-$41~7733 we detect the weak secondary \citep{Sanaetal07a} in a \lili\ spectrogram and assign a classification of
O9~IV~+~B to the system.

The identification of the \VO{022} population in the {\it Gaia}~EDR3 data is somewhat paradoxical. On the one hand, this is such a rich and concentrated cluster that we obtain the highest value of $N_\star$ of all groups. On the 
other hand, its sightline has low extinction (for its Galactic coordinates) for several kpc (being quite empty of dust until its location and more patchy for longer distances) 
and there is little change in proper motion as a function of distance,
making it hard to distinguish between the cluster and background populations. For that reason, both a maximum $\Delta(\GBPmGRPc)$ and a \GGcmax\ are needed in Table~\ref{filters} for this group. WR~79 and all of the O~stars in 
\VO{022} but one are selected by the membership algorithm, the exception being HD~\num{152219} due to its high RUWE and \sigmae. Another two very bright early-B supergiants, HD~\num{152234} (B0.5~Ia) and HD~\num{152235} (B0.7~Ia) 
are also excluded by poor-quality astrometry.
Three objects in \VO{022}, HDE~\num{326331}~A, HD~\num{152314}~Aa,Ab, and HDE~\num{326330}, are among the eleven stars included in the second step of the membership selection process (see above).

The earlier distance determinations to \VO{022} placed the cluster at values between 1.8~kpc and 2.4~kpc \citep{Reip08d} but most determinations after the 1980s placed it somewhat closer. Our value of 1551$^{+25}_{-24}$~pc is
not too different from the average value of $\sim$1.6~kpc from the last decades but we note that the two {\it Gaia}~DR2 determinations by \citet{CanGetal18} and \cite{Kuhnetal19} yield somewhat longer distances than our 
{\it Gaia}~EDR3 value, as it also happens for \VO{020} (NGC~2244). We also note that the CMD hints of a double sequence for mid-B-to-early-A stars, with one that follows the isochrone that corresponds to the massive stars in the
cluster and another with a lower extinction. The second sequence also has a somewhat higher average parallax that corresponds to a $\sim$20~pc difference, so one possible interpretation is that it corresponds to foreground members of
Sco~OB1, in which case a significant fraction of the extinction we see for \VO{022} would be associated with the cluster itself.

\subsubsection{\VO{023} = Orion~nebula cluster = M42+M43 = NGC~1976+1982}

\begin{figure}
 \centerline{\includegraphics[width=\linewidth]{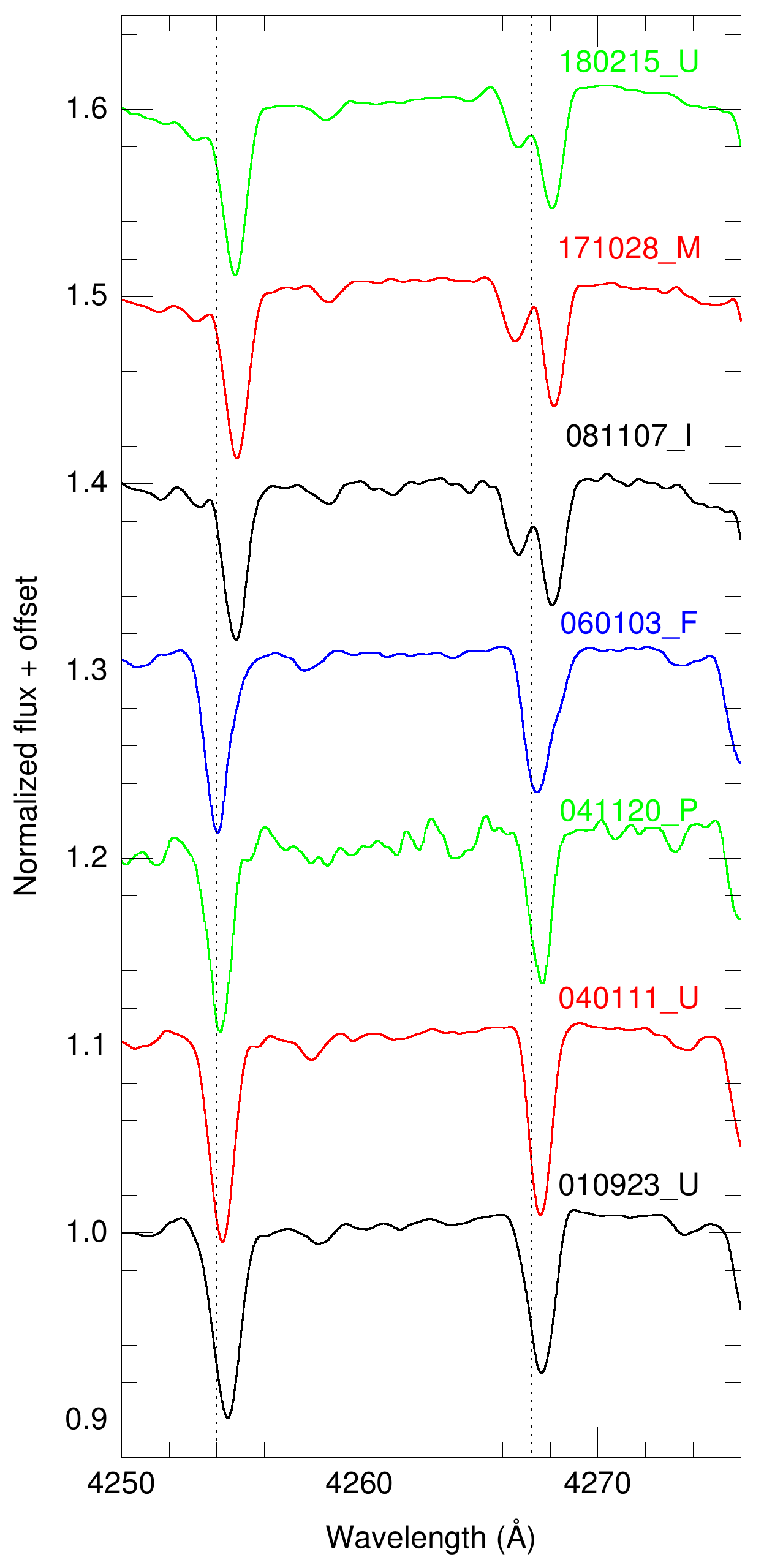}}
 \caption{Six \lili\ epochs of $\theta^1$~Ori~D degraded to a common $R\sim\num{10000}$. The two dotted lines indicate the rest wavelengths of \OII{4254} and \CII{4267}. The code for each spectra indicates the evening date as 
          YYMMDD and the source (U, P, F, I, and M stand for UVES, OHP, FEROS, FIES, and Mercator, respectively).}
 \label{theta1OriD}
\end{figure}

$\,\!$\indent The Orion nebula is arguably the most famous and studied \HII\ region due to its proximity, low foreground extinction, filled-in morphology, and favorable orientation with respect to the parent molecular 
cloud \citep{ODeletal08}. The nebula has the designation M42 (NGC~1976) and the underlying stellar group extends to the adjacent Mairan's nebula (M43 or NGC~1982). The cluster is usually called the Orion nebula cluster (ONC) and 
that is what we will refer to as \VO{023} here. \VO{023} is part of Ori~OB1d, the youngest subgroup of the Ori~OB1 association \citep{Ball08}, and has a rich and well-studied young stellar population \citep{Muenetal08}. The
core of the ONC is the Trapezium, which receives its name from the shape formed by the four bright systems $\theta^1$~Ori~A+B+C+D, each one of them a multiple \citep{Karletal18}. In the central region the dominant ionizing 
source is the O-type system $\theta^1$~Ori~Ca,Cb while further out a second O-type system, $\theta^2$~Ori~A, also contributes to the mix \citep{ODeletal17} and M43 is ionized by its central early-type-B system, NU~Ori 
\citep{Shuletal19}. $\theta^1$~Ori~Ca,Cb is a double (possibly triple) system \citep{Krauetal09b} whose primary (Ca) is a magnetic O7~f?p~var star \citep{Maizetal19b}. $\theta^2$~Ori~A is an SB2 system for which in 
\citet{Maizetal19b} we gave a GOSSS classification of O9.2~V~+~B0.5:~V(n) and a \lili\ classification of O9.5~V~+~B0.2:~V and also has a faint visual companion.

We present GOSSS and \lili\ spectrograms for nine B-type stars in \VO{023} in Figs.~\ref{GOSSS_spectra_13},~\ref{GOSSS_spectra_16},~\ref{LiLi_spectra_04},~and~\ref{LiLi_spectra_05}. Note that all seven \lili\ spectrograms have
GOSSS equivalents, something that is done in purpose to compare circular aperture \'echelle vs. long-slit single-order data. Everything else being equal, the first has the advantages of the better original resolution and the 
higher S/N at the (degraded) resolution of the single-order spectrograph. However, the second one has other advantages: easier rectification, possible separation of close components with profile fitting, and (most importantly 
in this case) better nebular subtraction. Amidst the strong nebulosity of the region, \lili\ spectra display \OIIId{4959,5007} and Balmer emission lines and, in some cases, partial He\,{\sc i} line infilling. Given the existence
of conflicting spectral types for some of the famous stars in the region we attempt to provide some clarity by giving details about the spectral classifications below:

\begin{itemize}
 \item $\theta^1$~Ori~D (= HD~\num{37023}) has a surprisingly broad range of spectral classifications, from O9.5~V: \citep{Slet63} to B1.5~Vp \citep{LeveLeis06}. Here we classify it as B0~V based on \HeII{4542} being weaker
       than \SiIII{4552} but not by much (if they were similar, the spectral classification would be O9.7, see \citealt{Sotaetal11a}) and on \HeII{4686} being stronger than \SiIII{4552}. Previous classifications 
       as O9.5 or B0.5 may be understandable (but still wrong) based on the diversity of data quality and classification criteria and on the different degrees of nebular contamination. However, anything later than B0.5
       is an obvious mistake, given the clear presence of He\,{\sc ii} absorption lines. Given that this system has at least three components \citep{Karletal18} and that one of them (Db) could be bright enough to contribute to the 
       spectral classification, do we see signs of its presence in the spectrograms? Indeed, in Fig.~\ref{LiLi_spectra_04} and even more so in Fig.~\ref{GOSSS_spectra_13} we see that \CII{4267}, a line that peaks around B2, 
       is enhanced with respect to what one would expect for a B0~V star. That prompted us to search for variations among the \lili\ epochs and the result is seen in Fig.~\ref{theta1OriD}: \OII{4254}, which originates in the
       primary, is always a single line that moves with a peak-to-peak amplitude of $\sim$60~km/s. \CII{4267}, on the other hand, becomes double with the two components having amplitudes of $\sim$60~km/s and $\sim$110~km/s
       (the Mercator epoch is the one in Fig.~\ref{LiLi_spectra_04} but the double lines are not seen there because of the lower spectral resolution of the plot). \citet{Vitr02} identified $\theta^1$~Ori~D as an SB1 (not an SB2, 
       as here) but his velocity amplitude does not agree with our measurements. If the double lines correspond to the two objects in the astrometric orbit determined by \citet{Karletal18}, the semi-major axis and period are in the
       correct ballpark but the low inclination is incompatible with the high velocity amplitudes. Our mass ratio is $\sim$0.55, which is not far from the \citet{Karletal18} value, and the deduced mass for the secondary (assuming 
       their total mass) would be 9-10~M$_\odot$, which is a reasonable value for a B2~V star. In any case, further data are needed to understand this system completely.
 \item $\theta^1$~Ori~A (= HD~\num{37020}) is classified here as B0.2~V, which is changed from the previous one of B0.5~V by \citet{SimDetal06} based on the reorganization of spectral types in \citet{Sotaetal11a} that moved 
       $\upsilon$~Ori from B0~V to O9.7~V and $\tau$~Sco from B0.2~V to B0~V (B0.2 stars of luminosity class V now have \HeII{4686}~$\sim$~\SiIII{4552}). The system includes two additional components (one astrometric and one 
       eclipsing) that are too faint to contribute to the spectrogram \citep{Karletal18}. We also note that some references (e.g. \citealt{Kreletal16c}) claim that this star is of O type, which is usually caused by a confusion 
       between the order in which components are named in most multiple systems (from bright to faint) and the way they are ordered in $\theta^1$~Ori (from west to east for A to D). 
 \item $\theta^1$~Ori~Ba,Bb (= HD~\num{37021}) is the faintest of the four systems in the Trapezium and the one with the most complex multiplicity, having at least six components \citep{Karletal18}. The subject of its (combined) 
       spectral classification is complicated by the fact that the two primary components (Ba and Bb in the WDS, B$_{1,5,6}$ and B$_{2,3}$ in \citealt{Karletal18}) are separated by 0\farcs9 and their relative contribution to 
       a spectrogram may depend on the setup and observing conditions. Nevertheless, that is not sufficient to explain the wide range of spectral classifications in the literature, which go from B1~V? \citep{Slet63} to B6n
       \citep{Trum31}. Here we classify $\theta^1$~Ori~Ba,Bb as B3~Vn using either GOSSS or \lili, though the latter has significant nebular contamination. The classification is consistent with a high-order multiple system with a 
       primary with $\sim$7~M$_\odot$ and the $\Delta G \sim 1.5$ magnitude difference observed with respect to A or D.
 \item NU~Ori (= HD~\num{37061}) is a quadruple system with three components (Aa, Ab, and C) that can be distinguished in high resolution, high S/N spectra moving one relative to each other \citep{Shuletal19}. NU~Ori~C is 
       interesting on its own as a magnetic star but the integrated light is dominated by Aa. Previous spectral classifications for the integrated spectrum are diverse but they concentrate around B0.5~V (with one claiming it
       is an O9~V star, \citealt{Bragetal12}). The GOSSS and \lili\ spectrograms clearly show that the He\,{\sc ii} lines are too strong for B0.5 (and way too weak for O9) and we classify it as B0~V(n), making it the third 
       earliest-type system in \VO{023}, tied with $\theta^1$~Ori~D. 
 \item $\theta^2$~Ori~B (= HD~\num{37042}) is the third brightest star in \GGc\ in the field, in part because of the higher extinction that affects the central part of the nebula \citep{ODeletal20}. \citet{SimDetal06} classified
       it as B0.5~V. We find instead a slightly later type of B0.7~V and we note how the \Teff\ sequence in their Table~4 from $\theta^1$~Ori~D (32~kK) to $\theta^1$~Ori~A (30~kK) to $\theta^2$~Ori~B (29~kK) corresponds in our
       spectral classifications to a sequence B0~V to B0.2~V to B0.7~V, which is the right direction. The spectral classification of $\theta^2$~Ori~B as B2/5 by \citet{Hill97} is clearly discarded.
 \item $\theta^2$~Ori~C (= HD~\num{37062}) is classified here as B4~V. However, at the original \lili\ resolution, some spectra have double lines. The two components appear to have similar spectral types but different values
       of $v\sin i$ (one is a slow rotator and the other an intermediate one). The mass ratio must be close to one, as both components have a peak-to-peak amplitude of $\sim$60~km/s. We thus confirm the spectroscopic binary nature
       detected by \citet{CorpLagr99} but detecting two components instead of just one. Note the residual nebular emission lines characteristic of the low-excitation region of the Orion nebula where $\theta^2$~Ori~C is located and
       which are difficult to eliminate in a faint star like this one.
\end{itemize}

\begin{figure*}[h!]
 \centerline{\includegraphics[width=\linewidth]{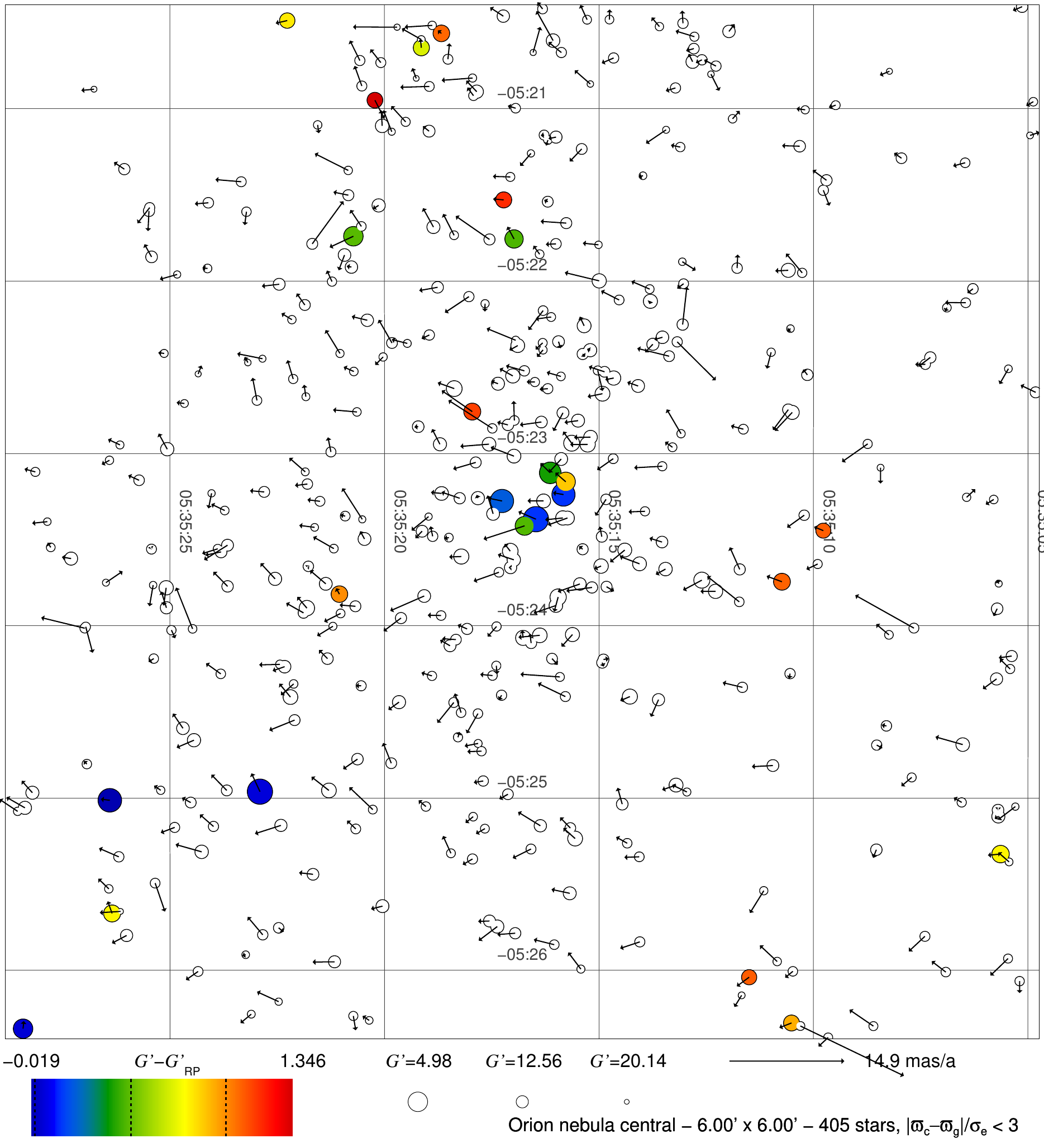}}
 \caption{{\it Gaia}~EDR3 chart of the central $6\arcmin\times6\arcmin$ of \VO{023} (Orion nebula cluster). Plotted sources are those with normalized parallaxes within 3 sigmas independently of other filters i.e. a lax selection 
           criterion for cluster membership rather than the strict one used to select the sample in this paper.
           Symbol color, symbol size, and arrows encode $\GGc-\GRPc$, \GGc, and proper motion, respectively, according to the caption. Objects without color either have no $\GGc-\GRPc$ or (more commonly) have $\Cstar > 0.4$
           due to nebular contamination. The Trapezium is at the field center.}
 \label{Orion_nebula_selected}
\end{figure*}

The {\it Gaia}~EDR3 source density is concentrated around the Trapezium and the parallax histogram of the whole field has a strong peak at the expected parallax dominated by sources with \GGc\ in the 15-17~mag range. The
membership algorithm, however, rejects most of those sources because the strong nebular contamination produces large \Cstar\ values (see Fig.~\ref{Orion_nebula_selected}, changing the maximum \Cstar\ to 3.0, for example, increases 
the number of members from 114 to 158
while changing the cluster distance by just 0.6~pc). Therefore, the selected members are not centrally concentrated and are spread over most of the selected region. Assuming a spherical distribution, this effect should lead to a
dispersion in distances of $\sim$0.7~pc, which is a factor of 3 lower than our uncertainty for the distance to \VO{023} (see below). In addition, most of the bright stars in \VO{023} have a bad RUWE and/or a large \sigmae\ that 
causes their exclusion by the membership algorithm. The three brightest stars that are selected by the algorithm are $\theta^1$~Ori~D, HD~\num{36981}, and HD~\num{36982}, with $\theta^1$~Ori~Ca,Cb and $\theta^2$~Ori~B 
excluded only because of their large \sigmae.
Three objects in \VO{023}, $\theta^1$~Ori~D, HD~\num{36982}, and HD~\num{36939}, are among the eleven stars included in the second step of the membership selection process (see above).

There is a long history of distance measurements to objects in \VO{023}. The values up to 2008 are listed in Table~2 and plotted in Fig.~6 of \citet{Muenetal08}, who noted that ``most measurements are accompanied by a large error 
bar, 15-20\%'' and that ``there seems to be some emerging convergence at 400~pc from the many varied techniques used in recent years''. A more recent analysis with {\it Gaia}~DR2 data \citep{Kounetal18} yields $389\pm 3$~pc and 
is in excellent agreement with the $390\pm 2$~pc value we derive here. Here we see a direct effect of the {\it Gaia} revolution. Thirteen years ago we were talking about distance uncertainties of 15-20\% and now we have two 
independent measurements that agree with uncertainties of less than 1\%.

The literature on runaway (and walkaway) stars from \VO{023} reaches to the 1950s, as the prototype dynamical ejection that expelled AE~Aur, $\mu$~Col, and $\iota$~Ori \citep{BlaaMorg54,Hoogetal00,Maizetal18b} took place there 
2.5~Ma ago. Recent analyses have been carried out using {\it Gaia}~DR2 \citep{Schoetal20,Farietal20} and HST \citep{Platetal20} data. We have searched the {\it Gaia}~EDR3 astrometry and found five walkaway/runaway candidates. Two 
of them, V1321~Ori and Brun~334, are listed as K~stars in \citet{Skif14} and were already identified as runaways by \citet{Schoetal20} and \citet{Farietal20}. For another two, BD~$-$05~1322 (with different spectral classifications 
listed in \citealt{Skif14}, some of them indicating broad lines) and HU~Ori we have found no prior identifications as walkaways/runaways. 
In a separate paper \citep{Maizetal21g}
we discuss the case of $\theta^1$~Ori~F and we propose that it has been very recently ejected from the $\theta^1$~Ori~Ca,Cb system.
%Perhaps the most interesting object in this category is $\theta^1$~Ori~F, 4\farcs5 away from 
%$\theta^1$~Ori~Ca,Cb, and classified as B8 by \citet{Herb50}. $\theta^1$~Ori~F has a large relative (with respect to $\theta^1$~Ori~Ca,Cb) proper motion of 4.1~mas/a towards the SE in a direction close to the line that joins them. 
%The parallax of $\theta^1$~Ori~F is consistent with \VO{023} membership and the data above implies a very close approach (and a possible ejection) $\sim$1.1~ka ago. Such an ejection would have taken place after another claimed 
%ejection from $\theta^1$~Ori~Ca,Cb, that of the highly extinguished Becklin-Neugebauer object, that took place 4~ka ago and in the opposite direction \citep{Tan04}.

\subsubsection{\VO{024} = $\gamma$~Velorum~cluster = Pozzo~1 = Brandt~1}
%     - gamma^2 Vel             - S06 - WC               + O7/8.5 III-II((f)) - not in Gaia EDR3, also LiLiMaRlin: WC8 + O7/8.5 III-II((f))
%   1 - gamma^1 Vel A,B         - S07 - B1.5  V                               - spi=0.338, G=4.2 so not used for k analysis.
%   3 - HD 68 092               - S10 - B3    V                               - dispm=1.99,SPI=0.125
%   4 - HD 67 820               - S10 - B8    IV                              - dispm=4.80
%   5 - CPD -46 2202 C          - S10 - B6    V                               - OK
%  15 - HD 68 157               - S10 - B9.5  V                               - OK
%  17 - HD 68 009               -     - B9/A0 V                               - OK
%  19 - CPD -46 2202 D          - S11 - F0                                    - OK
%  20 - HD 68 158               - S10 - A0    V                               - OK
%  24 - HD 67 927               -     - A1    V                               - npi=-6.74
%  25 - HD 68 186               -     - A1/2  V                               - OK
%  26 - CPD -47 1915            -     -                                       - OK, PMS?
%  38 - HD 68 577               -     - A2    V                               - OK
%  39 - HD 68 159               -     - A0/1  V                               - OK
%  43 - Tyc 8140-06513-1        -     -                                       - OK

\begin{figure*}[h!]
 \centerline{\includegraphics[width=\linewidth]{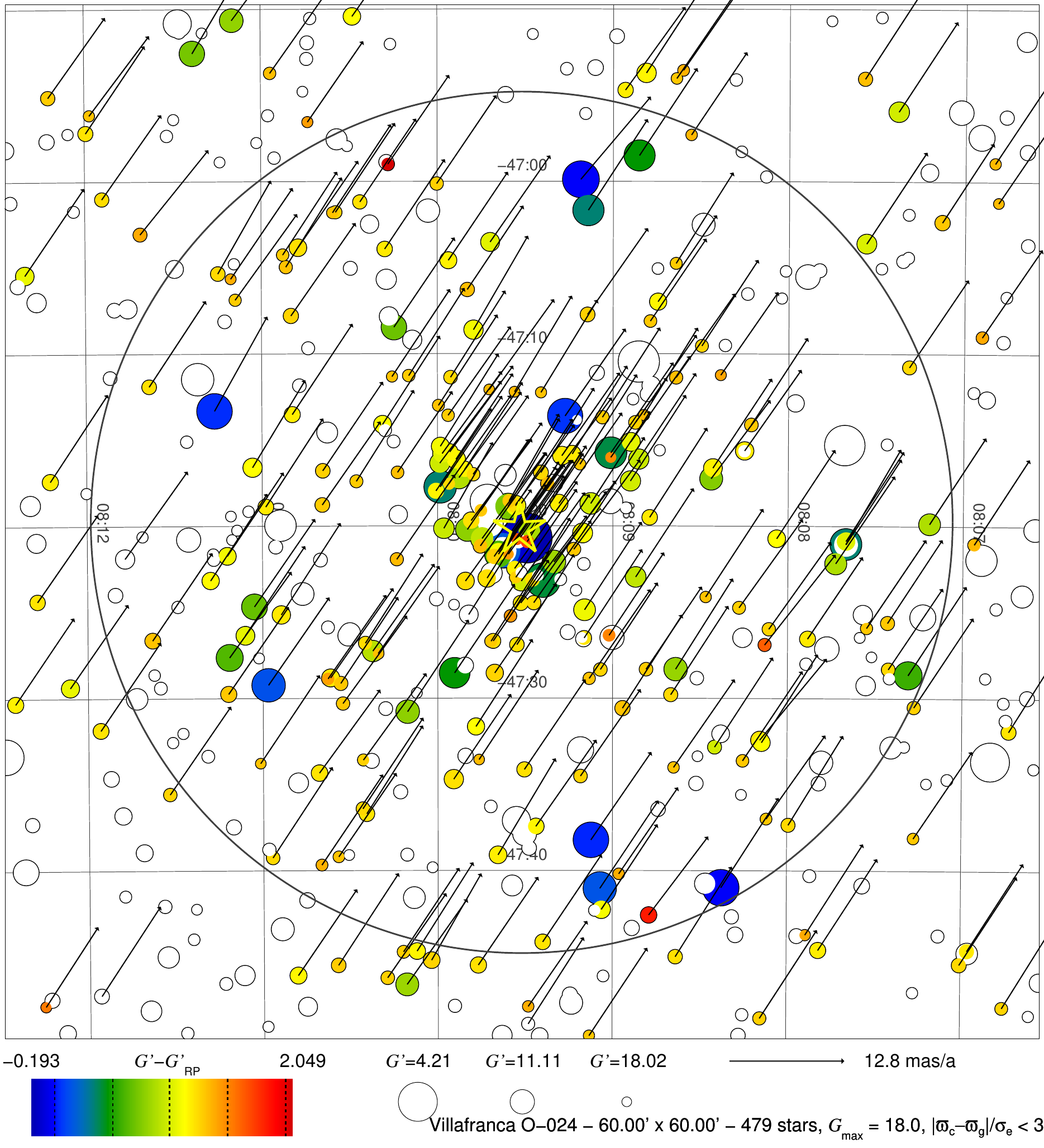}}
 \caption{{\it Gaia}~EDR3 chart of the $1^{\rm o}\times1^{\rm o}$ field centered on \VO{024} ($\gamma$~Vel cluster). Plotted sources are those with $\GG < 18$ and with normalized parallaxes within 3 sigmas independently of other 
           filters i.e. a lax selection criterion for cluster membership rather than the strict one used to select the sample in this paper. Symbol color, symbol size, and arrows encode $\GGc-\GRPc$, \GGc, and proper motion, 
           respectively, according to the caption. Objects without color and arrows are those with a proper motion separation larger than 1.35~\microas/a. The yellow star marks the location of $\gamma^2$~Vel (not a {\it Gaia}~EDR3 
           source) and the large circle the circular aperture used to select cluster sources.}
 \label{gamma_Vel}
\end{figure*}

$\,\!$\indent This cluster was surprisingly not identified until two decades ago by \citet{Pozzetal00} and its discovery was prompted by the detection of the X-rays originating in its population of PMS stars. We say surprisingly
because \VO{024} includes $\gamma^2$~Vel, a WR+O system (WC8~+~O7.5~V-III according to \citealt{DeMaSchm99}) in which the primary is the closest Wolf-Rayet and the only one\footnote{Using {\it Gaia}~DR2 data, \citet{RateCrow20} 
gave a distance of 950$\pm$60~pc for WR~94 but the {\it Gaia}~EDR3 parallax places it slightly beyond 1~kpc.} within 1~kpc. \citet{Jeffetal09} studied the PMS stars in the cluster and determined their age at 5-10~Ma and 
\citet{Jeffetal14} identified two distinct kinematic populations, with the more likely scenario being that one corresponds to \VO{024} (the cluster itself) and the second one to the surrounding Vela~OB2 association \citep{Saccetal15}.
\cite{Prisetal16} determined that the low-mass stars follow a canonical IMF with a total of 100~M$_\odot$ in the 0.16~M$_\odot < M < $~1.3~M$_\odot$ range.

We present GOSSS and \lili\ spectrograms of $\gamma^2$~Vel in Figs.~\ref{GOSSS_spectra_14}~and~\ref{LiLi_spectra_06}. The primary is classified as WC using GOSSS data and WC8 using \lili\ data, with the difference (as for WR~79) 
arising from the different wavelength ranges available. The secondary is classified as O7/8.5~III-II((f)) in both cases. We cannot provide a more detailed classification because we currently do not have a comparison WC8
spectrum and under such circumstances it is difficult to evaluate the contribution of each component \citep{DeMaSchm99}. 
We also present a GOSSS spectrogram of $\gamma^1$~Vel~A,B in Fig.~\ref{GOSSS_spectra_16}. We classify it as B1.5~V, in agreement with \citet{Abtetal76}. We point out that 
\citet{HernSaha80} identified $\gamma^1$~Vel~A,B as an SB1 but there are no 
recent studies of this system and we have no \lili\ data. The WDS lists a visual component with a $\Delta m$ of 1.5~mag located at a separation of 37~mas, hence the A,B designation.
% Need comparison spectra of WR 135.

$\gamma^2$~Vel is so bright that it has no entry in {\it Gaia}~EDR3. The second brightest star in \VO{024} is $\gamma^1$~Vel~A,B, which has a parallax and proper motion consistent with those of the cluster but its \sigmae\ is 
0.338~mas, so it is excluded by the selection algorithm. The next two bright stars with compatible distances are HD~\num{68092} and HD~\num{67820} but in both cases their proper motions are incompatible with that of the cluster 
(and do not point away from it, so they are not likely walkaways/runaways). The first stars selected by the algorithm are CPD~$-$46~2002~C~and~D (note that A and B are $\gamma^2$~Vel and $\gamma^1$ Vel, respectively), 
HD~\num{68157}, and HD~\num{68009}. 
%The information regarding those three stars indicates they have masses of 2-4~M$_\odot$ so we are left with a highly peculiar IMF in \VO{024}: two massive stars in $\gamma^2$~Vel with likely 
%initial masses of 30~M$_\odot$ each, another massive star with 10-15~M$_\odot$ ($\gamma^1$~Vel~A) and a companion (or two) of uncertain but lower mass, and then nothing until the 2-4~M$_\odot$ range. This peculiarity of the 
%IMF was already noted by \cite{Prisetal16} when they indicated that the total mass of the PMS stars was too low for a cluster with an O and a WR star. 
An interesting feature of the \VO{024} CMD in Fig.~\ref{CMDs} is the
presence of a nearly-equal-mass binary PMS located $\sim$0.75~mag above the main one for \GBPmGRPc\ between 2.2 and 3.5. We do not detect any runaways escaping from \VO{024}.

Given its proximity, \VO{024} is the easiest cluster to detect in our sample with respect to the dominant background population, located at distances of 2-3~kpc, using proper motions (Fig.~\ref{gamma_Vel}). The (mostly PMS) stars 
in the cluster are also relatively easy to distinguish in the CMD (Fig.~\ref{CMDs}), 
as already seen in Fig.~6 of \citet{Jeffetal09}. Indeed, the most likely contaminants in the sample would be from the B~population (association members not in the 
cluster) of \citet{Jeffetal14}. Stars with the proper motion of the cluster show a well defined core around $\gamma^2$~Vel and a halo while stars with other proper motions (many of them likely Vela~OB2 members) are more uniformly 
scattered across the field.

Literature distances include measurements to $\gamma^2$~Vel and to the cluster itself. Among the first, \citet{Milletal07} give $368^{+38}_{-13}$~pc and \citet{Nortetal07a} $336^{+8}_{-7}$~pc with interferometry while
\citet{Maizetal08a} give $278^{+50}_{-37}$~pc with Hipparcos data using the \citet{vanL07a} reduction and the same prior as here. As for cluster distances, \citet{Jeffetal09} used isochrone fitting to obtain 356$\pm$11~pc and
\citet{Franetal18} used {\it Gaia}~DR2 parallaxes to obtain 345$\pm$12~pc. Our new value of 336$\pm$1~pc is much more precise but otherwise within 1.2~sigmas of the previous ones with the exception of that of \citet{Milletal07}.

\subsubsection{\VO{025} = Trumpler~16~E}

$\,\!$\indent The Carina nebula association (Car~OB1) is the region with more O~stars within 3~kpc of the Sun. It is a complex region that was described in Villafranca~I, where we analyzed \VO{002} (Trumpler~14) and
\VO{003} (Trumpler~16~W), see also the previous subsection. Trumpler~14 is a well defined cluster while Trumpler~16 is more extended and has no clear boundaries. In the previous paper we analyzed Trumpler~14 and the western part
of Trumpler~16, as they are the two regions in Car~OB1 with the earliest-type O~stars. Here we analyze the eastern part of Trumpler~16 (Trumpler~16~E or \VO{025}), which we define as the region around its most famous denizen,
$\eta$~Car. $\eta$~Car is an eccentric binary system \citep{Damietal00,Damietal08,Ipinetal05} with an extremely luminous and massive LBV as a primary that has a historical record of extraordinary eruptions 
\citep{SmitFrew11,Kimietal16}. There are at least twelve O-type systems\footnote{The census should be complete in the region barring the unknown classification of $\eta$~Car~B and (unlikely) objects hidden by large extinction and
source confusion \citep{Smit06b,Preietal21}.}, of which the earliest ones are HDE~\num{303308}~A,B (O4.5~V((fc)), \citealt{Sotaetal14}) and CPD~$-$59~2641 (O6~V((fc)), \citealt{Sotaetal14}).

% Possible new LBV/B stars: eta Car, CPD -59 2598, CPD -59 2640, CPD -59 2616,
%                           2MASS J10452265-5942596, 2MASS J10445201-5939322
We present GOSSS spectra of $\eta$~Car (Fig.~\ref{GOSSS_spectra_15}) and of four B-type stars: CPD~$-$59~2598, CPD~$-$59~2640, CPD~$-$59~2616, and 2MASS~J10452265$-$5942596 (Figs.~\ref{GOSSS_spectra_16}~and~\ref{GOSSS_spectra_17}). 
$\eta$~Car is classified simply as an LBV because the companion, whose spectrum has never been directly observed, leaves no trace in the spectrogram. We also present \lili\ spectra of
four O-type SB2/SB3 binaries (V572~Car, CPD~$-$59~2635, HD~\num{93343}, and V573~Car, Fig.~\ref{LiLi_spectra_05}) that already had GOSSS classifications in either \citet{Sotaetal14} or \citet{Maizetal16} but have been 
observed again at a more favorable phase for velocity separation. For V572~Car we show two epochs taken in consecutive nights of this interesting SB3 system, with the evening date encoded as YYMMDD in Fig.~\ref{LiLi_spectra_05}. 
Two stars of spectral type O6.5~Vz and B0~V orbit each other in 2.15~d \citep{Rauwetal01a} and both trace a longer orbit around a B0.2~V star. Our classifications for the inner pair are slightly different from those of 
\citet{Rauwetal01a}, of earlier type for the primary and of later type for the secondary. CPD~$-$59~2635 is classified with \lili\ data as O8~Vz~+~O9.2~V, which is similar but not identical to the GOSSS classification or
to that of \citet{Albaetal01}. For HD~\num{93343}, in \citet{Maizetal16} we could only provide an O8~V combined classification, even though the presence of a
secondary spectrum had already been detected by \citet{Walb82a} and has been classified as O8~+~O8 by \cite{Rauwetal09} and as O7.5~Vz~+~O7.5V~(n)z by \citet{Putketal18}. With the \lili\ spectrogram
we classify it as O7.5~Vz~+~O7.5:~V(n), noting the uncertainty in the secondary caused by its moderately high rotation. 
V573~Car is classified with \lili\ data as O9.5~IV~+~B0.5~V, which is similar to the GOSSS classification but with narrower lines.

Not being a true cluster, \VO{025} does not show a central concentration and the limits traditionally assigned to Trumpler~16 are established actually by the V-shaped dust lane immediately to its south, which is placed in the
foreground with respect to the Carina nebula (top left plot in Fig.~7 of \citealt{MaizBarb18}, see also \citealt{Preietal21}). 
Nevertheless, the stellar population of \VO{025} is relatively easy to identify thanks to its richness and to its relatively low 
extinction in comparison with the background stars. Of the twelve O stars in \VO{025}, the membership algorithm selects nine, the six mentioned above plus CPD~$-$59~2644, CPD~$-$59~2627, and CPD~$-$59~2629. The other three stars 
are excluded by their RUWE or lack of parallax: CPD~$-$59~2636~A,B, CPD~$-$59~2624, and CPD~$-$59~2626~A,B. 
CPD~$-$59~2635 is one of the eleven stars included in the second step of the membership selection process (see above). We do not detect any new runaways.

We measure a distance to \VO{025} that is nearly identical to that of \VO{003} and very similar to that of \VO{002}, When we also consider that the separation between Trumpler~14 and Trumpler~16 is large enough for the 
effects of the checkered pattern in the {\it Gaia}~EDR3 zero point to be of the same order as the differences in parallax we observe, we conclude that the {\it Gaia}~EDR3 results are compatible with all three groups being at the
same distance. Differences of 50-100~pc cannot be accurately detected for stellar groups at 2.0-2.5~kpc with the current data. Nevertheless, it would be interesting to do a further study of Car~OB1 with {\it Gaia}~EDR3 surveying 
the whole association and trying to detect different kinematical components (see \citealt{Damietal17a} for some hints of this). In this respect, we note that the proper motion of \VO{025} is not identical to that of \VO{003} and 
that the two groups we have defined in Trumpler~16 appear to be separating from each other. 

\subsubsection{\VO{026} = $\sigma$~Orionis~cluster = Escorial~7}
%   1 - sigma Ori AaAb          - Ma  - O9.5  V          + B0.2  V        - RUWE,spi, B not in Gaia EDR3
%   3 - sigma Ori E             - S07 - B2    Vp                          - spi=0.17
%   4 - sigma Ori D             - S07 - B2    V(n)                        - spi=0.17
%   6 - HDE 294 271             - S07 - B5    V                           - RUWE=1.44,spi=0.275
%   7 - HD 37 525               - S10 - B5    V                           - spi=0.106
%   9 - HD 37 564               -     - A8    V                           - OK
%  10 - HDE 294 272 A           - S10 - B9.5  III                         - OK
%  12 - HDE 294 272 B           - S10 - B8    V                           - spi=0.1001
%  20 - sigma Ori C             - S11 - A2    Vn                          - OK
%  47 - Mayrit 968 292          -     - F7                                - RUWE=3.84,spi=0.167,dispm=2.962,disr=979, runaway?
%  49 - StHA 50                 -     - B6                                - npi=-42.41
%  57 - Mayrit 240 322          -     -                                   - RUWE=5.95,spi=0.276,dispm=11.11, possible member with bad astrometry
%  64 - Mayrit 114 305          -     - K2                                - OK

$\,\!$\indent The $\sigma$~Orionis cluster is part of the Ori~OB1d association \citep{Caba07,Ball08} and is the cluster with O-type stars (together with \VO{024})
with the lowest foreground extinction \citep{MaizBarb18}. The natal cloud at the 
location of the cluster has already dispersed but its radiation reaches the nearby Horsehead nebula $\sim$30\arcmin\ (3.5~pc) away and is the source of its formation and ionization. The central system of \VO{026} is
$\sigma$~Ori itself\footnote{Following \citet{Caba07}, we recommend calling the cluster $\sigma$~Orionis and the stellar system $\sigma$~Ori.}, a high-order multiple whose most massive stars form a hierarchical triple
whose nature remained hidden until recently. Traditionally, the system was composed of an optical pair A,B with a long period (at this point it has not completed one revolution since the first observed epoch) of 160~a
but a decade ago \citet{SimDetal11b} discovered that the A~component is a spectroscopic binary with a 143~d period and later \citet{Schaetal16} resolved the pair with interferometry and determined a combined visual and
spectroscopic orbit. In \citet{Maizetal18a,Maizetal19b,Maizetal21b} we used lucky spectroscopy to separate (spatially or in velocity) the three components and derive spectral types of O9.5~V for Aa, B0.2~V for Ab, and 
B0.2~V(n) for B.

We present GOSSS and \lili\ spectrograms for three stars in \VO{026} in Figs.~\ref{GOSSS_spectra_17}~and~\ref{LiLi_spectra_06}. $\sigma$~Ori~D is classified as B2~V(n) and we note that prior 
classifications did not include the (n) qualifier (it could be classified even as n). $\sigma$~Ori~E is a well-known peculiar, magnetic, and He-rich B2~Vp star \citep{LandBorr78}. HDE~\num{294271} is also a moderately fast 
rotator classified as B5~V(n).

The combination of a very low extinction with a well defined PMS makes this cluster easy to separate from the other populations in the CMD. Surprisingly, \VO{026} has very similar proper motions compared to the main background
population despite the factor of $\sim$10 in distance between the two, an effect likely caused by the proximity of the solar antapex. 
%One interesting feature of the \VO{026} CMD is the large gap (over two magnitudes in \GGc\ and over 1~mag in \GBPmGRPc) that exists between the brightest PMS 
%stars (Mayrit~\num{114305} or possibly Mayrit~\num{240322}) and the faintest star already in the MS ($\sigma$~Ori~C). Such a jump in luminosity is expected around 2~M$_\odot$ for a 3-5~Ma old cluster (Fig.~14 in 
%\citealt{Penaetal12}) and is indeed observed here.
The brightest objects are not selected by the membership algorithm: $\sigma$~Ori~Aa,Ab,B (one source in {\it Gaia}~EDR3) has no parallax; $\sigma$~Ori~E, $\sigma$~Ori~D, 
HDE~\num{294271}, HD~\num{37525}, and HDE~\num{294272}~B have $\sigmae > 0.1$ (HDE~\num{294271} also has RUWE=1.44). The brightest selected members are HD~\num{37564} and HDE~\num{294272}~A. The cleaned CMD in Fig.~\ref{CMDs}
shows a large gap between the MS and the PMS starting at a main-sequence mass of 2~M$_\odot$, a feature that is discussed below. We detect one possible runaway, 
Mayrit~\num{968292}, that could have been ejected from the cluster core $\sim$330~ka ago.

The value we derive for the distance to \VO{026} of 397$\pm$2~pc is in good agreement with the {\it Gaia}~DR2 measurement of $403^{+8}_{-7}$~pc of \citet{Zarietal19} for their group B$_2$. However, it has a significant tension 
with the 387.5$\pm$1.5~pc distance by \citet{Schaetal16} from the orbital parallax of $\sigma$~Ori~Aa,Ab, a 9.5~pc difference that amounts to 2.4\% or 3.8 sigmas. Given the cluster size \citep{Caba08c}, this cannot be a depth 
effect, so systematics must be present. One possible issue is the tension between the radial velocity curves derived by \citet{SimDetal15a} and \citet{Schaetal16} using only spectroscopic data: in \citet{Schaetal16} 
$K_{\rm Aa}$ is smaller and $K_{\rm Ab}$ and $\gamma$ are larger than in \citet{SimDetal15a}. Such discrepancies when combining radial velocities from different instruments is, unfortunately, relatively common for O and 
early-B stars \citep{Trigetal21} and that could be the cause of the lower value for the orbital parallax, which is based on a combination of both.

%\subsubsection{\VO{027} = Westerlund~1}
%
%$\,\!$\indent 

\section{Analysis and future work}

\subsection{Comparison between \textit{Gaia} DR2 and EDR3 distances}

\begin{figure}
 \centerline{\includegraphics[width=\linewidth]{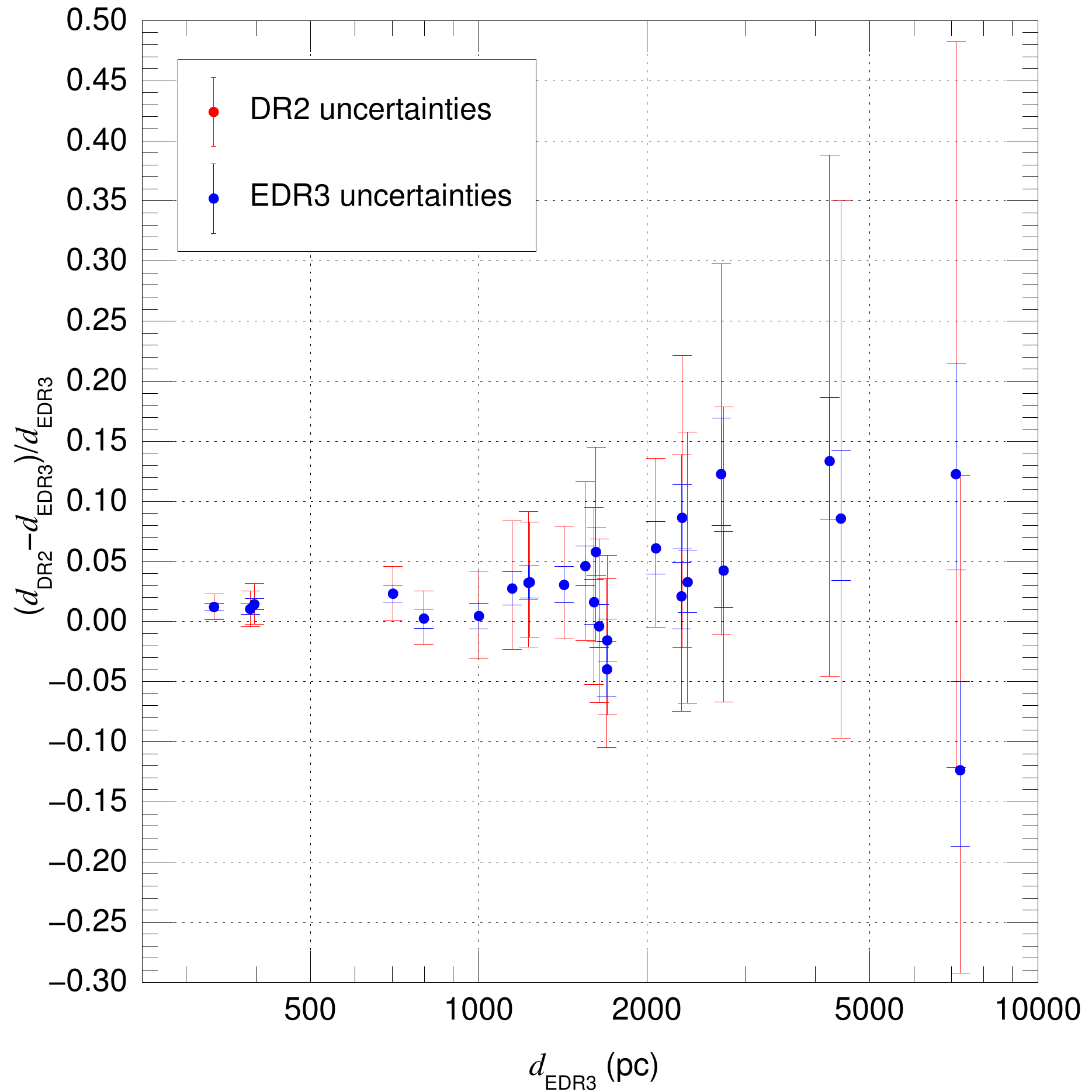}}
 \caption{Comparison between distances for the 26 Villafranca groups in this paper using {\it Gaia}~DR2 and EDR3 data. The {\it Gaia}~DR2 distances for the 16 groups in Villafranca~I were recomputed to use the same selection
          parameters as in this paper. Red and blue are used to represent the {\it Gaia}~DR2 and EDR3 uncertainties, respectively.}
 \label{distances}
\end{figure}

$\,\!$\indent As described in \citet{Maizetal21c}, the uncertainties derived from our method are a combination of random and systematic uncertainties, the latter being a consequence of the uncorrected parallax bias, \sigmas. 
In the limit of a large number of sources located within a small region in the sky, Eqn.~7 in that paper leads to a limit of \sigmas\ for the parallax uncertainty of the stellar group, \spig. That is indeed what we see in 
Table~\ref{results} and its equivalents in Villafranca~I (Table~3) and in \citet{Maizetal21c} (Table~5), with a value of 43~\microas\ for {\it Gaia}~DR2 \citep{Lindetal18b}\footnote{The value is 21~\microas\ for $\GG < 13$, but
most stars in a given group are fainter than that.} 
and of 10.3~\microas\ for {\it Gaia}~EDR3 \citep{Maizetal21c}.
The small deviations from those values arise either because the number of stars is relatively small, in which case \spig\ is larger
(e.g. \VO{013} and \VO{014}~SE in Table~\ref{results}) or because the group has a size large enough for the angular covariance effects to kick in, in which case \spig\ is smaller (e.g. $\omega$~Cen or 47~Tuc in 
\citealt{Maizetal21c}). Therefore, our group parallaxes (and the associated distances) are dominated by systematic uncertainties.

Going from parallaxes to distances implies the use of a prior and, in general, (assumed) Gaussian parallax uncertainties lead to asymmetric distance uncertainties \citep{Maiz05c}. Nevertheless, in the case of objects with small 
relative parallax uncertainties, $\sigma_d/d \approx \sigma_\varpi/\varpi$ is a good approximation and the asymmetry is small. For a constant parallax uncertainty (see above), this leads to the relative distance uncertainty being 
proportional to the distance, i.e.:

\begin{equation}
 \frac{\sigma_d}{d} \approx \sigmas d,
 \label{distance_uncertainty}
\end{equation}

\noindent where \sigmas\ is expressed in mas and $d$ is expressed in kpc. For {\it Gaia}~EDR3, distances to the Villafranca groups have uncertainties around 1\% at 1~kpc and 3\% at 3~kpc while for {\it Gaia}~DR2 the values are
about four times higher. This approximation holds relatively well for the distances in Table~\ref{results} and in Fig.~\ref{distances}. Comparing to the literature values in Villafranca~I and in this paper leads us to conclude that
{\bf the method in this paper with {\it Gaia}~EDR3 values is the most precise to date to calculate distances to stellar groups in the solar neighborhood.}

The calculated uncertainties using {\it Gaia}~EDR3 are significantly lower than for {\it Gaia}~DR2 but we should also ask ourselves if there is a systematic trend. Figure~\ref{distances} indicates there is indeed such an 
effect, with most {\it Gaia}~DR2 uncertainties being larger than their {\it Gaia}~EDR3 counterparts, but it is not a large one. Taking the {\it Gaia}~EDR3 values as the exact ones, the {\it Gaia}~DR2 distances are, on average,
0.5~sigmas larger but none are larger than 1.2~sigmas. A single distance calculated from {\it Gaia}~DR2 astrometry will be OK when judged by its uncertainty but a collection of them will be, on average, larger by $\sim$3\% at 
1~kpc and by $\sim$8\% at 3~kpc with respect to {\it Gaia}~EDR3 distances. This is likely a consequence of the improved calibration of {\it Gaia}~EDR3:
as for {\it Gaia}~DR2 parallaxes a single quasar-based parallax zero point was used for all sources, it is possible that a small offset exists for DR2~parallaxes sources brighter than the average quasar.
Such an effect would be compatible with the trend seen in Fig.~\ref{distances}.

Another issue when validating a method to calculate distances is whether uncertainties are correctly estimated or not. In the case of the method here, we provide in Table~\ref{results} for $t_\varpi$, the normalized $\chi^2$ test
for the group parallax. The 26 values are between 0.78 and 1.22 and quite symmetrically distributed about the expected value of 1.00, indicating that the individual parallaxes have correctly estimated uncertainties (see the 
discussion in the companion paper on the multiplicative constant $k$). Note that the equivalent $\chi^2$ tests for the proper motions are significantly larger than one in all cases. As discussed in Villafranca~I, this is a consequence
of the internal motions of the stellar groups and the values are larger when the targets are more massive or compact (larger physical velocities) and/or closer to us (larger proper motions due to the $1/d$ factor).

History teaches us that most distance measurements in astronomy are plagued by systematic uncertainties and, indeed, that is also the case here. However, our analysis in \citet{Maizetal21c} indicates that the systematic effects
are well characterized, in part by correcting for them and in part by accounting for them in the error budget. Furthermore, the uncorrected systematic effects are of different sign in different parts of the sky, as they form a
checkered pattern (Fig.~14 in \citealt{Lindetal21a}), so, on average, distances to a large number of stellar groups scattered across the sky should still be unbiased. The ultimate answer to the possible presence of remaining 
systematic effects in the form of the distances in this paper being systematically higher or lower should come from a comparison with (yet inexistent) more accurate distances: only time will tell, as is often the case. However, we 
can still compare our results with the few high-accuracy geometric distances to some of the groups analyzed here and in previous papers: those of \citet{Thometal20} for 47~Tuc \citep{Maizetal21c}, \citet{Hensetal00} for \VO{020},
\citet{Nortetal07a} for \VO{024}, and \citet{Smit06a} for $\eta$~Car in \VO{025}. The excellent agreement with those values in combination with the analysis in \citet{Maizetal21c} leads us to conclude that {\bf our distances are
accurate as systematic effects have been properly characterized.}

\subsection{Orphan clusters}

$\,\!$\indent One result of this series of papers is the observational demonstration of the existence of ``orphan clusters'', stellar groups in which the most massive stars have been ejected through a dynamical interaction and have left 
a system in which the present-day mass function (PDMF) is capped at a value significantly lower than that of the IMF \citep{Ohetal15}. 
There are two clear examples in the current Villafranca sample: In \VO{012}~S (Haffner~18), HD~\num{64568} (O3~V((f*))z) and HD~\num{64315}~A,B (O5.5~V~+~O7~V) were 
ejected $\sim$400~ka ago, leaving the cluster with two later O-type stars, CPD~$-$26~2704 and CPD~$-$26~2711. In \VO{014}~NW in the North America nebula, the two more massive objects, the Bajamar (O3.5~III(f*)~+~O8:) and Toronto 
(O6.5~V((f))z + B) stars were expelled in two events 1.5-1.6~Ma ago as walkaways. Other OB stars were expelled in those two events and in another earlier one, leaving the cluster with apparently just two possible late-O-type stars 
\citep{Maizetal21f}.
Both cases are similar in that one of the ejected systems contains a very-early-O star (hence, very massive) and the other one contains another O star earlier 
than O7. Another common characteristic is the complexity involved in the interactions that led to the ejections: in \VO{014}~NW both O-type systems are SB2s and were ejected in two different episodes that involved other stars and in 
\VO{012}~S HD~\num{64315}~A,B is a SB2+SB2 \citep{Loreetal17} and HD~\num{64568} appears to be single, implying a minimum of five stars.

Massive stars have been known to experience ejections from clusters by dynamical interactions since it was first proposed for AE~Aur by \citet{BlaaMorg53}. 
The novelty of orphan clusters lies in that the interaction takes place among the most massive stars in the system and is violent enough to unbind all of the participants. Are \VO{012}~S and \VO{014}~NW unique in our sample? Apparently so 
but it is worth discussing two other related cases. In \VO{004} (Westerlund~2) three very-early type stars, THA~35-II-42 (WR~21a), SS~215 (WR~20aa), and WR~20c appear to have been ejected but the cluster is significantly more 
massive and includes several other very massive stars (Villafranca~I and references therein). The ejection of AE~Aur, which also
involved $\mu$~Col and $\iota$~Ori, is interesting in that it involved five or six objects\footnote{$\iota$~Ori is a complex multiple system \citep{MaizBarb20} and it is unclear whether the B component is associated with the other 
three. At the time of \citet{MaizBarb20}, only {\it Gaia}~DR2 was available and the A component had no parallax there. There is one available in {\it Gaia}~EDR3 but its \sigmae\ is too large to be useful.} and in that it took
place 2.5~Ma ago \citep{Hoogetal00}, significantly farther in the past than for \VO{012}~S or \VO{014}~NW. \citet{Hoogetal00} traced the three involved systems back to the same position where \VO{023} was located at the time, suggesting
that they were ejected from the Orion nebula cluster. As \VO{023} currently contains $\theta^1$~Ori~Ca,Cb, which is of earlier type than any of the ejected systems (of which $\iota$~Ori contains the star with the earliest spectral type, 
O8.5~III), this would not be truly an orphan cluster. However, the situation may be more complicated as the current Trapezium is younger than 2.5~Ma ($\theta^1$~Ori~Ca,Cb should be located farther away from the 
ZAMS if it was of that age or older, the nebula itself is located very close to their ionizing sources, and the cluster itself is compact but in a hot dynamical state, \citealt{deGretal08,AlliGood11}) and we are witnessing two 
generations of massive-star formation within \VO{023}: a first one that may have led to the ejection and left an orphan cluster and a second one that produced the current Trapezium \citep{Krouetal18}. If we also consider that the current 
26 Villafranca groups are a small fraction of the known clusters with O stars in the Galaxy (which remain mostly unexplored with {\it Gaia}), cluster orphanization may be a significant phenomenon.

{\bf The existence of orphan clusters has important consequences in several fields of astrophysics.} The most obvious one, as already mentioned, is that it introduces a large difference between the PDMF and the IMF, 
forcing a study of the proper motions and/or radial velocities in the cluster surroundings to detect the real maximal star mass \citep{WeidKrou06}. The availability of {\it Gaia} astrometry is undoubtedly a game changer but any such study
should consider the possibility of the walkaway or runaway stars having already exploded as SNe. A related issue is the apparent age of the cluster as determined by the earliest-type star present among MS stars: if the most massive stars 
have already been ejected, we may think that the cluster is older than what it really is. We may also detect that the age as determined from PMS stars does not agree with the MS value. This effect is especially important when one 
considers that ejections take place preferentially right after cluster birth \citep{OhKrou16}. The disappearance of the most massive stars from the core also reduces the ionization of the gas there and may favor the reignition of star
formation there and the existence of multiple generations \citep{Krouetal18}. Another effect is the spatial distribution of core-collapse SNe in the Galaxy, as an enhanced rate of massive-star ejections will scatter the 
explosions far from star formation regions (a star with an ejection velocity of 100~km/s travels 102~pc in 1~Ma). Finally, the distribution and dynamics of compact objects is significantly altered: there would be fewer than expected in 
clusters and more in the diffuse population and some of the runaway neutron stars or black holes would be produced in binary systems previously ejected, leading to a double-runaway scenario (first by dynamical interaction and later by 
explosion) where the object is not moving away directly from any cluster \citep{PflaKrou10}. 

\subsection{CMDs and group properties}

\begin{figure*}[h!]
\centerline{\includegraphics[width=0.49\linewidth]{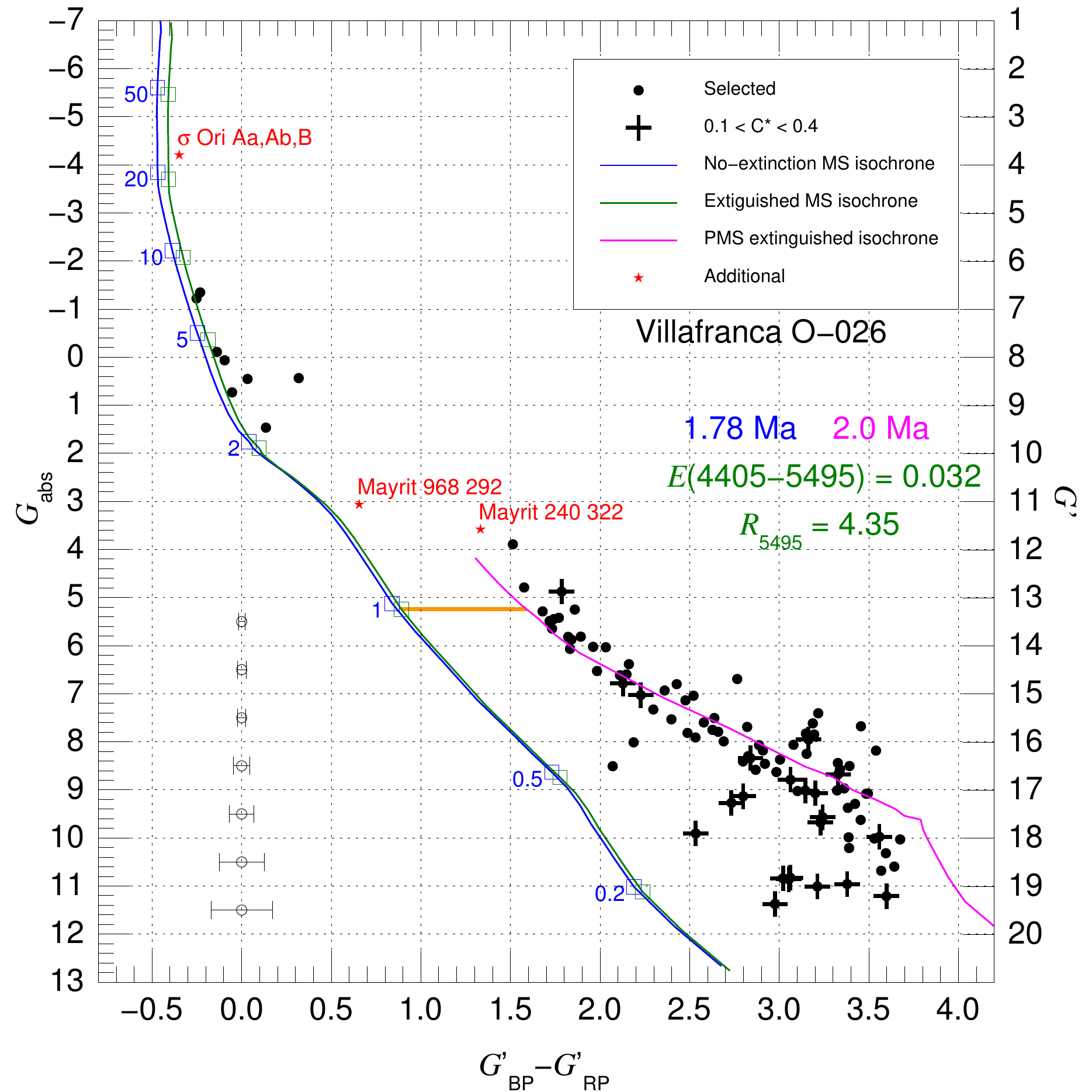} \
            \includegraphics[width=0.49\linewidth]{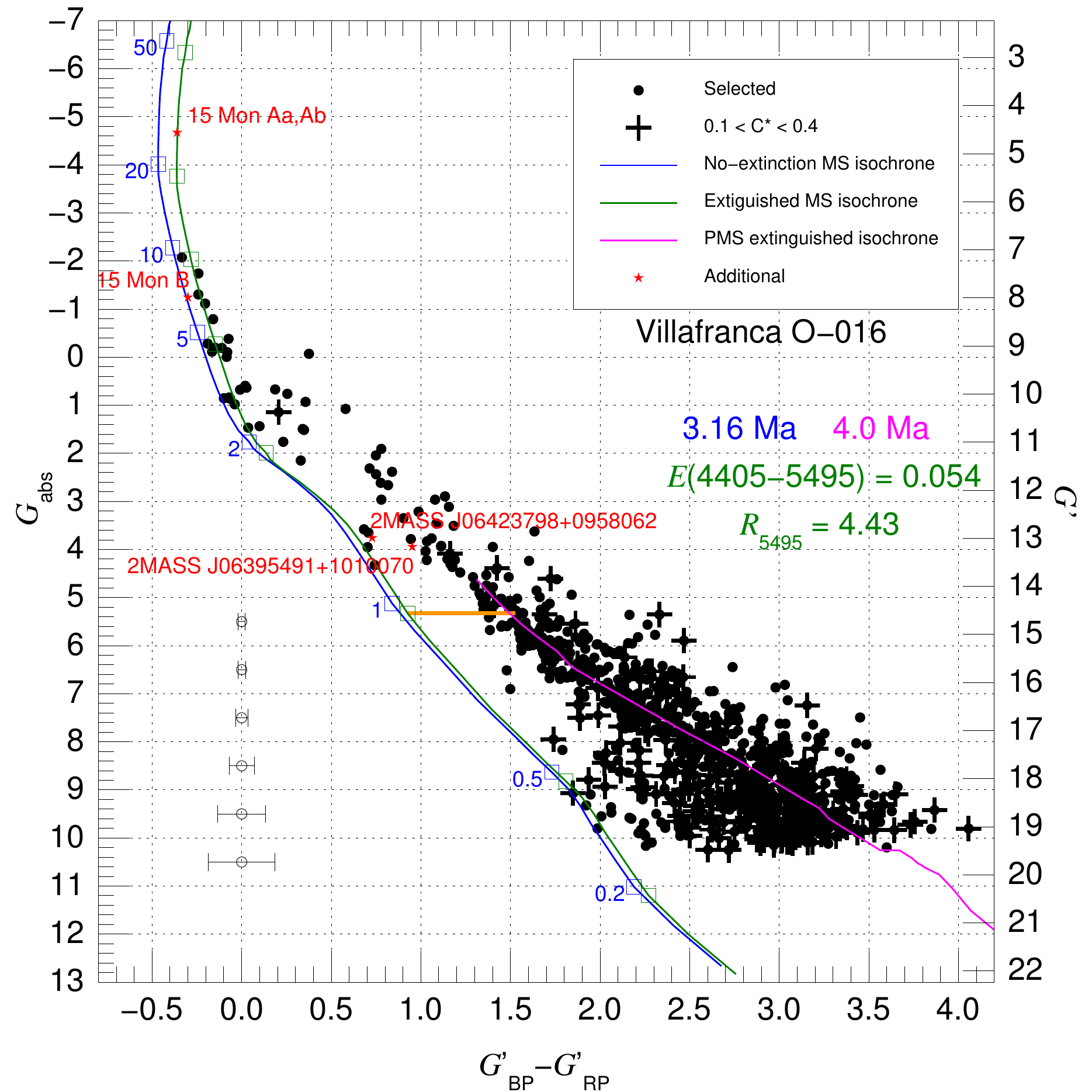}}
\centerline{\includegraphics[width=0.49\linewidth]{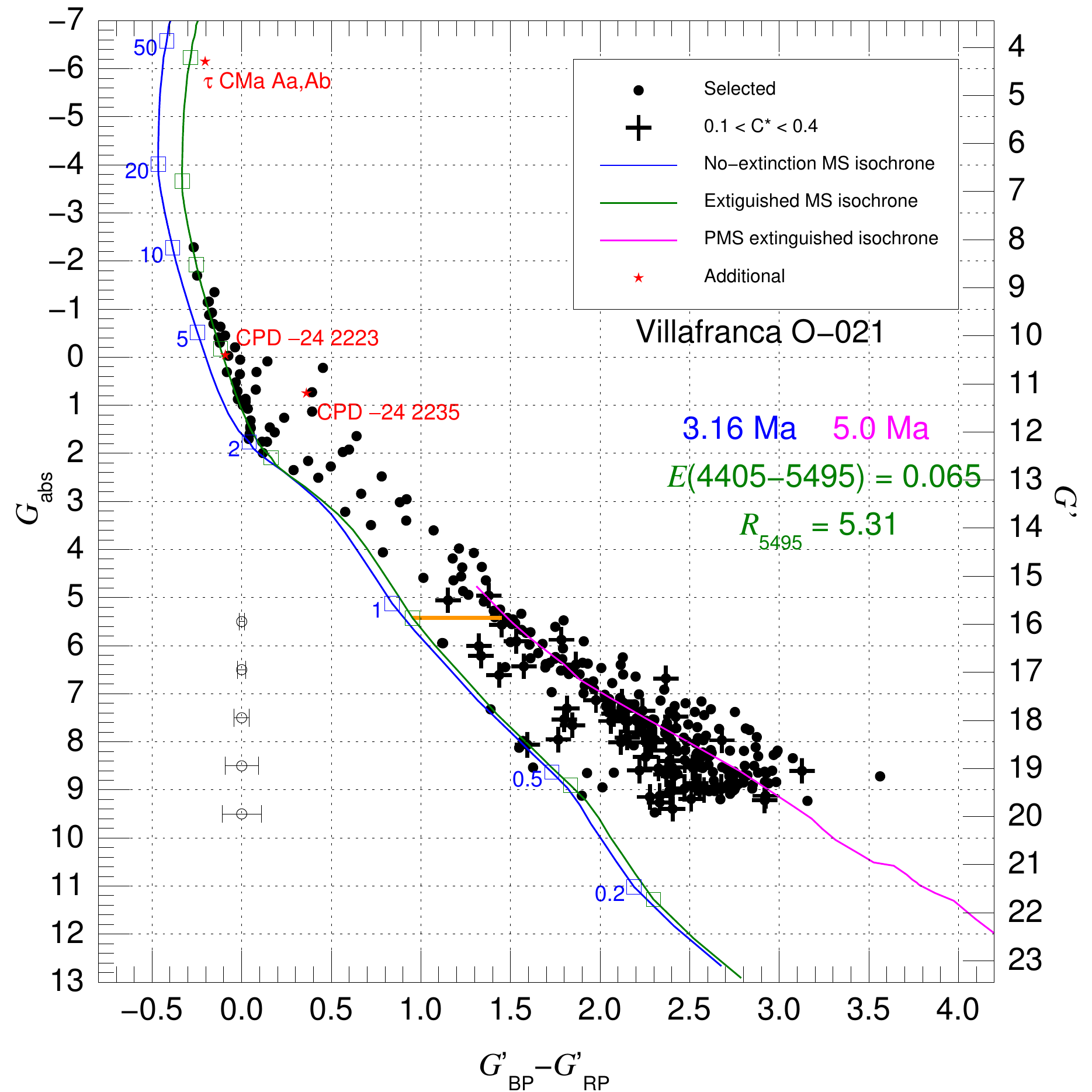} \
            \includegraphics[width=0.49\linewidth]{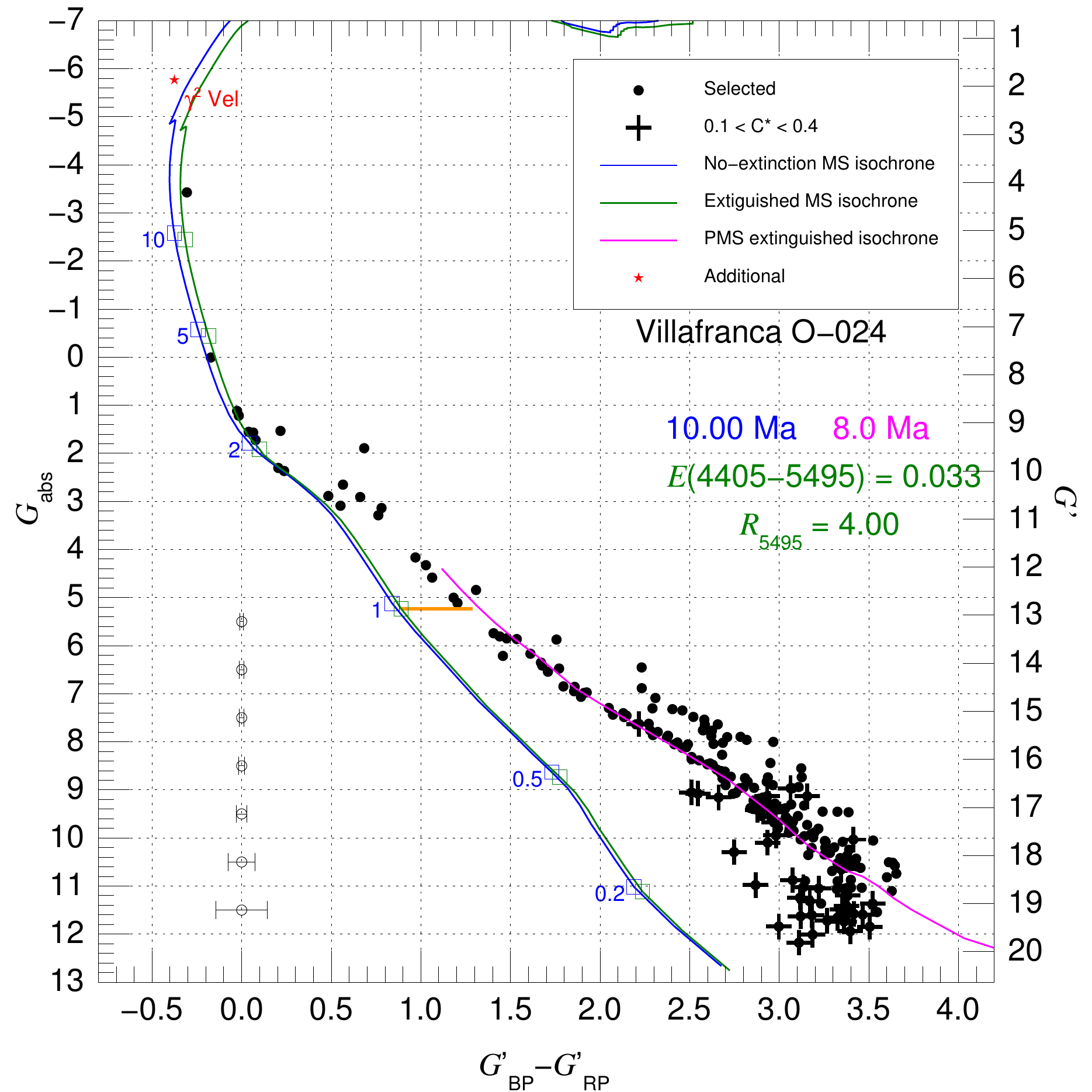}}
 \caption{{\it Gaia}~EDR3 CMDs for four clusters in this paper: \VO{026} ($\sigma$ Orionis cluster, top left), \VO{016} (NGC~2264, top right), \VO{021} (NGC~2362, bottom left), and \VO{024} ($\gamma$~Velorum cluster, bottom right). 
          In each case the sample plotted was selected using less restrictive criteria than for the sample used for the distance (the \sigmae\ restriction was changed to $< 0.5$~mas and no RUWE filter was applied). Also, likely 
          additional members and possible walkaways/runaways are plotted and labelled in red. Crosses indicate objects in the sample with likely contaminated \GBPmGRPc. The error bars in the lower left quadrant of each plot give
          typical color uncertainties for different $G_{\rm abs}$ bins. The blue and green lines show MS Geneva isochrones \citep{LejeScha01} with no extinction and with an average extinction appropriate to the cluster, respectively. 
          Initial masses (in solar units) are labelled along the isochrones and the blue and green text at the right side of each panel gives the age and extinction parameters of the isochrones. The magenta line shows the extinguished 
          PMS isochrone from \citet{Baraetal15} of the age indicated by the text of the same color at the right side of the panel. The PMS isochrones reach up to 1.4~M$_\odot$. The orange line marks the gap in \GBPmGRPc\ between the MS 
          and the PMS at the location of a 1~M$_\odot$~main-sequence star.}
\label{CMDs}    
\end{figure*}

$\,\!$\indent As previously discussed, the immediate goal in this paper of measuring distances to OB groups with precision and accuracy has been achieved. Of course, in the end we are interested in the properties of 
the groups themselves: masses, age, IMF, dynamical state, and extinction. Some of those properties can be obtained through their CMDs and they are the ultimate goal of the Villafranca project but also require additional information. 
In this subsection we discuss the challenges and present some first results. Regarding the challenges, there are several to consider:

\begin{itemize}
 \item The first one is the one presented in the previous subsection. Clusters can eject OB stars and we have to search for possible runaways and walkaways.
 \item The selection criteria for cluster membership has to be less restrictive than the one used to obtain the distances. Regarding astrometry, our strategy when analyzing cluster properties will be to start with the known distance
       and make an initial cut based on normalized parallax of e.g. 3 sigmas and a larger \sigmae\ than 0.1~mas. Stars with bad RUWE or with missing astrometry should be included on a one-by-one basis with the help of additional
       information.
 \item {\it Gaia} photometry (and future spectrophotometry) is compromised in crowded fields (e.g. \VO{001}) and nebular regions (e.g. \VO{013} and \VO{023}). In those cases it is useful to complement the bad-quality (as determined
       by \Cstar) or missing photometry with ground-based data. In our case, we plan to use photometry from GALANTE \citep{Maizetal21d} in future analyses of Galactic OB groups.
 \item Complementary studies using spectroscopy for massive stars and IR or X-ray for PMS stars are needed to constrain the properties. In the case of spectroscopy we will use the results from GOSSS in both
       hemispheres, from AstraLux \citep{Maiz10a} and MONOS \citep{Maizetal19b,Trigetal21} in the northern hemisphere, and from OWN \citep{Barbetal10,Barbetal17} in the southern hemisphere to characterize the multiplicity of the O stars.
 \item In addition to binarity, massive stellar evolution can be strongly affected by rotation, to the point of rapid rotators evolving towards the left of the CMD after settling in the ZAMS \citep{DeMietal10}. Such rapid rotators 
       have to be identified in order to correctly estimate their masses from a CMD.
 \item Many OB group studies assume that the extinction in front of the system is constant in both amount and type but, as shown by \citet{MaizBarb18}, important differences in both can be present and that is indeed what we see
       in some of the Villafranca groups. {\it Gaia}~EDR3 in the form of a spread in the width of the isochrone. To tackle the issue, our future analyses in this project will use the Bayesian photometric tool CHORIZOS \citep{Maiz04c}
       to calculate extinction on a star-by-star basis.
\end{itemize}

Given those challenges and that in some aspects we do not have all the necessary information at this point, here we present preliminary results just for the four nearby OB groups with the lowest extinction: \VO{026} ($\sigma$ Orionis 
cluster), \VO{016} (NGC~2264), \VO{021} (NGC~2362), and \VO{024} ($\gamma$~Velorum cluster). To select the sample in each cluster we used the previously obtained distances, eliminated the RUWE requirement, and included stars with 
$\sigmae < 0.5$~mas, leaving the rest of the filters unchanged. The resulting CMDs are shown in Fig.~\ref{CMDs}, where we also included additional members and runaways/walkaways discussed in the text above. We have also plotted the 
Geneva MS isochrones \citep{LejeScha01}, both without and with extinction, choosing the closest age (among 1.78~Ma, 3.16~Ma, and 10.0~Ma) to the one estimated for the cluster and the extinguished PMS isochrone from \citet{Baraetal15} 
that is a best fit for the cluster\footnote{\label{foot}The photometry for the MS isochrones is calculated from the SEDs using the synthetic photometry package used for CHORIZOS. Such SEDs are not available for the PMS isochrones, so 
we use their {\it Gaia} photometry downloaded from \url{http://perso.ens-lyon.fr/isabelle.baraffe/BHAC15dir/}
and extinguished it using the values for a typical star. As a consequence, there may be small offsets at the level of a few 
hundredths of a magnitude which would correspond to age differences of $\sim$0.5~Ma. This would affect all clusters in a similar way, so the relative ages should not change.}. The photometry is not corrected from extinction but, 
given the low values measured (shown in Fig.~\ref{CMDs} and taken from the data for a reference star from \citealt{MaizBarb18}), its effect is small as evident from the two isochrones in each plot.

The observed CMDs (Fig.~\ref{CMDs}  are similar for the four clusters in showing a MS that extends from $\sim$2~M$_\odot$ upwards and a well defined PMS that extends downwards. The PMS shows an equal-mass binary sequence 0.75~mag above the 
primary one that is especially well visible for \VO{024}. The large dynamic range in magnitude provided by {\it Gaia} is at full display
in Fig.~\ref{CMDs}. The first difference arises from the number of stars detected in each group, with \VO{016} being the richest one and \VO{026} the poorest one. A second difference is caused by the different distances: members are 
detected for all four clusters down to $\GGc\sim 19.5$~mag (right vertical axes) but that value corresponds to different $G_{\rm abs}$ (left vertical axes), allowing us to reach fainter absolute magnitudes 
for the two closest clusters, \VO{024} and \VO{026}. The spread in \GBPmGRPc\ for faint stars is mostly caused by large uncertainties in the measured colors. The third and most relevant difference is the separation between the MS (as 
extrapolated from that for massive stars) and the observed PMS. To determine it, we have measured the $\Delta(\GBPmGRPc)$ at the position expected for a 1~M$_\odot$ at the MS and marked it with orange lines in Fig.~\ref{CMDs}. 
{\bf The measured values determine an age sequence from the youngest, \VO{026} (0.7~mag), to \VO{016} (0.6~mag) to \VO{021} (0.5~mag) to the oldest, \VO{024} (0.4~mag), as the PMS stars approach the MS as they age. 
The corresponding ages from fitting the best PMS isochrone from \citet{Baraetal15} are 2.0$\pm$0.5~Ma, 4$\pm$2~Ma, 5.0$\pm$0.5~Ma, and 8$\pm$2~Ma, respectively}, with an estimated additional systematic uncertainty of 0.5~Ma due to the 
photometric calibration (see footnote~\ref{foot}). Those ages are in good agreement with the values (a) for \VO{026} from \citet{Zapaetal02} of 2-4~Ma based on \LiI{6708} absorption, (b) for \VO{016} from \citet{Sungetal04} of 
3.1$\pm$0.5~Ma based on PMS fitting, (c) for \VO{021} from \citet{Dahm05} of 3.5-5.0~Ma based on PMS fitting, and (d) for \VO{024} from \citet{Jeffetal09} of 5-10~Ma from a combination of methods. The large spread of the PMS (and 
corresponding uncertainty in its age) for \VO{016} may be caused by differential extinction, known to exist in NGC~2264, or by a mixture of ages.

We observe no significant discrepancies between the PMS ages and the ones constrained by the upper main sequence. However, the latter age constraints are weak (assuming low-rotation, single-star-like evolution), given that these clusters 
are relatively low mass and only have one or two O stars each. \VO{026} only has one late-O dwarf, so that only gives an upper age limit of $\sim$5~Ma. The earliest O-type in \VO{016} is an O7~V, constraining the age to be 4~Ma or less. 
\VO{021} contains two evolved late-O stars, giving an upper age limit of 6-7~Ma. Finally, the most massive system in \VO{024} is composed of an evolved WR+O binary, whose components are unlikely to have followed a single-star-like 
evolution, in which case they can be somewhat older than an isolated O star. 

\begin{table}
\caption{Cluster members by mass range. The last column is the expected result from a Kroupa IMF based on the measured number in the 2-50~M$_\odot$ range.}
\centerline{
\begin{tabular}{lccc}
\hline
Cluster  & 2-50 M$_\odot$ & 8-50 M$_\odot$ & 8-50 M$_\odot$ \\
         & measured       & measured       & expected       \\
\hline
\VO{026} & 11             & 3              & 1.7$\pm$1.2    \\ % sigma Ori Aa+Ab+B
\VO{016} & 33             & 6              & 5.0$\pm$2.1    \\ % 15 Mon Aa+Ab, HD 47 887, HD 47 961, HD 47 777, 15 Mon B
\VO{021} & 44             & 5              & 6.7$\pm$2.4    \\ % tau CMa Aa+Ab+E, tau CMa D, CPD -24 2213
\VO{024} & 11             & 4              & 1.7$\pm$1.2    \\ % gamma^2 Vel A+B, gamma^1 Vel A+B
\hline
\end{tabular}
}
\label{IMF}
\end{table}
% kk   = kroupaint([0.16, 1.30, 2. , 8., 50.],/M12,MINT=mint)
% norm = 100/mint[0]
% print, norm*kk
%      235.84646       11.788670       13.113382       2.3509586
% print, norm*mint
%      100.00000       18.775573       46.306451       37.974198
% p    = kk[3]/(kk[2]+kk[3])
% q    = 1-p
% n    = [11,33,44,10]
% print, n*p
%      1.6722695       5.0168084       6.6890778       1.5202450
% print, sqrt(n*p*q)
%      1.1908163       2.0625544       2.3816327       1.1353988

The single most interesting CMD in Fig.~\ref{CMDs} is that of \VO{026}, the youngest and less massive of the four. The only two objects found between the end of the PMS isochrone (at a mass of $\sim$1.4~M$_\odot$) and the first objects
close to the MS isochrone (at a mass of $\sim$2.0~M$_\odot$) had to be inserted by hand as one of them (Mayrit~\num{240322}) has poor-quality astrometry and the other one (Mayrit~\num{968292}) is outside the region but is a possible 
runaway. The gap is very prominent in the CMD: if we exclude the two doubtful stars, it amounts to 1.2~mag in \GBPmGRPc\ and 2.4~mag in $G_{\rm abs}$. This could indicate that stars with $\gtrsim$2~M$_\odot$ evolve towards the MS 
significantly faster than lower-mass objects, an effect already apparent in Fig.~14 of \citet{Penaetal12}. 
It would be interesting to test this hypothesis with other clusters of similar age, especially if they are more massive and better populated.

We finally make an estimate of the IMF for the upper MS of each of the four clusters by using the CMDs to count the number of stars with masses in the 8-50~M$_\odot$ range (massive stars) and in the 2-50~M$_\odot$ range (intermediate 
and massive stars), for which a Kroupa IMF predicts a ratio of 0.15. Known binary stars are separated to be properly accounted for and the results are listed in Table~\ref{IMF}. For the two most massive clusters (\VO{016} and \VO{021}) 
the measured number of stars in the 8-50~M$_\odot$ range is within one sigma of the expected value from a Kroupa IMF. The two less massive clusters (\VO{026} and \VO{024}) have 3 and 4 stars, respectively, 
in the 8-50~M$_\odot$ range, which in both cases 
implies a slight overrepresentation of massive stars (between one and two sigmas) with respect to a Kroupa IMF and is a likely manifestation of the increased stochasticity of the IMF for small clusters. In the case of the two older 
clusters it is possible that some stars have already exploded as SNe, which if true would be especially relevant for \VO{024}, as that would push its number of initial 8-50~M$_\odot$ stars beyond a two-sigma result for a Kroupa IMF.
For this cluster \citet{Prisetal16} measured a total mass of 100~M$_\odot$ in the 0.16-1.30~M$_\odot$ range distributed in a manner consistent with a Kroupa IMF there. An extrapolation into larger masses from those values assuming a 
Kroupa IMF yields 15.5 stars in the 2-50~M$_\odot$ range and 2.4 stars in the 8-50~M$_\odot$ range. Both results are approximately one sigma from the measured numbers of 11 and 4 but in opposite directions, indicating that the cluster is
slightly deficient in intermediate-mass stars and slightly overabundant in massive ones.

\subsection{Future work}

$\,\!$\indent Our immediate plans are to include more OB groups in the Villafranca sample using {\it Gaia}~EDR3, GOSSS, and \lili\ data. The project will be coordinated with the ALS catalog \citep{Reed03,Pantetal21}, which is 
compiling and analyzing the OB stars included in {\it Gaia}. The next Villafranca paper will concentrate on the solar neighborhood to complement the analysis of \citet{Pantetal21} on the mapping of the spatial distribution of massive
stars and will also include clusters with B but no O stars. In the long term, {\it Gaia}~DR4 data will be incorporated into our analysis.

\begin{acknowledgements}
We thank J. A. Caballero for his help with the Mayrit stars of Escorial~7.
J.~M.~A., R.~F.~A., M.~P.~G., P.~C.~B., and A.~S. acknowledge support from the Spanish Government Ministerio de Ciencia e Innovaci\'on through grant PGC2018-\num{095049}-B-C22. 
R.~H.~B. acknowledges support from ANID FONDECYT Regular Project \num{1211903} and the ESAC visitors program.
E.J.A. acknowledges support from the State Agency for Research of the Spanish Government Ministerio de Ciencia e Innovaci\'on through the ``Center of Excellence Severo Ochoa'' award to 
the Instituto de Astrof{\'\i}sica de Andalucía (SEV-2017-0709) and through grant PGC2018-\num{095049}-B-C21
This work has made use of data from the European Space Agency (ESA) mission \href{https://www.cosmos.esa.int/gaia}{\it Gaia}, 
 processed by the {\it Gaia} Data Processing and Analysis Consortium (\href{https://www.cosmos.esa.int/web/gaia/dpac/consortium}{DPAC}).
Funding for the DPAC has been provided by national institutions, in particular the institutions participating in the {\it Gaia} Multilateral Agreement. 
The {\it Gaia} data is processed with the the computer resources at Mare Nostrum and the technical support provided by BSC-CNS.
%Additionally, this paper includes data obtained with the MPG/ESO 2.2~m Telescope at the Observatorio~de~La~Silla, Chile; 
%the 2.5~m du Pont Telescope at the Observatorio de Las Campanas, Chile; the 10~m Hobby-Eberly Telescope at McDonald Observatory, Texas, U.S.A.; 
%the 4.2~m William Herschel Telescope and the 10.4~m Gran Telescopio Canarias at the Observatorio del Roque de los Muchachos, La Palma, Spain; and 
%the 2.2~m Telescope at the Centro Astron\'omico Hispano Andaluz, Almer{\'\i}a, Spain.
%and the JAST/T80 telescope at the Observatorio Astrof{\'\i}sico de Javalambre, Teruel, Spain (owned, managed, and operated by the Centro de 
%Estudios de F{\'\i}sica del Cosmos de Arag\'on). 
%We thank the staff at those observatories for their support.
This research has made extensive use of the \href{http://simbad.u-strasbg.fr/simbad/}{SIMBAD} and 
\href{https://vizier.u-strasbg.fr/viz-bin/VizieR}{VizieR} databases, operated at 
\href{https://cds.u-strasbg.fr}{CDS}, Strasbourg, France. 
\end{acknowledgements}

%------------------------------------------------------------------

\bibliographystyle{aa} % style aa.bst
\bibliography{general} % your references references.bib

\begin{appendix}

\section{Additional tables and figures}

\begin{figure*}
\centerline{\includegraphics[width=\linewidth]{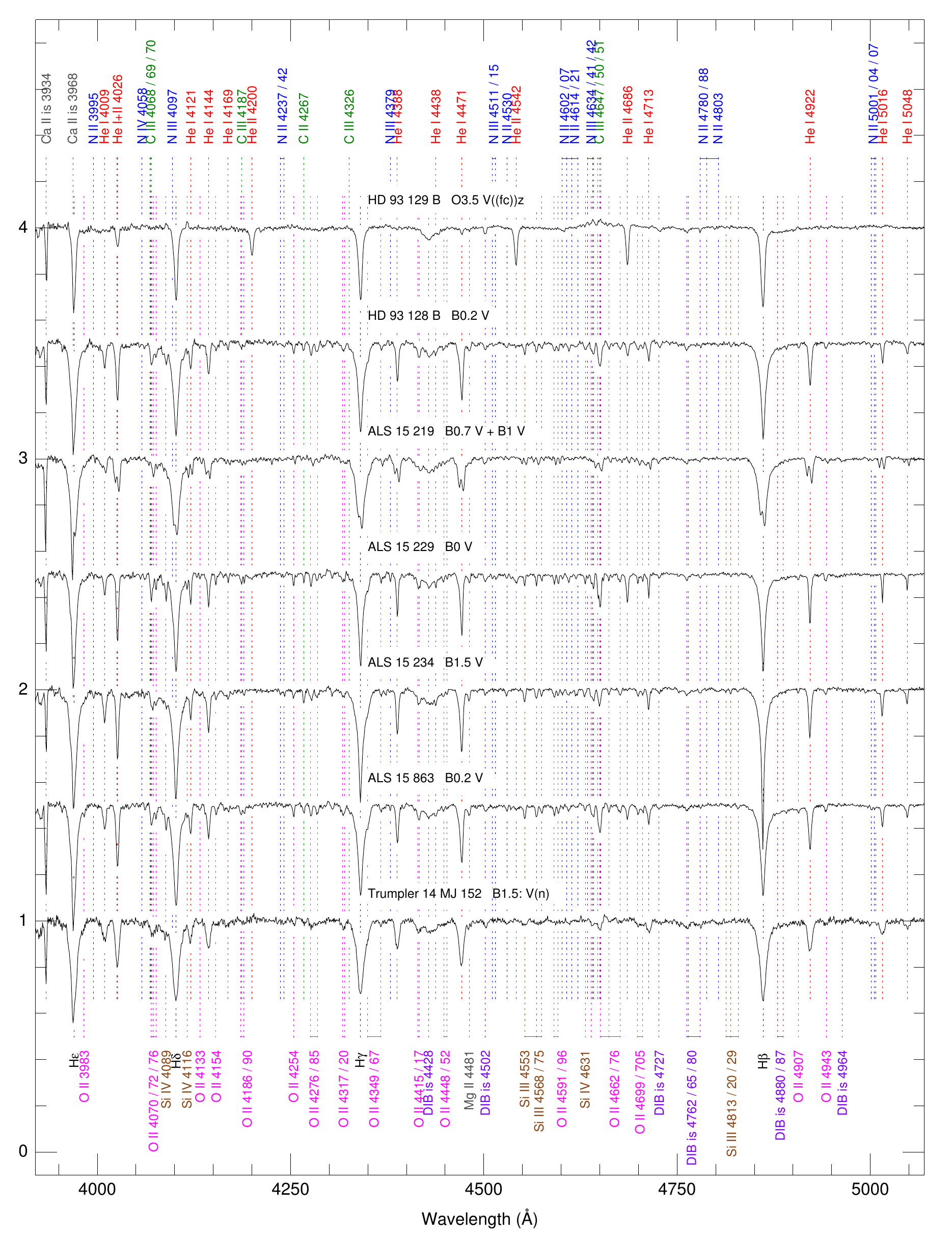}}
\caption{New GOSSS spectra for stars in \VO{002}.}
\label{GOSSS_spectra_01}
\end{figure*}

\begin{figure*}
\centerline{\includegraphics[width=\linewidth]{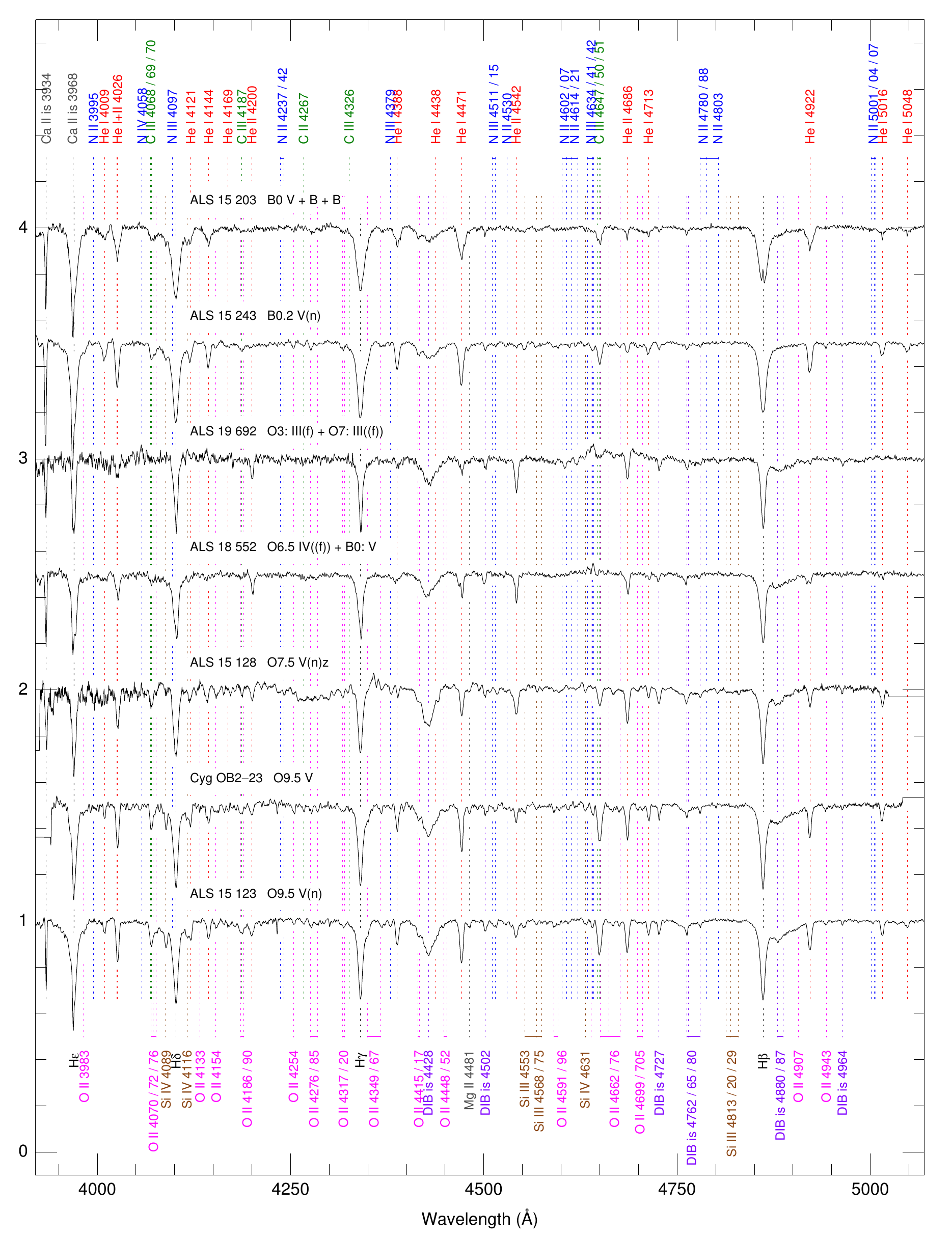}}
\caption{New GOSSS spectra for stars in \VO{002}, \VO{003}, \VO{005}, \VO{006}, \VO{007}, and \VO{008}.}
\label{GOSSS_spectra_02}
\end{figure*}

\begin{figure*}
\centerline{\includegraphics[width=\linewidth]{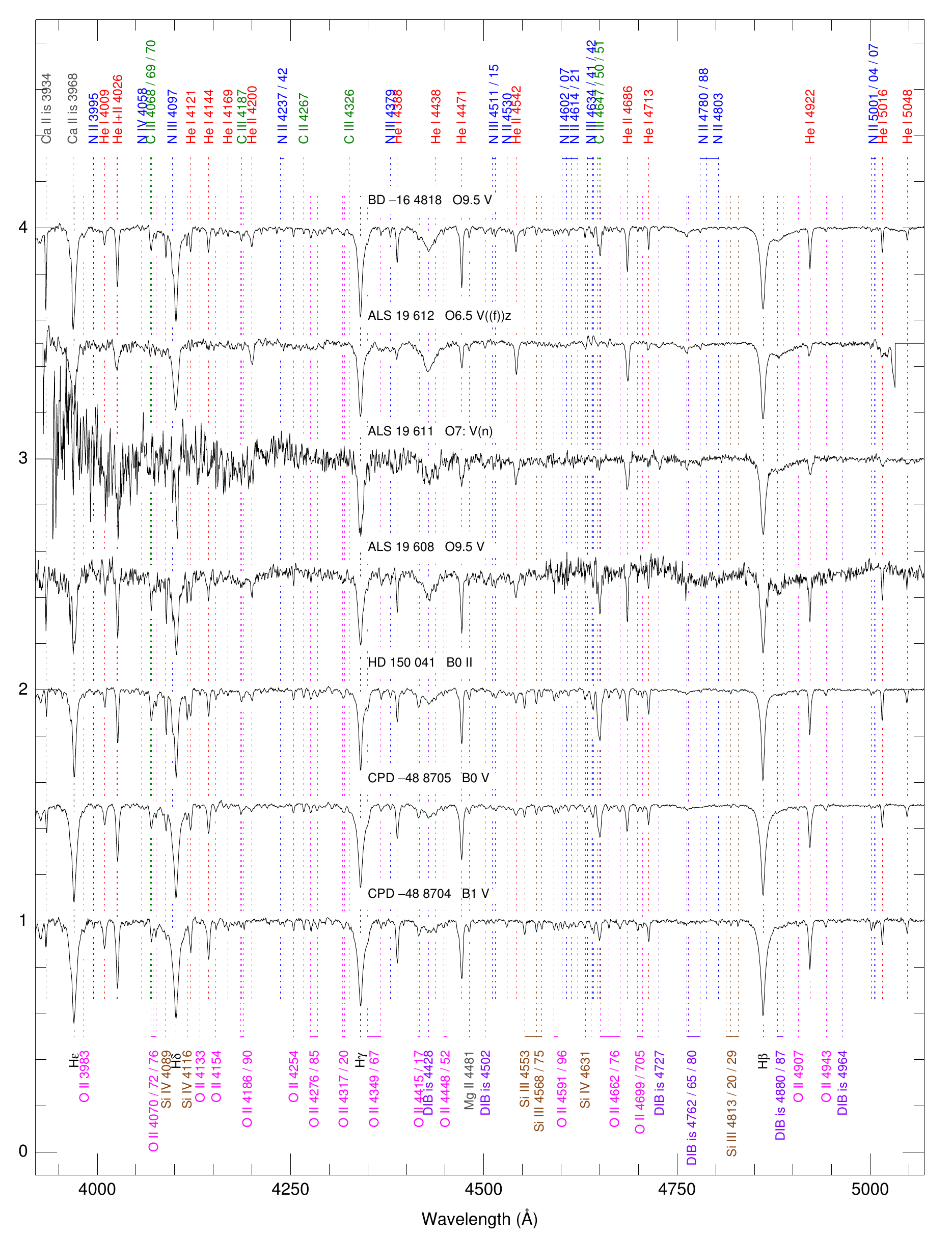}}
\caption{New GOSSS spectra for stars in \VO{009} and \VO{010}.}
\label{GOSSS_spectra_03}
\end{figure*}

\begin{figure*}
\centerline{\includegraphics[width=\linewidth]{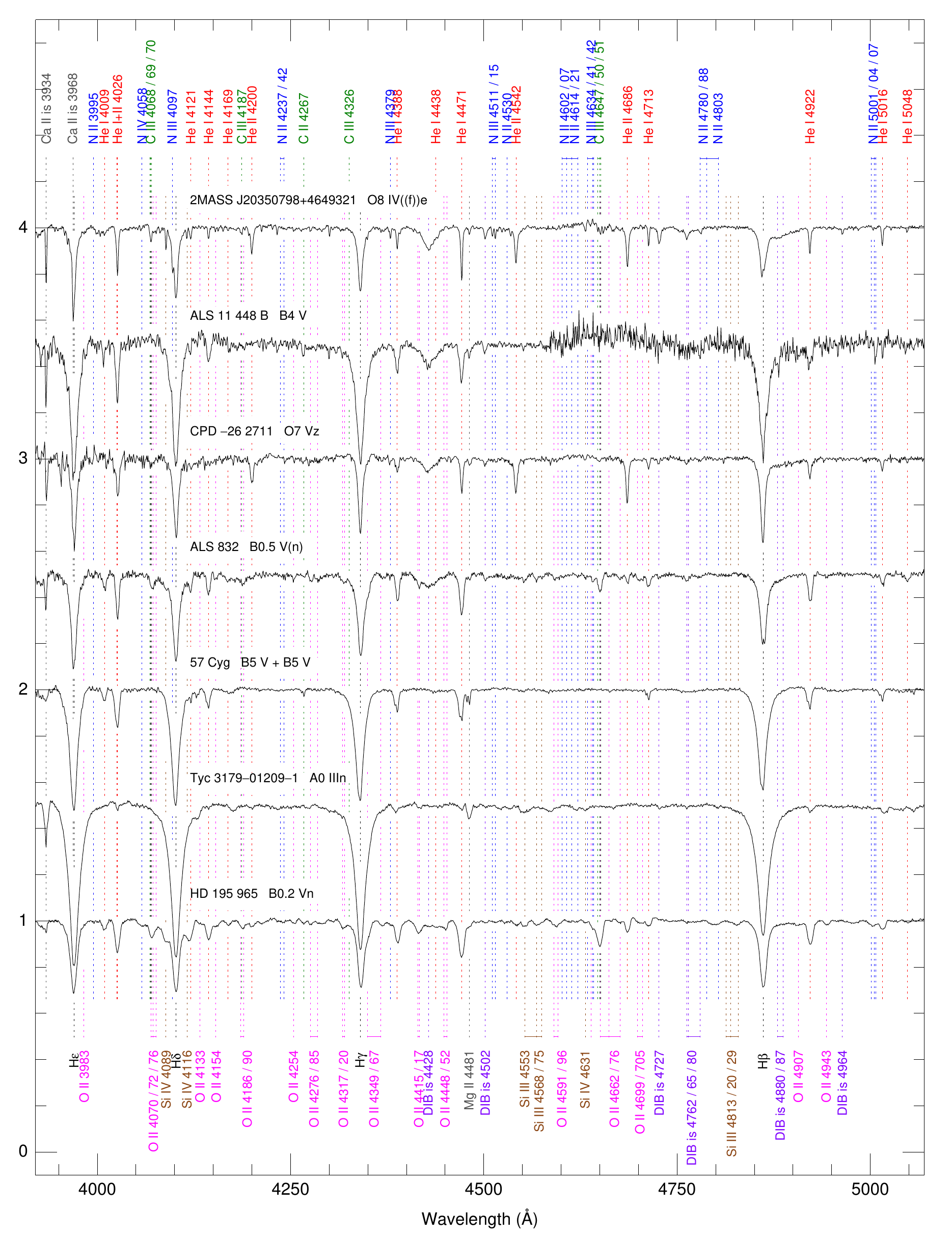}}
\caption{New GOSSS spectra for stars in \VO{011}, \VO{012}, and \VO{014}.}
\label{GOSSS_spectra_04}
\end{figure*}

\begin{figure*}
\centerline{\includegraphics[width=\linewidth]{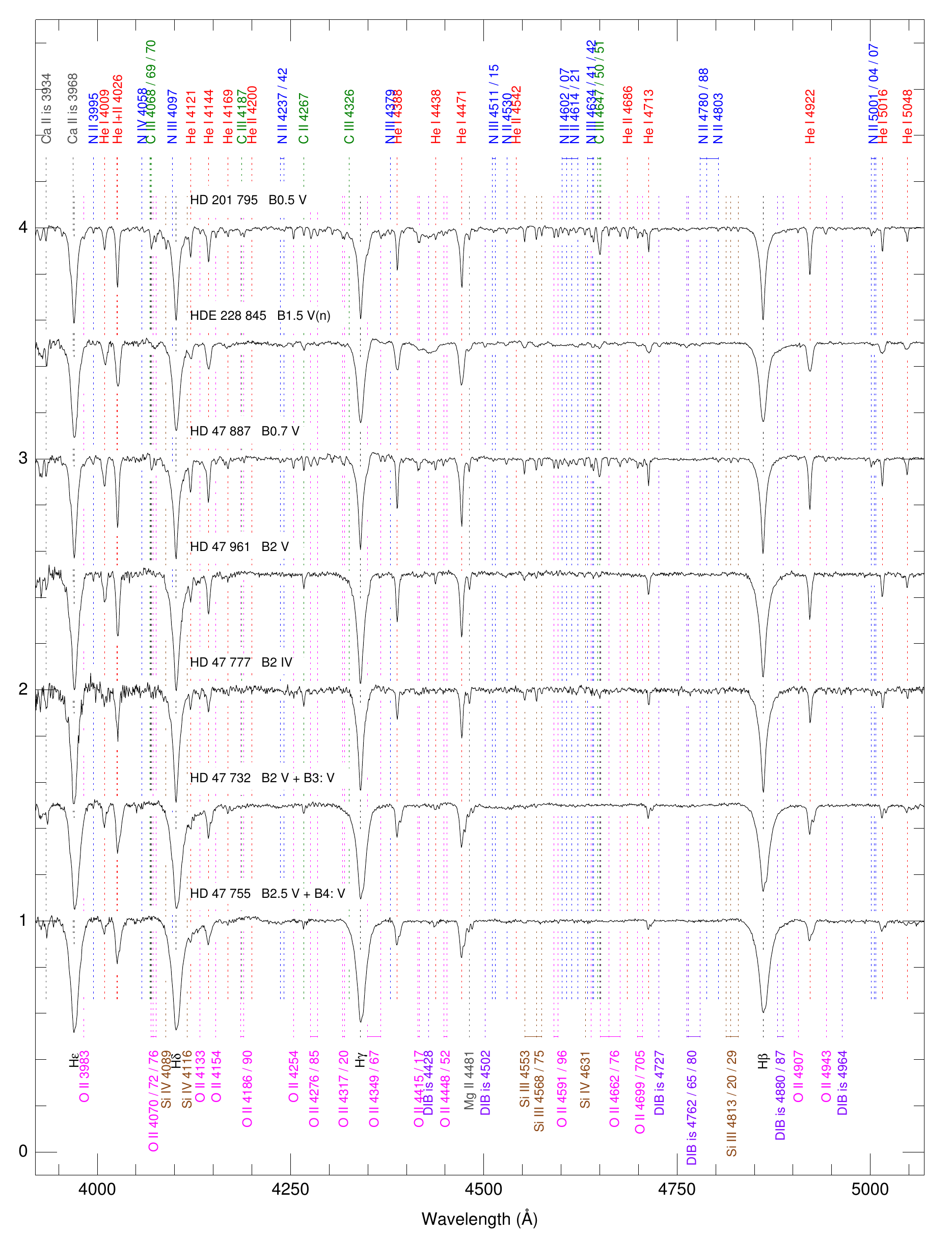}}
 \caption{New GOSSS spectra for stars in \VO{014}, \VO{015}, and \VO{016}.}
\label{GOSSS_spectra_05}
\end{figure*}

\begin{figure*}
\centerline{\includegraphics[width=\linewidth]{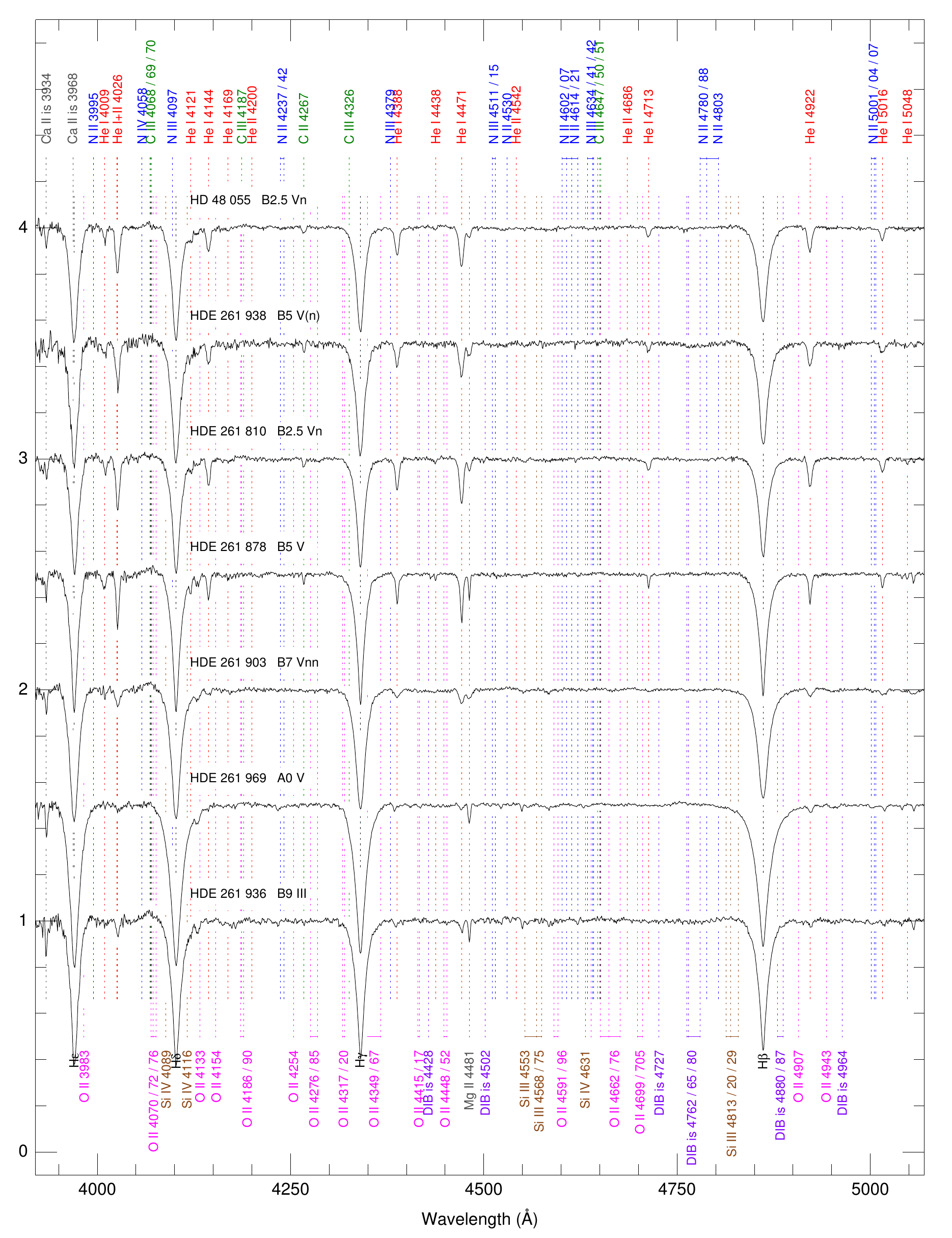}}
\caption{New GOSSS spectra for stars in \VO{016}.}
\label{GOSSS_spectra_06}
\end{figure*}

\begin{figure*}
\centerline{\includegraphics[width=\linewidth]{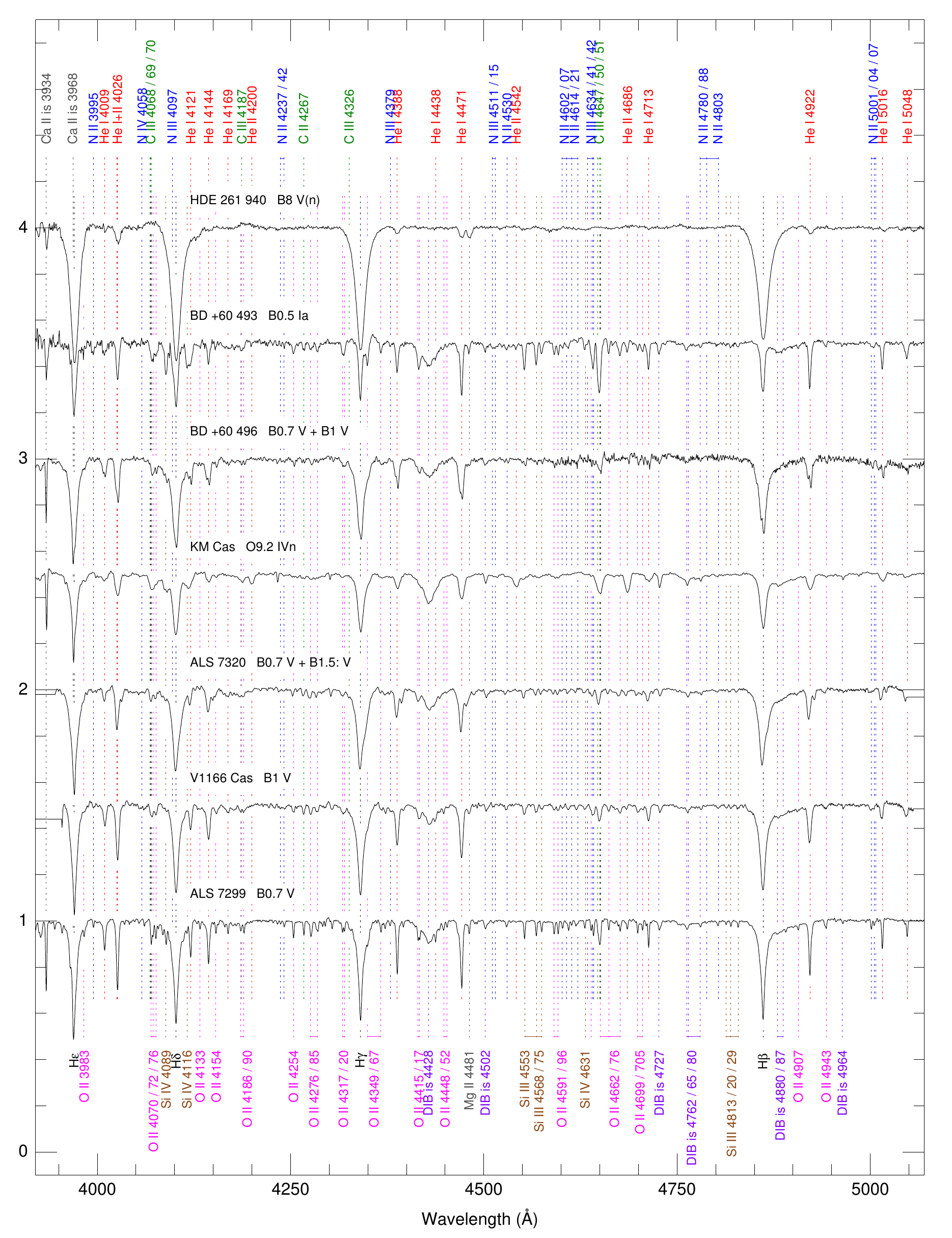}}
 \caption{New GOSSS spectra for stars in \VO{016} and \VO{017}.}
\label{GOSSS_spectra_07}
\end{figure*}

\begin{figure*}
\centerline{\includegraphics[width=\linewidth]{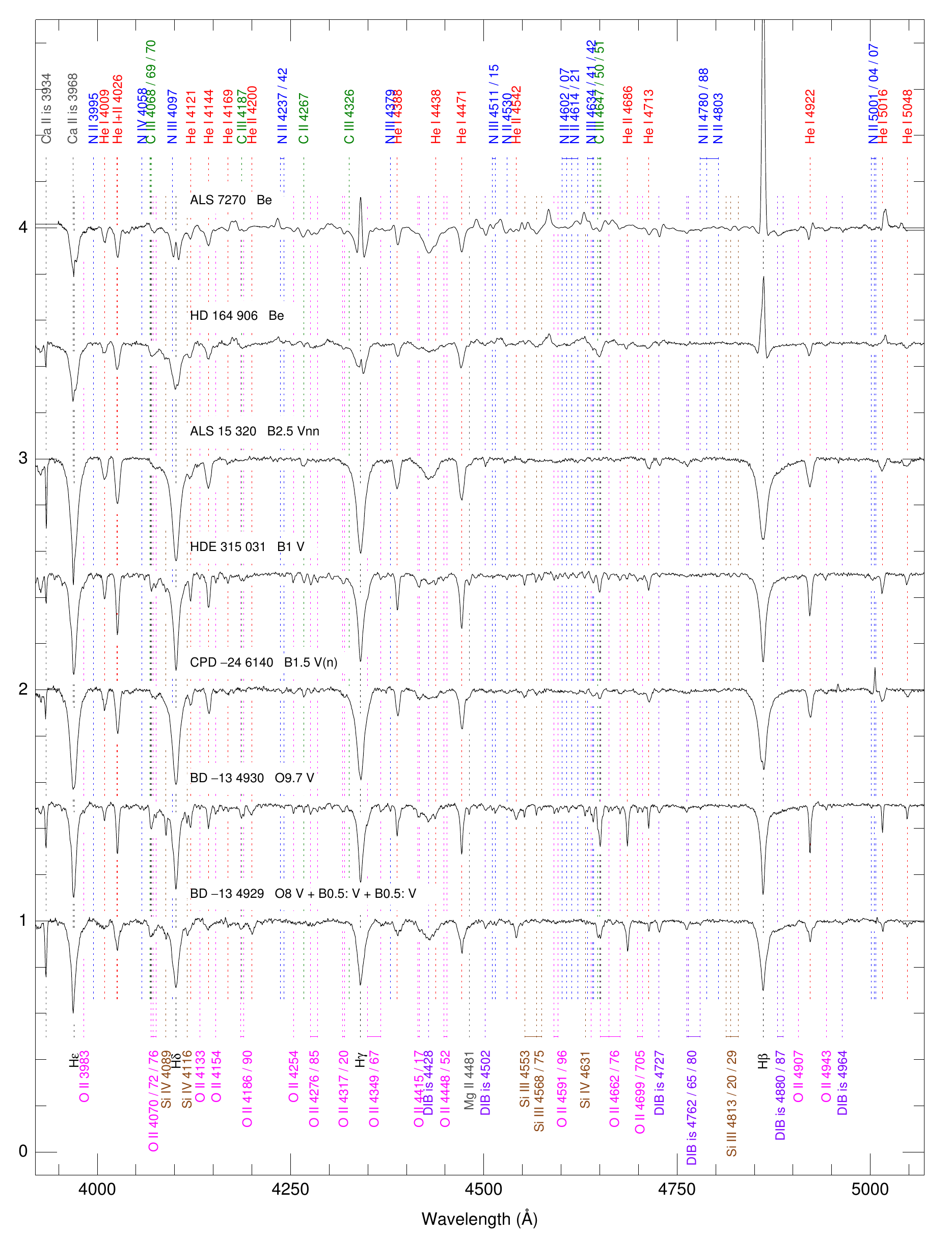}}
 \caption{New GOSSS spectra for stars in \VO{017}, \VO{018}, and \VO{019}.}
\label{GOSSS_spectra_08}
\end{figure*}

\begin{figure*}
\centerline{\includegraphics[width=\linewidth]{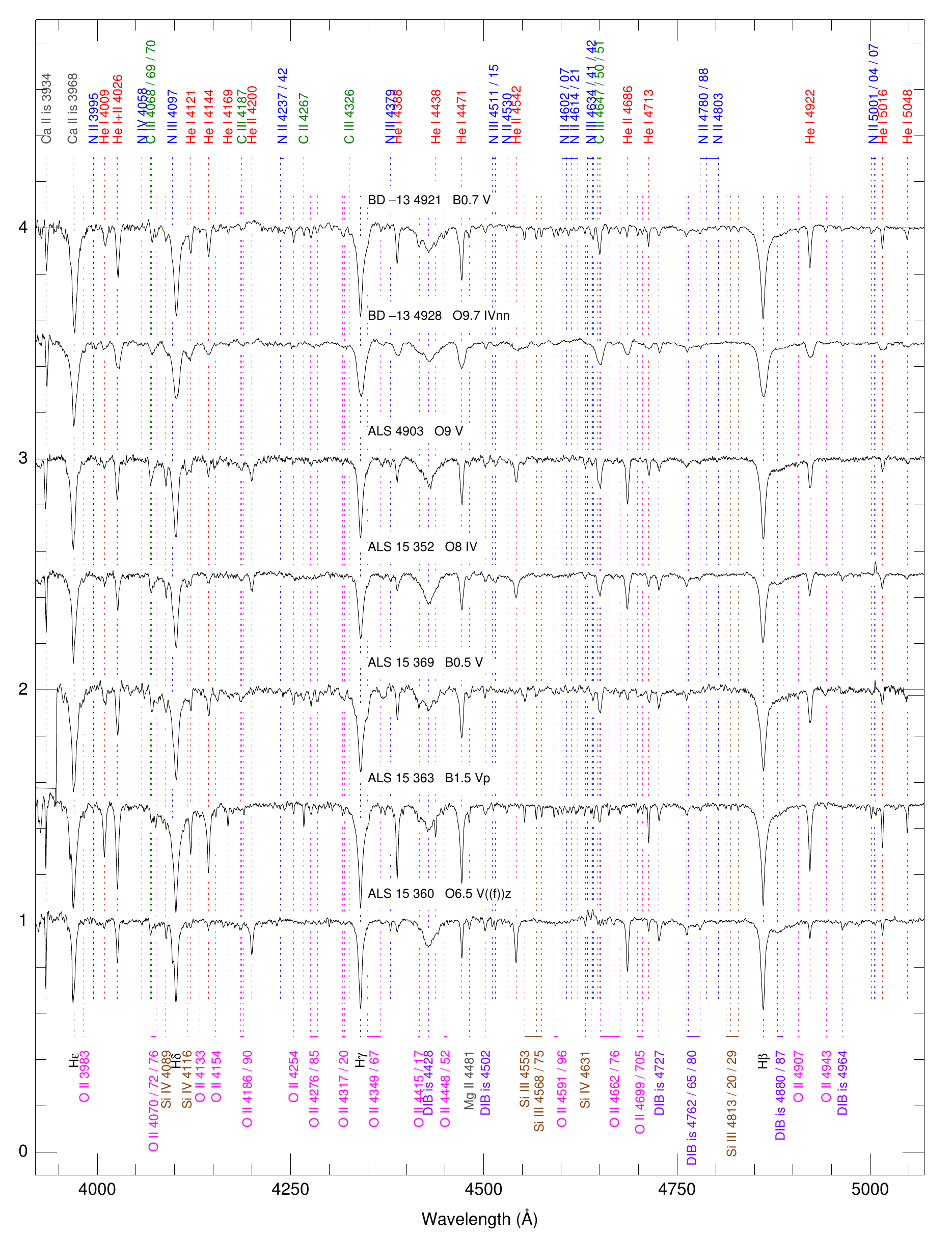}}
\caption{New GOSSS spectra for stars in \VO{019}.}
\label{GOSSS_spectra_09}
\end{figure*}

\begin{figure*}
\centerline{\includegraphics[width=\linewidth]{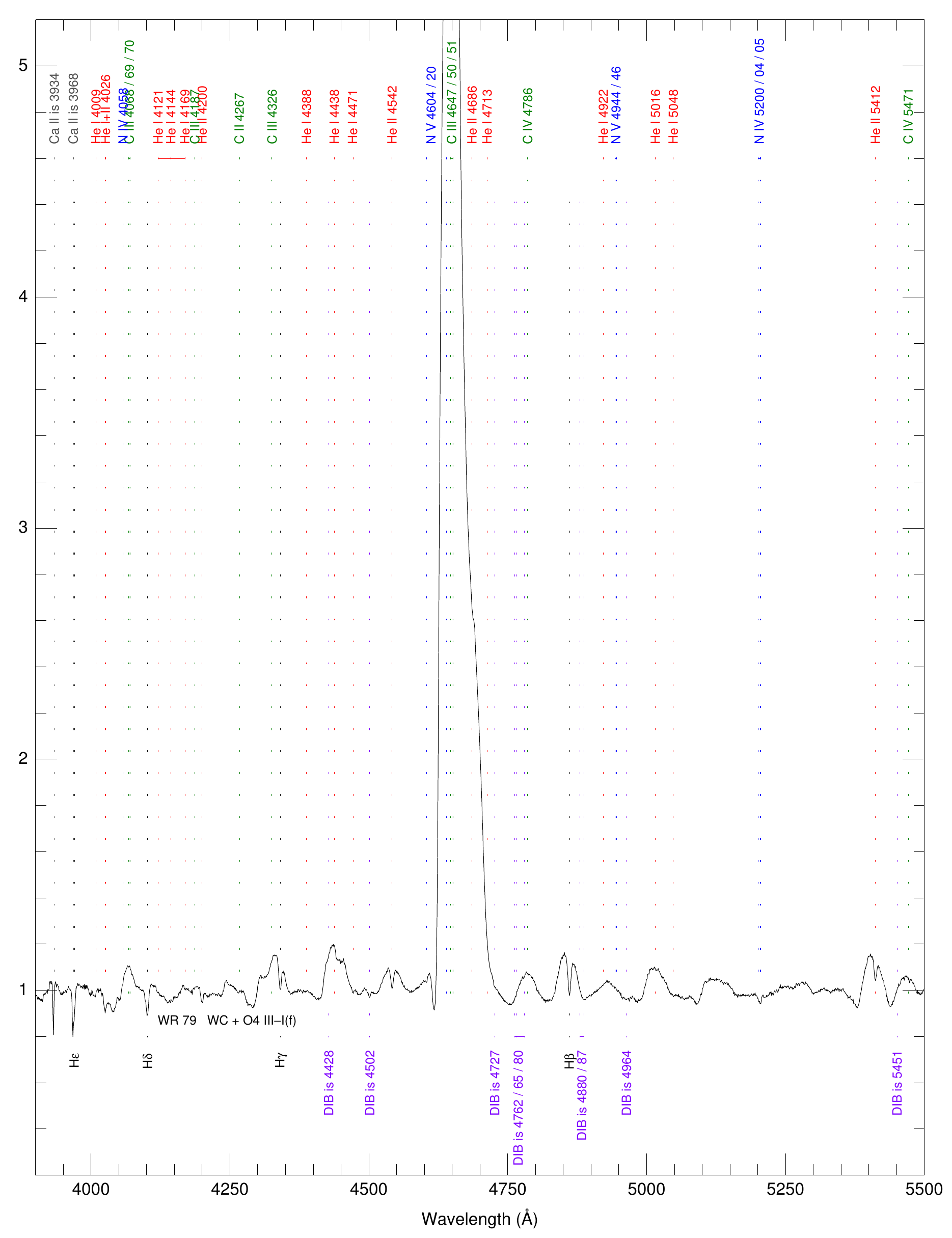}}
\caption{New GOSSS spectra for a star in \VO{022}. Note the expanded wavelength range.}
\label{GOSSS_spectra_10}
\end{figure*}

\begin{figure*}
\centerline{\includegraphics[width=\linewidth]{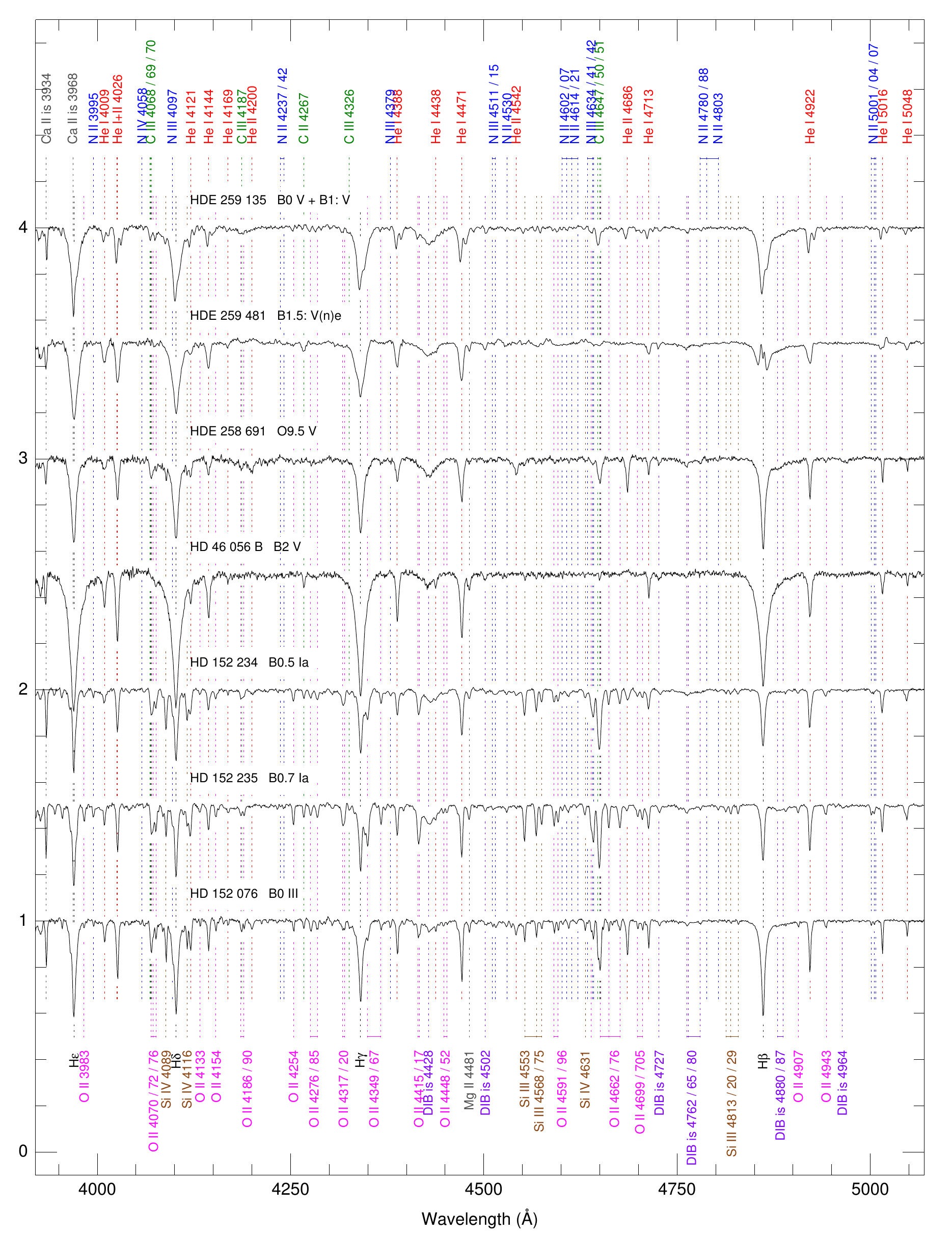}}
 \caption{New GOSSS spectra for stars in \VO{020} and \VO{022}.}
\label{GOSSS_spectra_11}
\end{figure*}

\begin{figure*}
\centerline{\includegraphics[width=\linewidth]{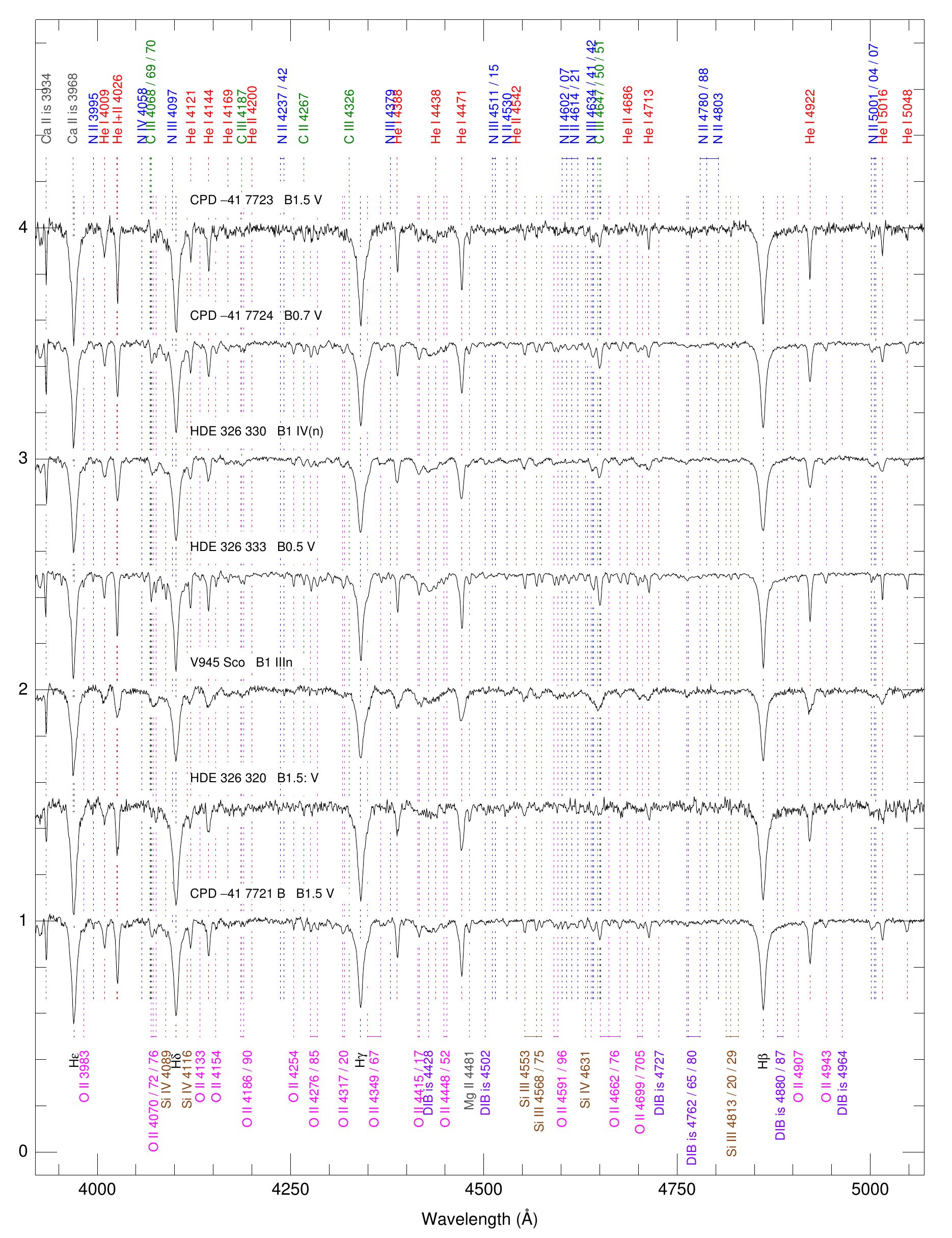}}
\caption{New GOSSS spectra for stars in \VO{022}.}
\label{GOSSS_spectra_12}
\end{figure*}

\begin{figure*}
\centerline{\includegraphics[width=\linewidth]{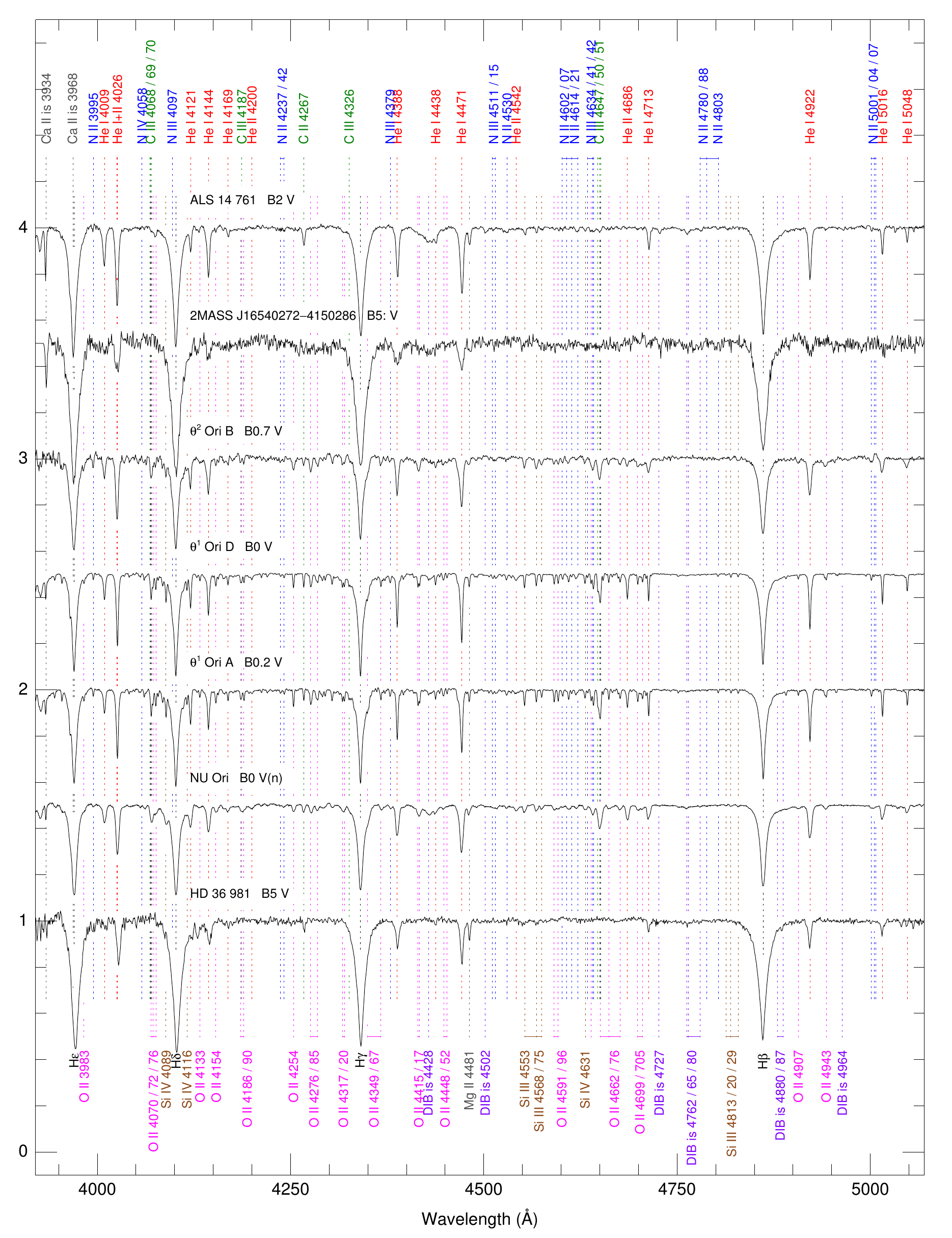}}
 \caption{New GOSSS spectra for stars in \VO{022} and \VO{023}.}
\label{GOSSS_spectra_13}
\end{figure*}

\begin{figure*}
\centerline{\includegraphics[width=\linewidth]{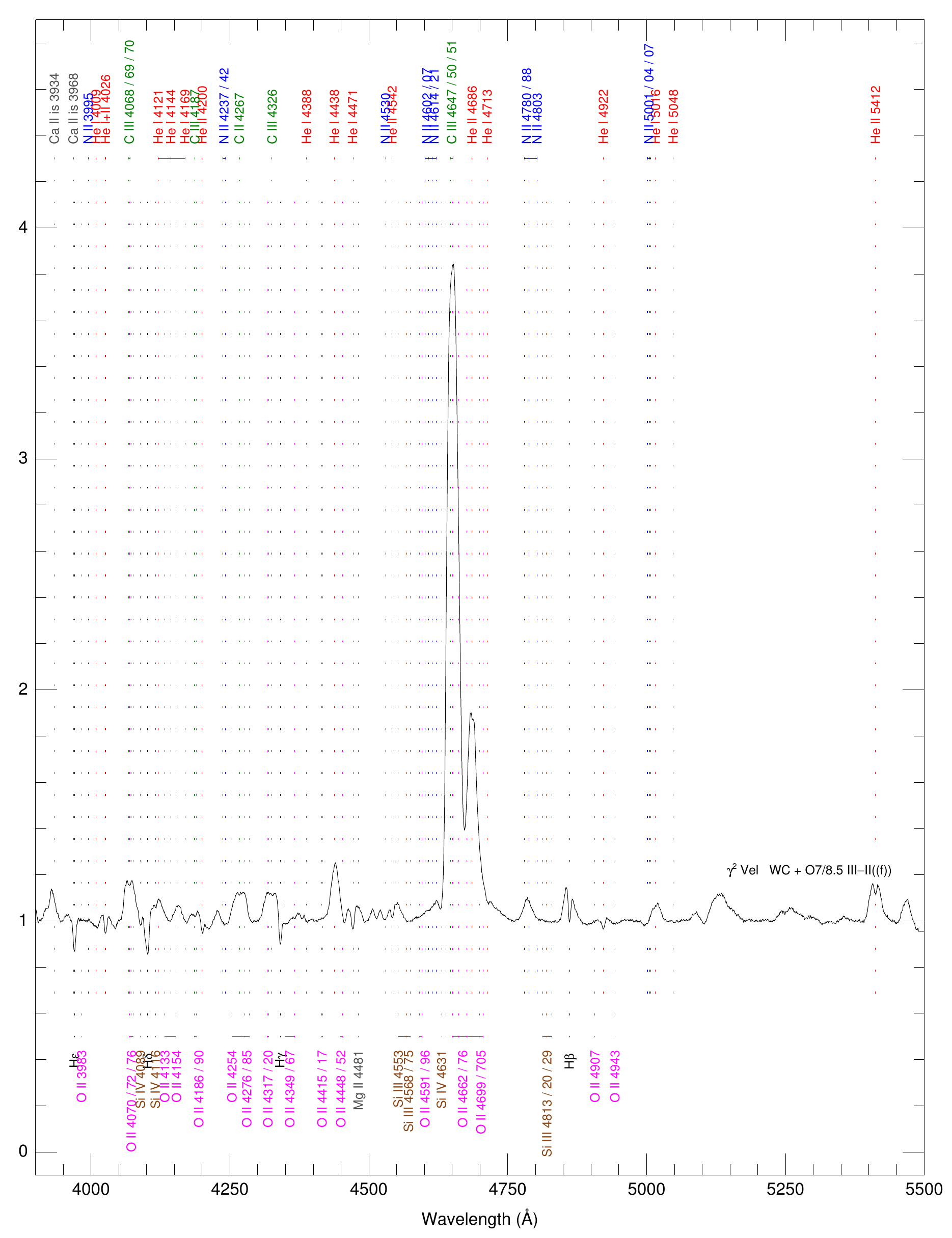}}
\caption{New GOSSS spectra for a star in \VO{024}. Note the expanded wavelength range.}
\label{GOSSS_spectra_14}
\end{figure*}

\begin{figure*}
\centerline{\includegraphics[width=\linewidth]{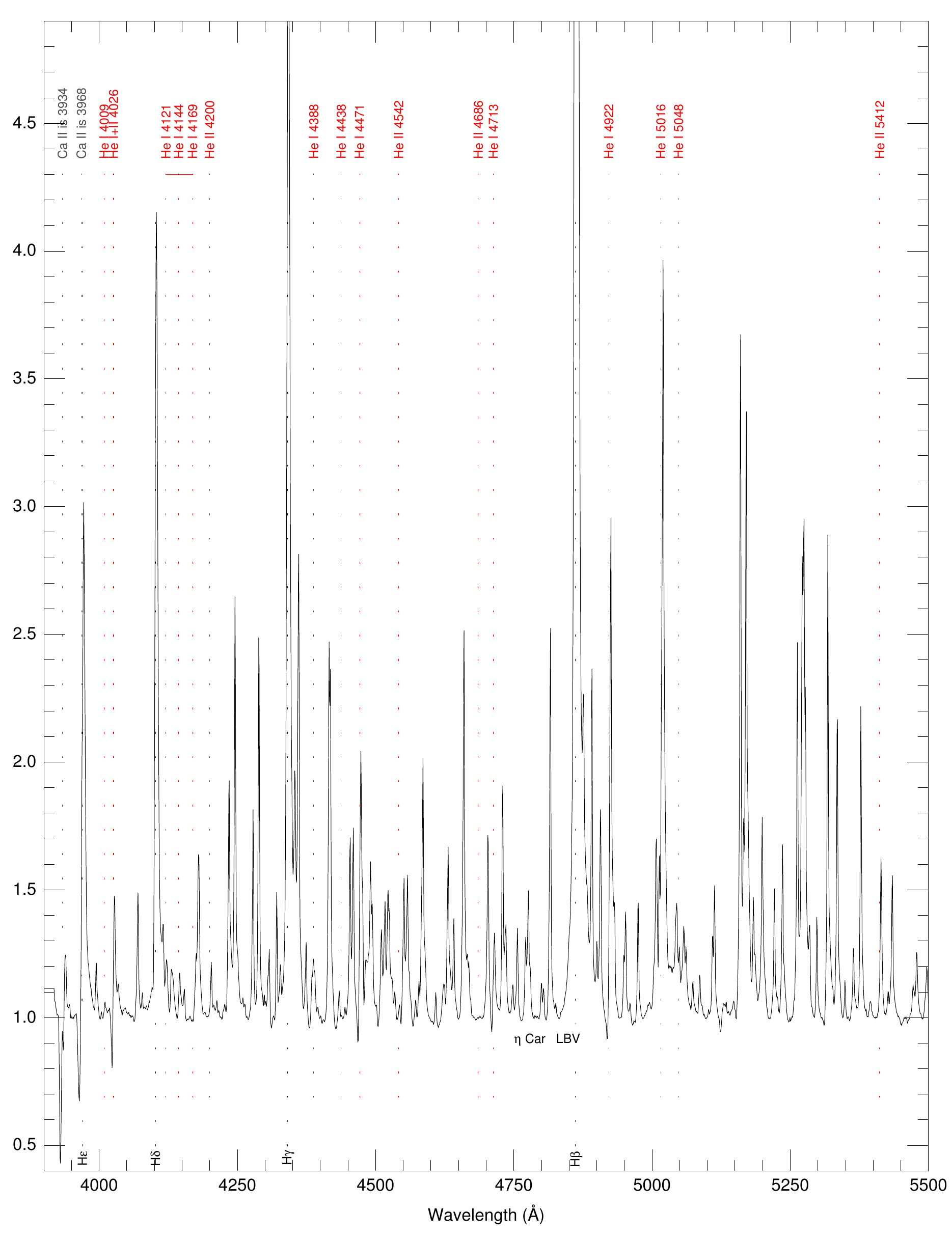}}
\caption{New GOSSS spectra for a star in \VO{025}. Note the expanded wavelength range.}
\label{GOSSS_spectra_15}
\end{figure*}

\begin{figure*}
\centerline{\includegraphics[width=\linewidth]{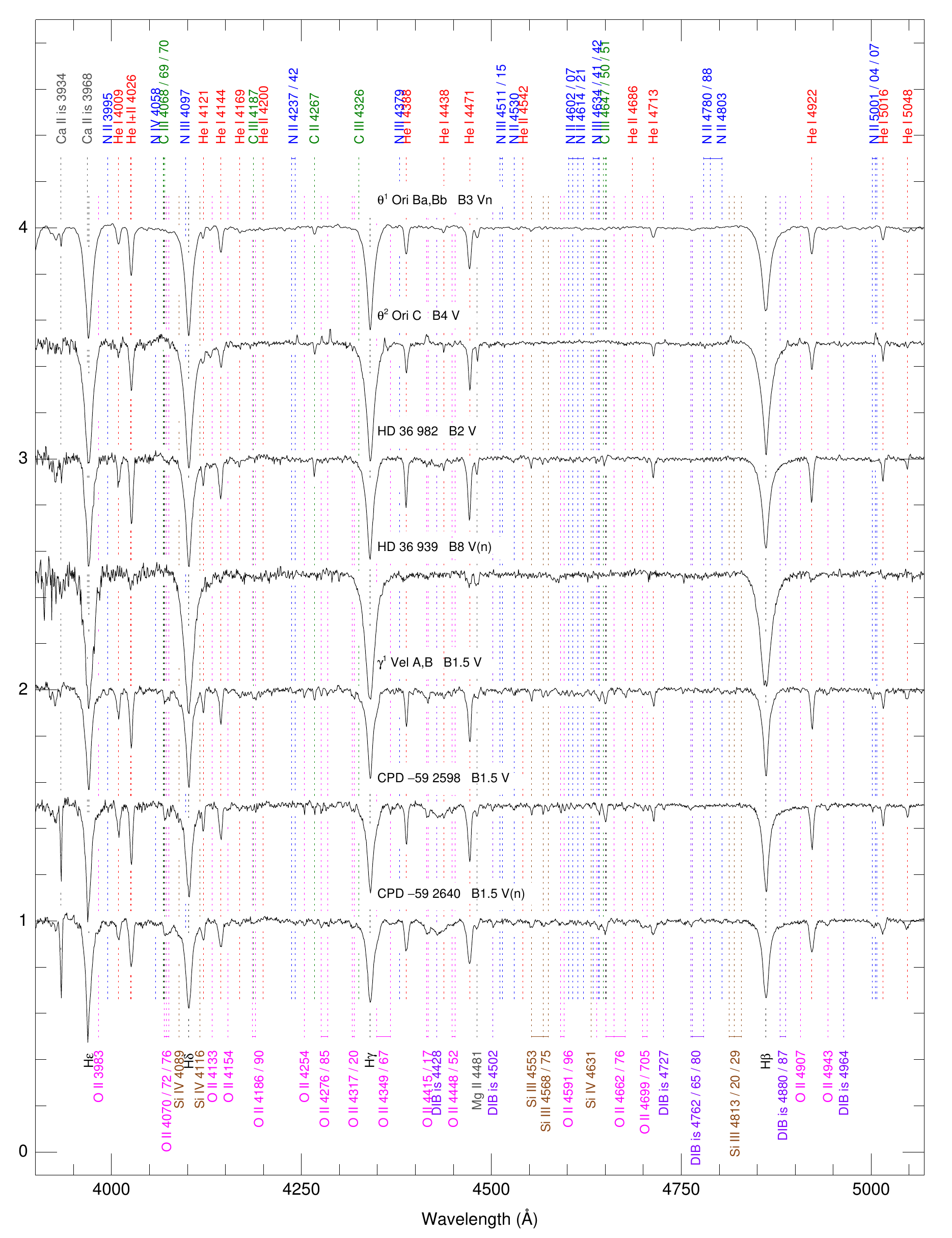}}
 \caption{New GOSSS spectra for stars in \VO{023}, \VO{024}, and \VO{025}.}
\label{GOSSS_spectra_16}
\end{figure*}

\begin{figure*}
\centerline{\includegraphics[width=\linewidth]{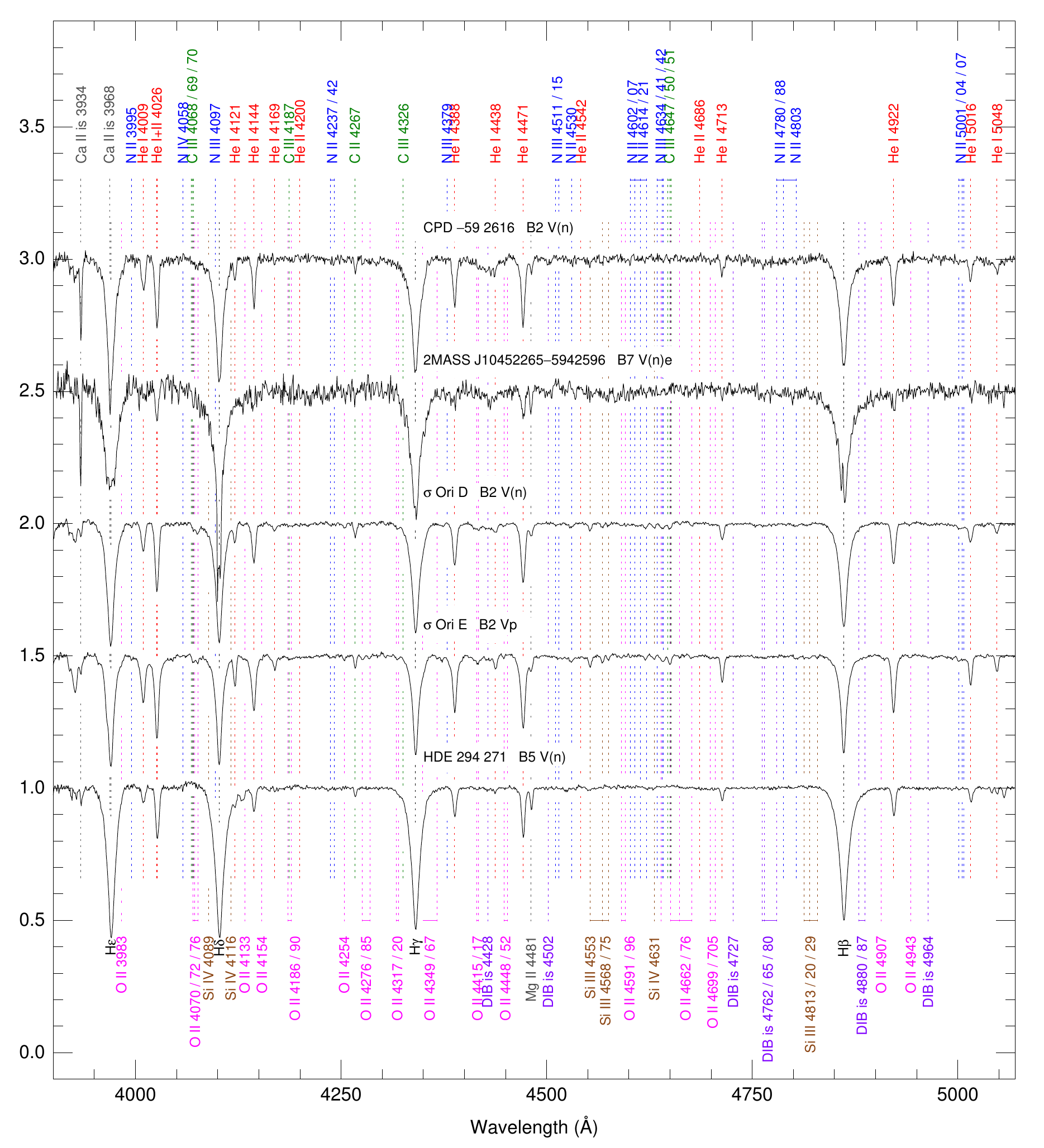}}
\caption{New GOSSS spectra for stars in \VO{025} and \VO{026}.}
\label{GOSSS_spectra_17}
\end{figure*}

\begin{figure*}
\centerline{\includegraphics[width=\linewidth]{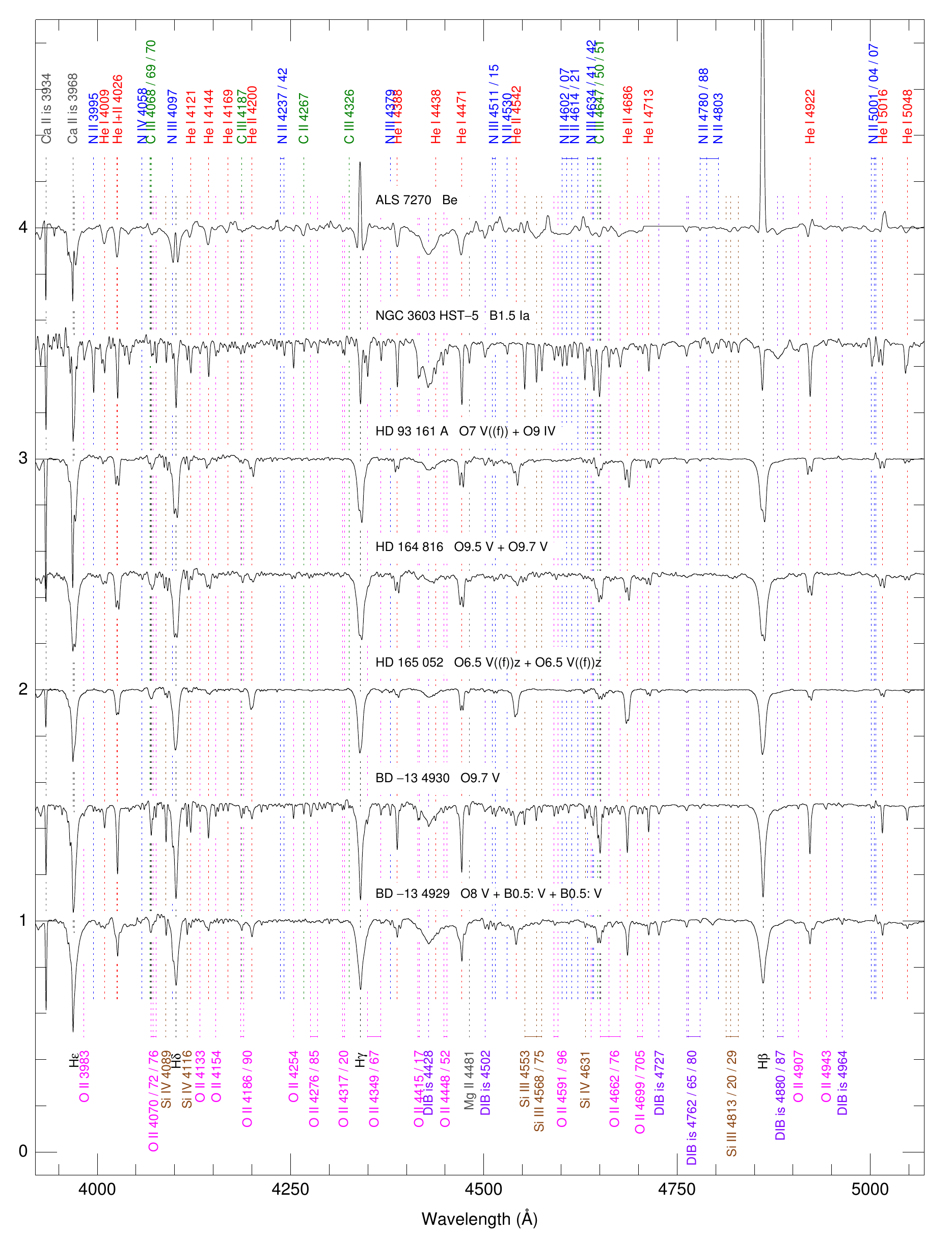}}
\caption{New \lili\ spectra at $R\sim 2500$ for stars in \VO{001}, \VO{002}, \VO{018}, and \VO{019}. Note that the ALS~7270 data does not include the 4700-4760~\AA\ region.}
\label{LiLi_spectra_01}
\end{figure*}

\begin{figure*}
\centerline{\includegraphics[width=\linewidth]{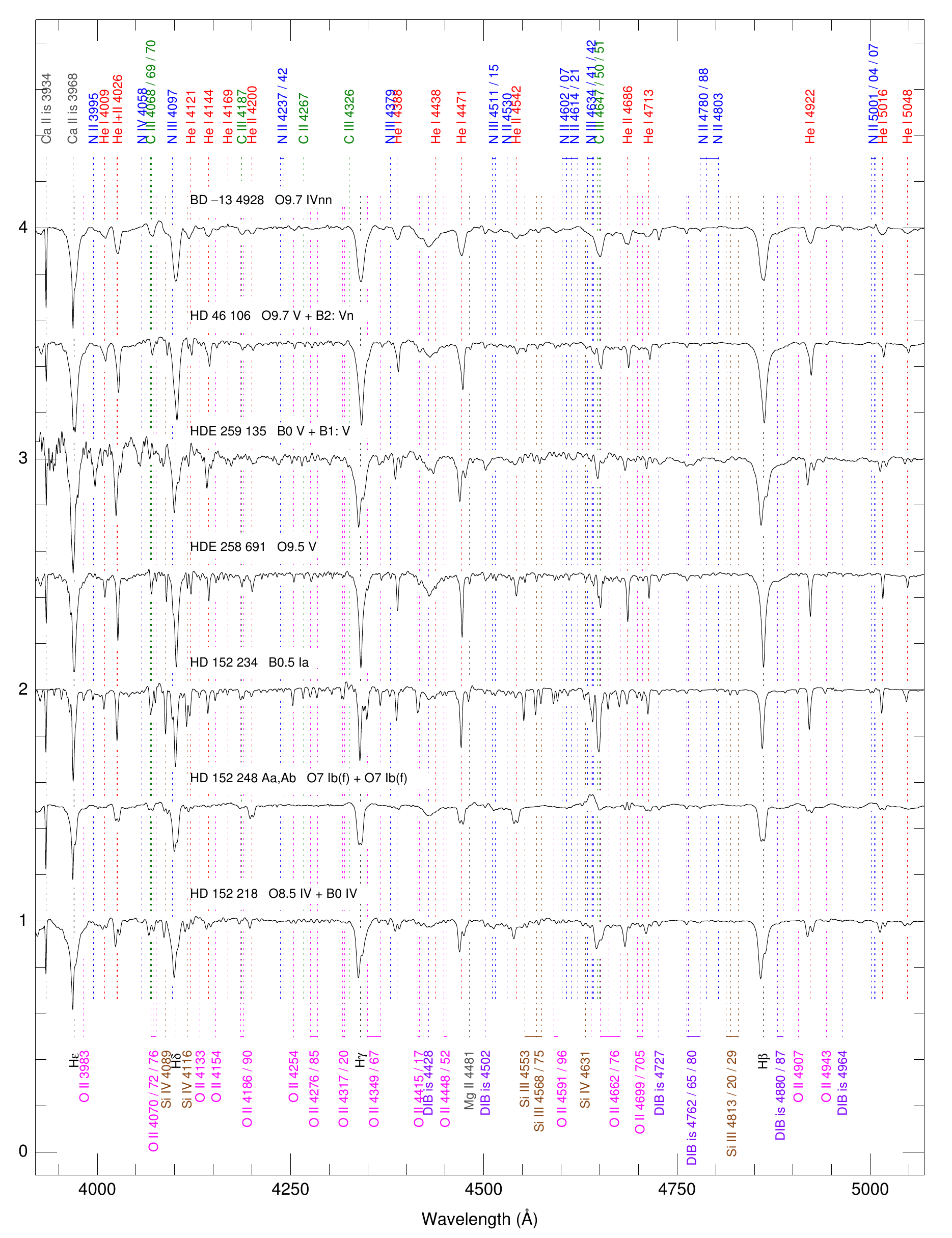}}
\caption{New \lili\ spectra at $R\sim 2500$ for stars in \VO{019}, \VO{020}, and \VO{022}.}
\label{LiLi_spectra_02}
\end{figure*}

\begin{figure*}
\centerline{\includegraphics[width=\linewidth]{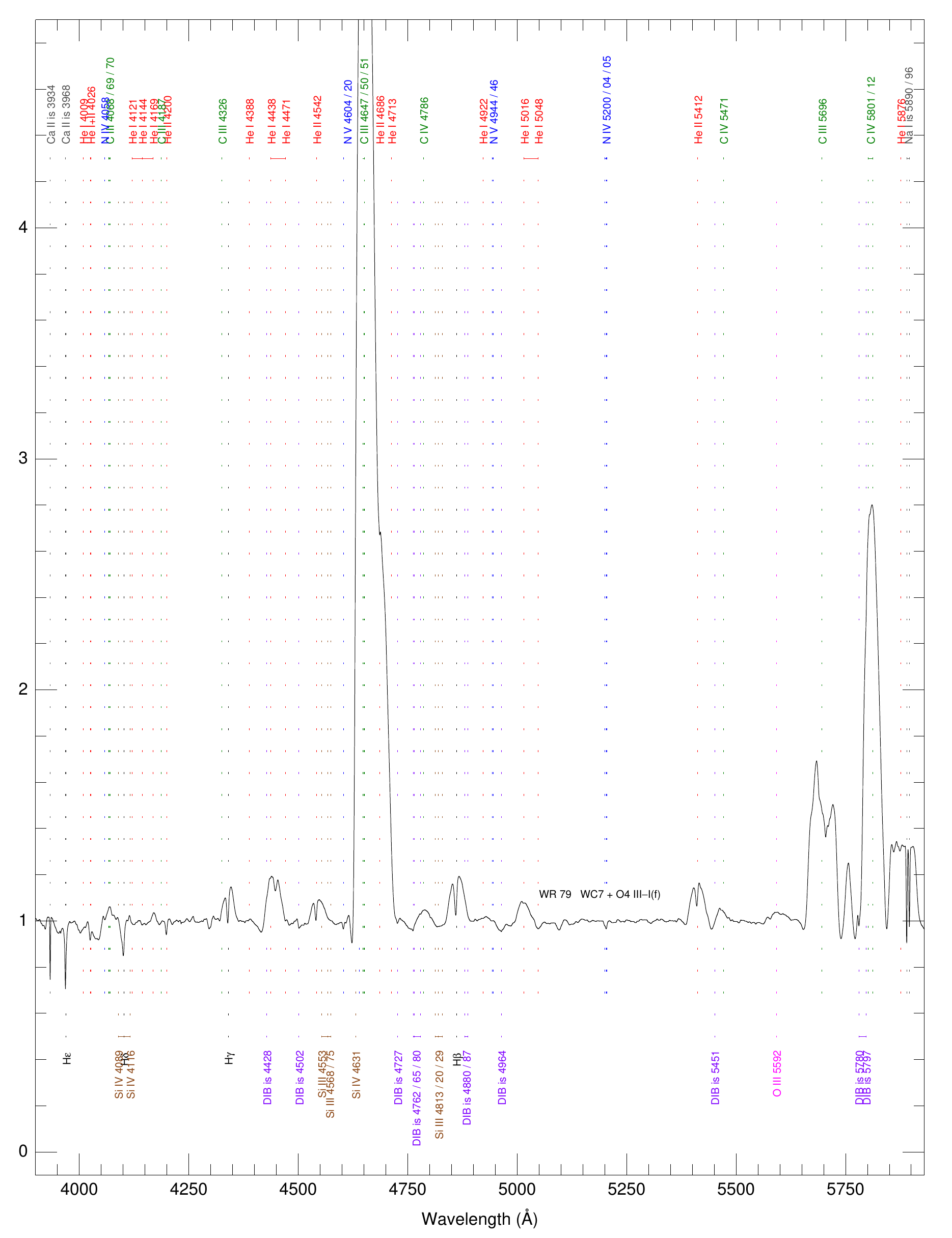}}
\caption{New \lili\ spectra at $R\sim 2500$ for a star in \VO{022}. Note the expanded wavelength range.}
\label{LiLi_spectra_03}
\end{figure*}

\begin{figure*}
\centerline{\includegraphics[width=\linewidth]{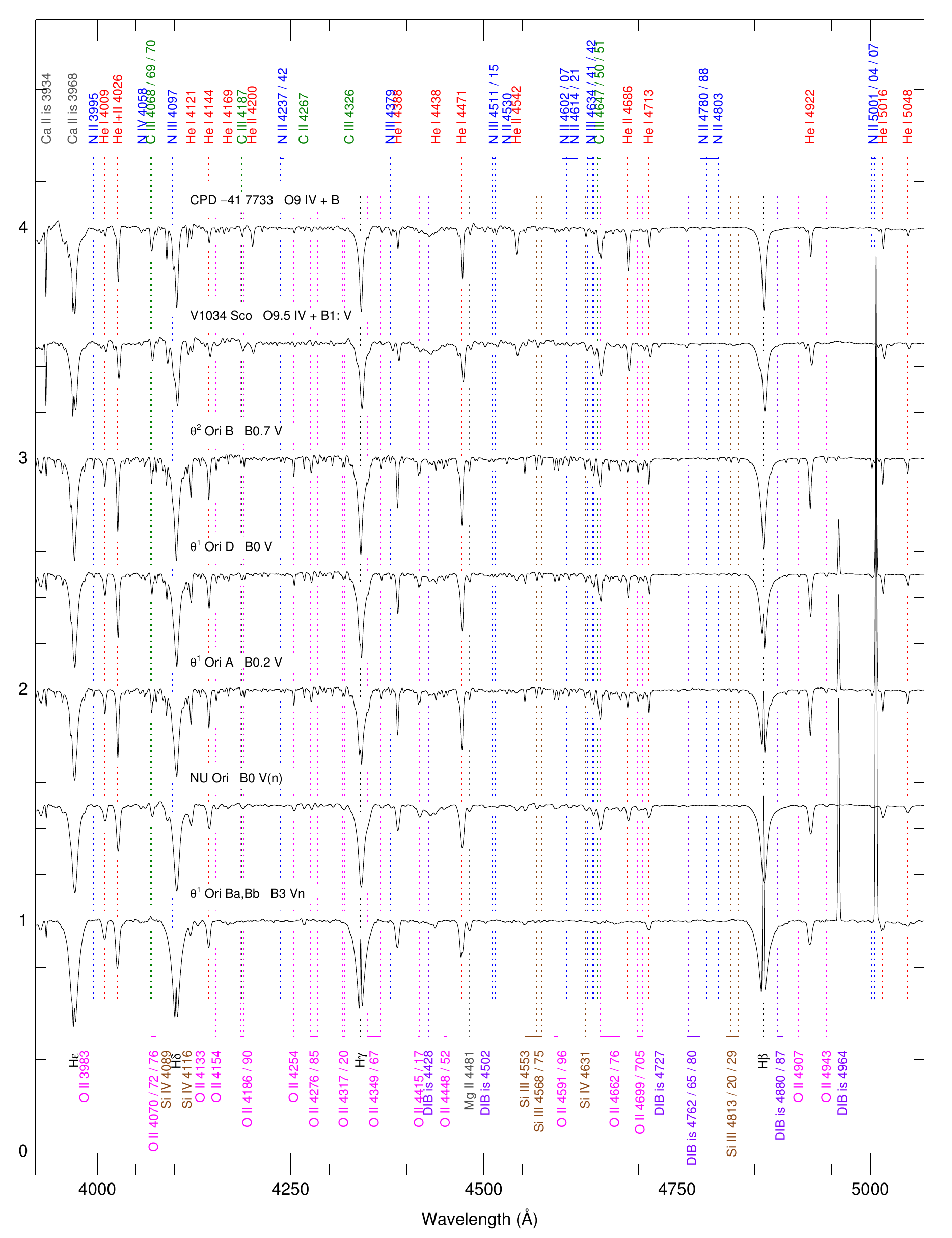}}
\caption{New \lili\ spectra at $R\sim 2500$ for stars in \VO{022} and \VO{023}.}
\label{LiLi_spectra_04}
\end{figure*}

\begin{figure*}
\centerline{\includegraphics[width=\linewidth]{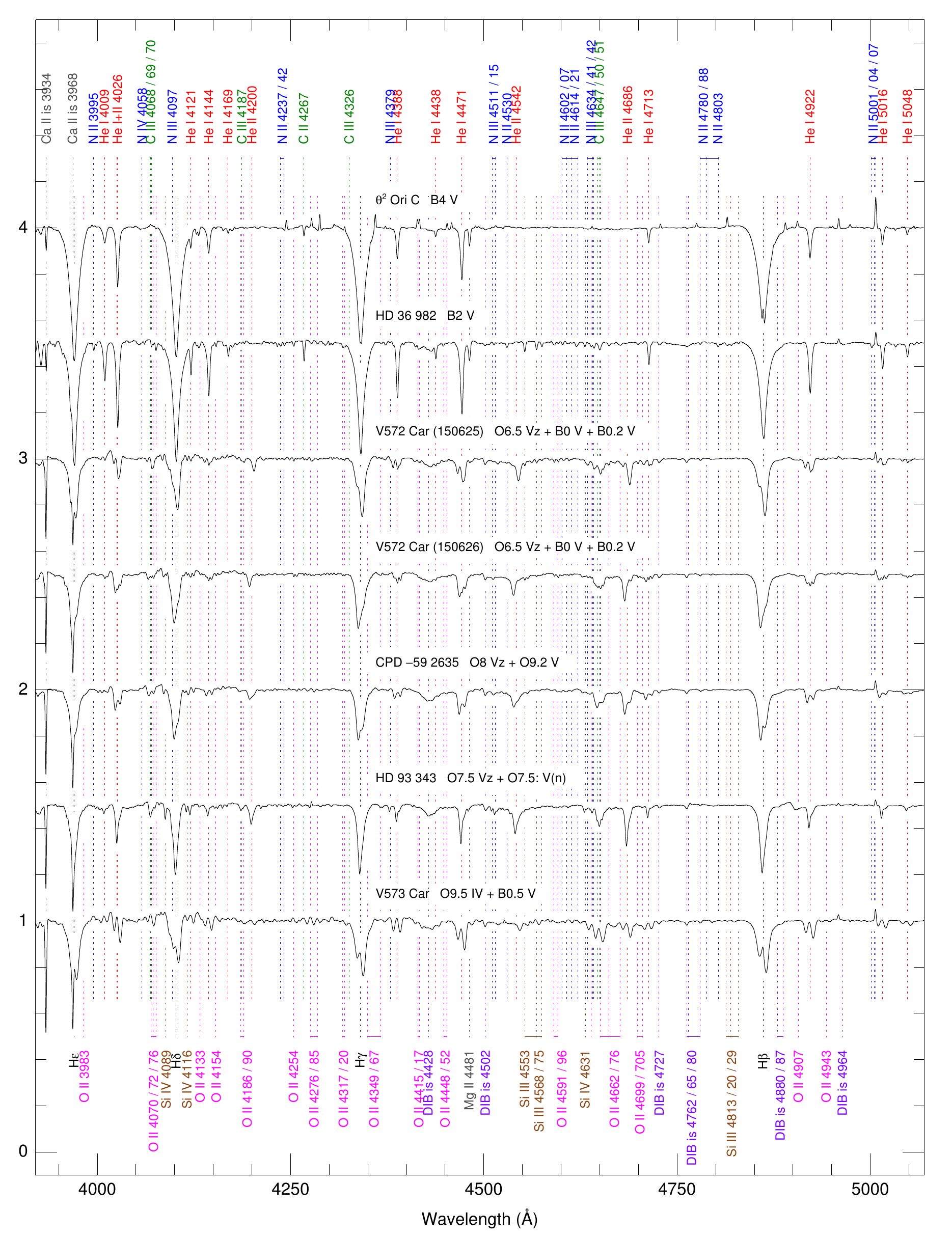}}
\caption{New \lili\ spectra at $R\sim 2500$ for stars in \VO{023}, \VO{025}, and \VO{026}.}
\label{LiLi_spectra_05}
\end{figure*}

\begin{figure*}
\centerline{\includegraphics[width=\linewidth]{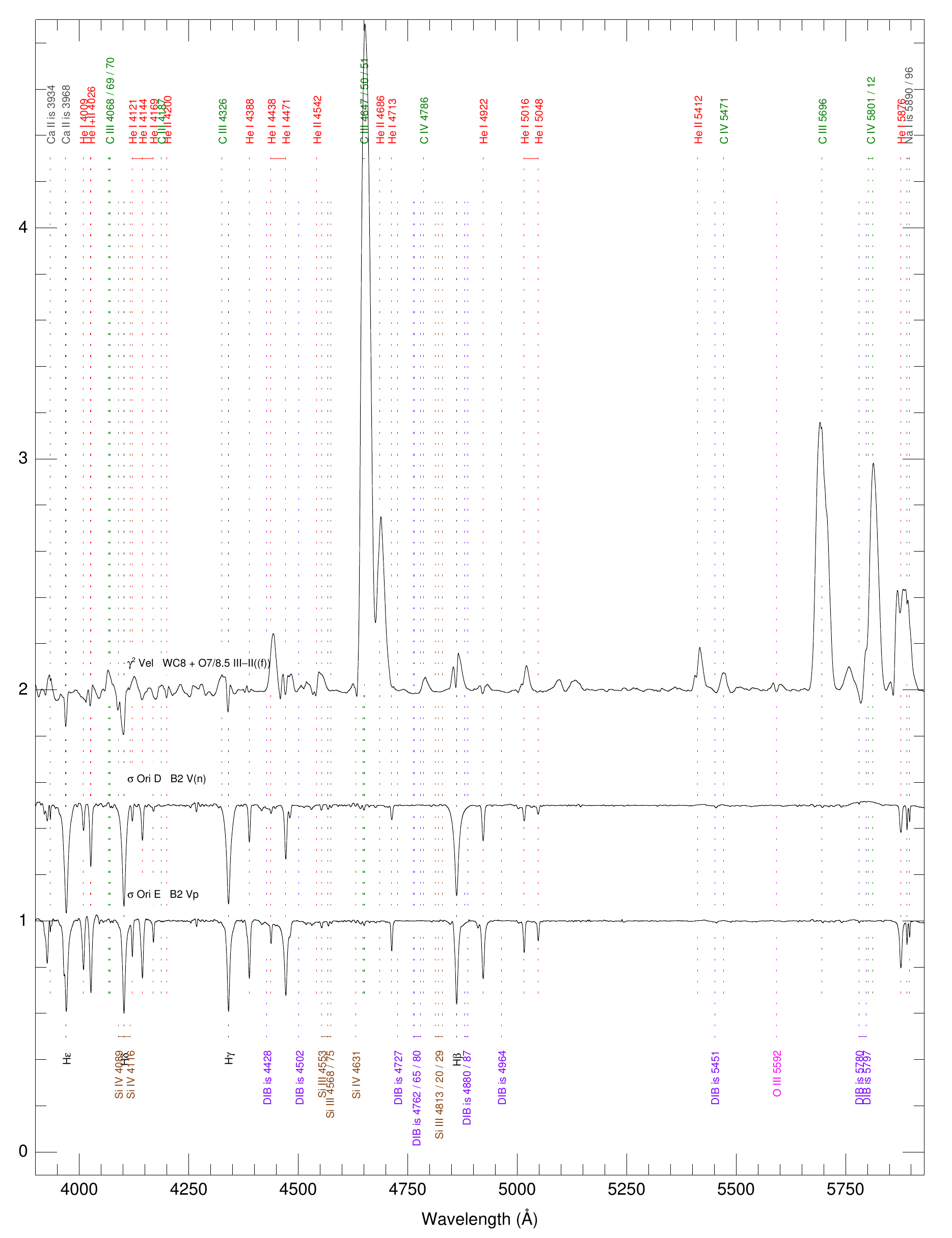}}
\caption{New \lili\ spectra at $R\sim 2500$ for stars in \VO{024} and \VO{026}. Note the expanded wavelength range.}
\label{LiLi_spectra_06}
\end{figure*}

$\,\!$\indent In this appendix we present:

\begin{enumerate}
 \item The field sizes and filters used to determine the parameters for the 26 stellar groups in this paper (Table~\ref{filters}).
 \item The main results for those 26 stellar groups (Table~\ref{results}).
 \item The spectrograms from GOSSS and \lili\ that have not appeared in previous papers. Figures~\ref{GOSSS_spectra_01}~to~\ref{GOSSS_spectra_17} show
       the first type and Figures~\ref{LiLi_spectra_01}~to~\ref{LiLi_spectra_06} the second type, in both cases sorted by the Villafranca group the belong to.
 \item The corresponding spectral classifications in Tables~\ref{GOSSS_spclas_01}, \ref{GOSSS_spclas_02}, and \ref{LiLi_spclas}.
\end{enumerate}
%
%\vfill
%
%\eject

\begin{table*}
\caption{Field sizes and filters applied to the O-type stellar groups and subgroups in this paper.}
\centerline{
\begin{tabular}{lrcrrrccccc}
\hline
 ID       & \mci{\Nf}    & field            & \mci{$\alpha$} & \mci{$\delta$} & \mci{$r$}       & \pmra   & \pmdec  & $r_\mu$ & $\Delta(\GBPmGRPc)$ & \GGcmax \\
          &              &                  & \mci{(deg)}    & \mci{(deg)}    & \mci{(\arcsec)} & (mas/a) & (mas/a) & (mas/a) &                     &         \\
\hline
 O-001    &  \num{28594} & $ 20\arcmin\times 20\arcmin$ &   168.79 & $-$61.26 &             206 & $-$5.61 & $+$1.97 &    0.44 &            $>-$0.34 & ---     \\ % NGC 3603
%O-001    &  \num{27873} & $ 20\arcmin\times 20\arcmin$ &   168.79 & $-$61.26 &             206 & $-$5.61 & $+$1.97 &    0.44 &            $>-$0.34 & ---     \\ %  Incorrect spimax=9.99
 O-002    &  \num{15183} & $ 20\arcmin\times 20\arcmin$ &   160.95 & $-$59.56 &             171 & $-$6.45 & $+$2.25 &    0.70 &            $>-$0.69 & ---     \\ % Trumpler 14
%O-002    &  \num{14340} & $ 20\arcmin\times 20\arcmin$ &   160.95 & $-$59.56 &             166 & $-$6.45 & $+$2.25 &    0.70 &            $>-$0.69 & ---     \\
 O-003    &  \num{19709} & $ 20\arcmin\times 20\arcmin$ &   161.09 & $-$59.73 &             103 & $-$7.10 & $+$2.80 &    0.73 &            $>-$1.30 & ---     \\ % Trumpler 16 W
%O-003    &  \num{18441} & $ 20\arcmin\times 20\arcmin$ &   161.09 & $-$59.73 &             103 & $-$7.10 & $+$2.80 &    0.73 &            $>-$1.30 & ---     \\ %  Incorrect pmra+pmdec
 O-004    &  \num{62124} & $ 40\arcmin\times 40\arcmin$ &   155.99 & $-$57.76 &             400 & $-$5.10 & $+$2.82 &    0.70 &            $>-$0.65 & ---     \\ % Westerlund 2
%O-004    &  \num{60144} & $ 40\arcmin\times 40\arcmin$ &   155.99 & $-$57.76 &             400 & $-$5.10 & $+$2.82 &    0.70 &            $>-$0.65 & ---     \\ %  Incorrect spimax=9.99
 O-005    &  \num{10999} & $ 30\arcmin\times 30\arcmin$ &   261.18 & $-$34.21 &             605 & $-$1.16 & $-$2.26 &    1.40 &            $>-$1.00 & ---     \\ % Pismis 24
%O-005    &  \num{10374} & $ 30\arcmin\times 30\arcmin$ &   261.18 & $-$34.21 &             605 & $-$1.10 & $-$2.20 &    1.40 &            $>-$1.00 & ---     \\ %  Incorrect spimax=9.99
 O-006    &  \num{49576} & $ 20\arcmin\times 20\arcmin$ &   164.68 & $-$61.18 &             200 & $-$5.50 & $+$2.30 &    0.30 &            $>-$0.90 & ---     \\ % Gum 35
%O-006    &  \num{48211} & $ 20\arcmin\times 20\arcmin$ &   164.68 & $-$61.18 &             200 & $-$5.50 & $+$2.30 &    0.30 &            $>-$0.90 & ---     \\ %  Incorrect spimax=9.99
 O-007    &         5361 & $ 20\arcmin\times 20\arcmin$ &   308.30 & $+$41.22 &             186 & $-$2.65 & $-$4.49 &    0.72 &            $>-$0.60 & ---     \\ % Cyg OB2-22
%O-007    &         5241 & $ 20\arcmin\times 20\arcmin$ &   308.30 & $+$41.22 &             186 & $-$2.65 & $-$4.49 &    0.72 &            $>-$0.60 & ---     \\ %  Incorrect spimax=9.99
 O-008    &         6299 & $ 20\arcmin\times 20\arcmin$ &   308.32 & $+$41.31 &             126 & $-$2.67 & $-$4.15 &    0.50 &            $>-$0.80 & ---     \\ % Cyg OB2-8
%O-008    &         6136 & $ 20\arcmin\times 20\arcmin$ &   308.32 & $+$41.31 &             126 & $-$2.67 & $-$4.15 &    0.50 &            $>-$0.80 & ---     \\ %  Incorrect spimax=9.99
 O-009    &         5987 & $ 20\arcmin\times 20\arcmin$ &   275.12 & $-$16.18 &             236 & $+$0.10 & $-$1.52 &    1.10 &            $>-$0.70 & ---     \\ % M17
%O-009    &         5671 & $ 20\arcmin\times 20\arcmin$ &   275.12 & $-$16.18 &             236 & $+$0.10 & $-$1.52 &    1.10 &            $>-$0.70 & ---     \\ %  Incorrect spimax=0.13
 O-010    &  \num{33947} & $ 20\arcmin\times 20\arcmin$ &   250.30 & $-$48.76 &             285 & $+$1.57 & $-$3.92 &    1.04 &            $>-$0.70 & ---     \\ % NGC 6193
%O-010    &  \num{33424} & $ 20\arcmin\times 20\arcmin$ &   250.30 & $-$48.76 &             285 & $+$1.57 & $-$3.92 &    1.04 &            $>-$0.70 & ---     \\ %  Incorrect spimax=9.99
 O-011    &         6806 & $ 20\arcmin\times 20\arcmin$ &   308.83 & $+$46.84 &             341 & $-$2.78 & $-$4.34 &    0.50 &            $>-$0.28 & ---     \\ % Berkeley 90
%O-011    &         6513 & $ 20\arcmin\times 20\arcmin$ &   308.83 & $+$46.84 &             341 & $-$2.78 & $-$4.31 &    0.50 &            $>-$0.20 & ---     \\ %  Incorrect Delta(GBP-GRP), spimax=9.99
 O-012    &  \num{26170} & $ 30\arcmin\times 30\arcmin$ &   118.18 & $-$26.33 &             650 & $-$2.50 & $+$2.60 &    0.40 &            $>-$0.15 & ---     \\ % NGC 2467
%O-012    &  \num{25632} & $ 30\arcmin\times 30\arcmin$ &   118.18 & $-$26.33 &             390 & $-$2.50 & $+$2.60 &    0.40 &            $>-$0.15 & ---     \\ %  Incorrect spimax=9.99
 O-012~N  &  \num{26170} & $ 30\arcmin\times 30\arcmin$ &   118.19 & $-$26.28 &             200 & $-$2.50 & $+$2.60 &    0.40 &            $>-$0.15 & ---     \\ % NGC 2467 N
%O-012~N  &  \num{25632} & $ 30\arcmin\times 30\arcmin$ &   118.19 & $-$26.28 &             200 & $-$2.50 & $+$2.60 &    0.40 &            $>-$0.15 & ---     \\ %  Incorrect spimax=9.99
 O-012~S  &  \num{26170} & $ 30\arcmin\times 30\arcmin$ &   118.18 & $-$26.42 &             320 & $-$2.50 & $+$2.60 &    0.40 &            $>-$0.15 & ---     \\ % NGC 2467 S
%O-012~S  &  \num{25632} & $ 30\arcmin\times 30\arcmin$ &   118.18 & $-$26.38 &             200 & $-$2.50 & $+$2.60 &    0.40 &            $>-$0.15 & ---     \\ %  Incorrect spimax=9.99
 O-013    &         3266 & $ 20\arcmin\times 20\arcmin$ &   348.43 & $+$61.50 &             109 & $-$3.70 & $-$2.14 &    0.73 &            $>-$0.20 & ---     \\ % Sh 2-158
%O-013    &         3235 & $ 20\arcmin\times 20\arcmin$ &   348.43 & $+$61.50 &             109 & $-$3.70 & $-$2.14 &    0.73 &            $>-$0.20 & ---     \\ %  Incorrect Delta(GBP-GRP), spimax=9.99
 O-014~SE & \num{291324} & $120\arcmin\times120\arcmin$ &   314.56 & $+$43.84 &             420 & $-$0.50 & $-$3.50 &    1.30 &            $>+$0.00 & ---     \\ % Gulf of Mexico
 O-014~NW & \num{291324} & $120\arcmin\times120\arcmin$ &   313.10 & $+$44.40 &            1600 & $-$1.75 & $-$3.05 &    0.70 &            $>+$0.60 & ---     \\ % Bermuda 
%O-014    & \num{396261} & $120\arcmin\times120\arcmin$ &   313.96 & $+$43.87 &             --- & $+$0.01 & $-$4.52 &    0.90 &            $>+$0.00 & ---     \\
 O-015    &  \num{19903} & $ 30\arcmin\times 30\arcmin$ &   304.60 & $+$40.78 &             800 & $-$2.60 & $-$6.40 &    0.75 &            $>-$0.30 & ---     \\ % Coliinder 419
%O-015    &  \num{19049} & $ 30\arcmin\times 30\arcmin$ &   304.60 & $+$40.78 &             800 & $-$2.60 & $-$6.40 &    0.75 &            $>-$0.30 & ---     \\
 O-016    &  \num{26017} & $ 60\arcmin\times 60\arcmin$ &   100.25 & $+$09.75 &            1500 & $-$1.80 & $-$3.70 &    1.50 &            $>-$0.20 & ---     \\ % NGC 2264
%O-016    &  \num{25177} & $ 60\arcmin\times 60\arcmin$ &   100.25 & $+$09.75 &            1500 & $-$1.80 & $-$3.70 &    1.50 &            $>-$0.20 & ---     \\
 O-016~N  &  \num{26017} & $ 60\arcmin\times 60\arcmin$ &   100.20 & $+$09.88 &             540 & $-$1.80 & $-$3.70 &    1.50 &            $>-$0.20 & ---     \\ % NGC 2264 N
%O-016~N  &  \num{25177} & $ 60\arcmin\times 60\arcmin$ &   100.20 & $+$09.88 &             540 & $-$1.80 & $-$3.70 &    1.50 &            $>-$0.20 & ---     \\
 O-016~S  &  \num{26017} & $ 60\arcmin\times 60\arcmin$ &   100.28 & $+$09.53 &             540 & $-$1.80 & $-$3.70 &    1.50 &            $>-$0.20 & ---     \\ % NGC 2264 S
%O-016~S  &  \num{25177} & $ 60\arcmin\times 60\arcmin$ &   100.28 & $+$09.53 &             540 & $-$1.80 & $-$3.70 &    1.50 &            $>-$0.20 & ---     \\
 O-017    &  \num{59219} & $ 60\arcmin\times 60\arcmin$ &    38.17 & $+$61.46 &            1400 & $-$0.90 & $-$0.50 &    0.75 &            $>-$0.30 & ---     \\ % Heart nebula
 O-018    & \num{243382} & $ 60\arcmin\times 60\arcmin$ &   271.10 & $-$24.40 &            1000 & $+$1.50 & $-$2.00 &    1.40 &  $-$0.30 to $+$2.50 & ---     \\ % Lagoon nebula
 O-019    &  \num{11353} & $ 30\arcmin\times 30\arcmin$ &   274.68 & $-$13.79 &             800 & $+$0.20 & $-$1.70 &    1.00 &            $>-$0.43 & ---     \\ % Eagle nebula
 O-020    &  \num{33691} & $ 60\arcmin\times 60\arcmin$ &    97.98 &  $+$4.94 &            1200 & $-$1.80 & $+$0.30 &    1.00 &            $>-$0.70 & ---     \\ % Rosette nebula
 O-021    &  \num{10521} & $ 20\arcmin\times 20\arcmin$ &   109.68 & $-$24.95 &             400 & $-$2.80 & $+$3.10 &    1.00 &            $>-$0.37 & ---     \\ % tau CMa cluster
 O-022    &  \num{39632} & $ 30\arcmin\times 30\arcmin$ &   253.54 & $-$41.83 &             800 & $-$0.40 & $-$2.40 &    1.00 &  $-$0.70 to $+$2.00 & 16.5    \\ % NGC 6231
 O-023    &   \num{6637} & $ 60\arcmin\times 60\arcmin$ &    83.82 &  $-$5.39 &             700 & $+$1.50 & $+$0.50 &    1.30 &            $>-$0.62 & ---     \\ % Orion nebula
 O-024    &  \num{82518} & $ 60\arcmin\times 60\arcmin$ &   122.38 & $-$47.33 &            1500 & $-$6.60 & $+$9.80 &    1.35 &            $>-$0.15 & ---     \\ % Pozzo 1
 O-025    &  \num{18153} & $ 20\arcmin\times 20\arcmin$ &   161.30 & $-$59.70 &             250 & $-$6.80 & $+$2.48 &    0.65 &            $>-$0.35 & ---     \\ % Trumpler 16 E
 O-026    &  \num{14576} & $ 60\arcmin\times 60\arcmin$ &    84.69 &  $-$2.60 &             600 & $+$1.10 & $-$0.60 &    1.00 &            $>-$0.07 & ---     \\ % sigma Ori
\hline
\multicolumn{11}{l}{RUWE is filtered as $<$1.4, \Cstar\ as $<$0.4, and \sigmae\ as $<$0.1~mas in all cases.}
\end{tabular}
}
\label{filters}                  
\end{table*}

\begin{table*}
\caption{Membership and distance results.}
\centerline{
\renewcommand{\arraystretch}{1.2}
\begin{tabular}{lrrrrrr@{$\pm$}lr@{$\pm$}lr@{$\pm$}lr@{}l}
\hline
 ID       & $N_{*,0}$ & $N_*$ & $t_\varpi$ & $t_{\mu_{\alpha *}}$ & $t_{\mu_{\delta}}$ & \mcii{\pmrag}  & \mcii{\pmdecg} & \mcii{\pig}  & \mcii{$d$}              \\
          &           &       &            &                      &                    & \mcii{(mas/a)} & \mcii{(mas/a)} & \mcii{(mas)} & \mcii{(pc)}             \\
\hline
 O-001    &       170 &   166 &       0.91 &                 2.93 &               2.21 & $-$5.581&0.023 & $+$2.004&0.023 & 0.141&0.011  & 7130&$^{+590}_{-500}$   \\
%O-001    &       208 &   202 &       0.99 &                 2.88 &               2.31 & $-$5.582&0.023 & $+$2.006&0.023 & 0.142&0.011  & 7080&$^{+570}_{-500}$   \\ % EDR3 uncorrected PM + k6
%O-001    &       212 &   208 &       0.99 &                 2.79 &               2.22 & $-$5.582&0.023 & $+$2.010&0.023 & 0.143&0.011  & 6990&$^{+560}_{-480}$   \\ % EDR3 orig
%O-001    &       143 &   137 &       1.13 &                 1.97 &               1.65 & $-$5.551&0.067 & $+$1.974&0.067 & 0.114&0.044  & 8000&$^{+2600}_{-1700}$ \\ % DR2 results with spimax = 9.99
 O-002    &       186 &   184 &       0.83 &                 4.51 &               4.49 & $-$6.534&0.023 & $+$2.076&0.023 & 0.424&0.011  & 2363&$^{+61}_{-58}$     \\
%O-002    &       210 &   207 &       0.92 &                 4.51 &               4.66 & $-$6.533&0.023 & $+$2.106&0.023 & 0.423&0.011  & 2369&$^{+61}_{-58}$     \\ % EDR3 uncorrected PM + k6
%O-002    &       211 &   207 &       0.91 &                 4.56 &               4.84 & $-$6.532&0.023 & $+$2.109&0.023 & 0.423&0.011  & 2371&$^{+61}_{-58}$     \\ % EDR3 orig
%O-002    &        93 &    91 &       0.88 &                 3.98 &               3.29 & $-$6.540&0.067 & $+$2.042&0.067 & 0.427&0.043  & 2430&$^{+290}_{-230}$   \\ % DR2
 O-003    &        31 &    29 &       0.78 &                 3.11 &               2.73 & $-$7.128&0.024 & $+$2.670&0.024 & 0.435&0.012  & 2305&$^{+64}_{-61}$     \\
%O-003    &        38 &    36 &       0.96 &                 3.02 &               3.19 & $-$7.125&0.024 & $+$2.699&0.024 & 0.436&0.012  & 2302&$^{+64}_{-60}$     \\ % EDR3 uncorrected PM + k6
%O-003    &        38 &    36 &       0.95 &                 3.62 &               3.36 & $-$7.136&0.024 & $+$2.710&0.024 & 0.435&0.012  & 2303&$^{+63}_{-60}$     \\ % EDR3 orig
%O-003    &        20 &    20 &       0.86 &                 2.42 &               1.88 & $-$7.073&0.067 & $+$2.641&0.066 & 0.441&0.043  & 2380&$^{+270}_{-220}$   \\ % DR2
 O-004    &       114 &   110 &       0.95 &                 3.27 &               3.68 & $-$5.297&0.023 & $+$2.890&0.023 & 0.226&0.011  & 4440&$^{+230}_{-210}$   \\
%O-004    &       129 &   125 &       0.96 &                 3.22 &               3.57 & $-$5.295&0.023 & $+$2.887&0.023 & 0.227&0.011  & 4430&$^{+230}_{-210}$   \\ % EDR3 uncorrected PM + k6
%O-004    &       132 &   128 &       0.92 &                 3.06 &               3.39 & $-$5.297&0.023 & $+$2.887&0.023 & 0.226&0.011  & 4450&$^{+230}_{-210}$   \\ % EDR3 orig
%O-004    &       178 &   174 &       1.05 &                 1.89 &               1.76 & $-$5.190&0.064 & $+$2.927&0.064 & 0.228&0.043  & 4730&$^{+1130}_{-780}$  \\ % DR2 results with spimax = 9.99
 O-005    &       118 &   111 &       1.15 &                 6.27 &               9.25 & $-$0.898&0.023 & $-$2.221&0.023 & 0.610&0.011  & 1642&$^{+30}_{-29}$     \\
%O-005    &       120 &   113 &       1.14 &                 6.27 &               9.16 & $-$0.907&0.023 & $-$2.221&0.023 & 0.610&0.011  & 1642&$^{+30}_{-29}$     \\ % EDR3 uncorrected PM + k6
%O-005    &       110 &   103 &       1.13 &                 6.50 &               8.62 & $-$0.907&0.023 & $-$2.179&0.023 & 0.606&0.011  & 1654&$^{+31}_{-30}$     \\ % EDR3 orig
%O-005    &       208 &   197 &       1.20 &                 2.45 &               3.71 & $-$0.918&0.059 & $-$2.216&0.059 & 0.602&0.041  & 1690&$^{+130}_{-110}$   \\ % DR2 results with spimax = 9.99
 O-006    &        97 &    96 &       0.97 &                 1.93 &               2.03 & $-$5.511&0.023 & $+$2.285&0.023 & 0.137&0.011  & 7260&$^{+610}_{-520}$   \\
%O-006    &       111 &   109 &       0.99 &                 1.90 &               2.09 & $-$5.508&0.023 & $+$2.295&0.023 & 0.138&0.011  & 7230&$^{+600}_{-520}$   \\ % EDR3 uncorrected PM + k6
%O-006    &       112 &   110 &       0.97 &                 1.84 &               2.03 & $-$5.510&0.024 & $+$2.300&0.023 & 0.139&0.011  & 7180&$^{+600}_{-510}$   \\ % EDR3 orig
%O-006    &        99 &    98 &       1.06 &                 1.38 &               1.52 & $-$5.513&0.067 & $+$2.270&0.067 & 0.152&0.044  & 6400&$^{+1800}_{-1200}$ \\ % DR2 results with spimax = 9.99
 O-007    &        51 &    51 &       1.10 &                 4.88 &               3.86 & $-$2.725&0.024 & $-$4.448&0.024 & 0.618&0.011  & 1620&$^{+30}_{-29}$     \\
%O-007    &        55 &    55 &       1.10 &                 4.76 &               3.80 & $-$2.720&0.024 & $-$4.456&0.024 & 0.618&0.011  & 1621&$^{+30}_{-29}$     \\ % EDR3 uncorrected PM + k6
%O-007    &        55 &    55 &       1.23 &                 4.70 &               3.77 & $-$2.715&0.024 & $-$4.462&0.024 & 0.613&0.011  & 1633&$^{+31}_{-30}$     \\ % EDR3 orig
%O-007    &        82 &    81 &       0.98 &                 2.10 &               1.67 & $-$2.697&0.067 & $-$4.482&0.068 & 0.594&0.044  & 1720&$^{+140}_{-120}$   \\ % DR2 results with spimax = 9.99
 O-008    &        40 &    40 &       0.93 &                 3.18 &               3.54 & $-$2.659&0.024 & $-$4.400&0.024 & 0.623&0.011  & 1608&$^{+30}_{-29}$     \\
%O-008    &        40 &    40 &       0.93 &                 3.19 &               3.66 & $-$2.655&0.024 & $-$4.414&0.024 & 0.623&0.011  & 1608&$^{+30}_{-29}$     \\ % EDR3 uncorrected PM + k6
%O-008    &        40 &    40 &       0.88 &                 3.10 &               3.85 & $-$2.651&0.024 & $-$4.417&0.024 & 0.622&0.011  & 1609&$^{+30}_{-29}$     \\ % EDR3 orig
%O-008    &        45 &    45 &       1.20 &                 1.62 &               1.92 & $-$2.667&0.067 & $-$4.371&0.067 & 0.621&0.044  & 1640&$^{+130}_{-110}$   \\ % DR2 results with spimax = 9.99
 O-009    &        27 &    25 &       0.98 &                 4.98 &               7.70 & $+$0.303&0.026 & $-$1.523&0.024 & 0.591&0.014  & 1696&$^{+41}_{-39}$     \\
%O-009    &        34 &    32 &       1.04 &                 4.65 &               7.87 & $+$0.321&0.025 & $-$1.558&0.024 & 0.595&0.014  & 1684&$^{+39}_{-38}$     \\ % EDR3 uncorrected PM + k6
%O-009    &        33 &    32 &       1.14 &                 4.59 &               7.74 & $+$0.334&0.025 & $-$1.525&0.024 & 0.604&0.013  & 1659&$^{+38}_{-36}$     \\ % EDR3 orig
%O-009    &        30 &    30 &       1.15 &                 3.26 &               3.12 & $+$0.166&0.069 & $-$1.618&0.069 & 0.627&0.044  & 1630&$^{+130}_{-110}$   \\ % DR2 results with spimax = 0.13
 O-010    &        71 &    68 &       1.00 &                 4.66 &               5.29 & $+$1.225&0.024 & $-$3.967&0.023 & 0.872&0.012  & 1148&$^{+16}_{-15}$     \\
%O-010    &        83 &    79 &       0.96 &                 4.46 &               5.12 & $+$1.206&0.024 & $-$3.972&0.023 & 0.874&0.012  & 1145&$^{+15}_{-15}$     \\ % EDR3 uncorrected PM + k6
%O-010    &        79 &    75 &       1.08 &                 5.35 &               5.34 & $+$1.218&0.024 & $-$3.968&0.023 & 0.866&0.012  & 1155&$^{+16}_{-15}$     \\ % EDR3 orig
%O-010    &       115 &   111 &       1.12 &                 2.39 &               3.00 & $+$1.307&0.066 & $-$3.972&0.066 & 0.853&0.044  & 1185&$^{+65}_{-59}$     \\ % DR2 results with spimax = 9.99
 O-011    &        68 &    66 &       0.97 &                 2.90 &               2.46 & $-$2.855&0.023 & $-$4.391&0.023 & 0.365&0.011  & 2741&$^{+86}_{-81}$     \\
%O-011    &        69 &    67 &       1.00 &                 2.95 &               2.57 & $-$2.853&0.023 & $-$4.397&0.023 & 0.365&0.011  & 2737&$^{+86}_{-81}$     \\ % EDR3 uncorrected PM + k6
%O-011    &        69 &    68 &       1.02 &                 2.90 &               2.55 & $-$2.849&0.024 & $-$4.403&0.024 & 0.367&0.011  & 2728&$^{+85}_{-80}$     \\ % EDR3 orig
%O-011    &        82 &    79 &       1.22 &                 1.67 &               1.38 & $-$2.840&0.066 & $-$4.316&0.066 & 0.327&0.043  & 2990&$^{+390}_{-340}$   \\ % DR2 results with spimax = 9.99
 O-012    &       258 &   228 &       0.99 &                 3.01 &               3.00 & $-$2.499&0.022 & $+$2.605&0.022 & 0.237&0.011  & 4241&$^{+200}_{-180}$   \\
%O-012    &       280 &   249 &       1.02 &                 2.92 &               2.95 & $-$2.497&0.022 & $+$2.609&0.022 & 0.237&0.011  & 4240&$^{+200}_{-180}$   \\ % EDR3 uncorrected PM + k6
%O-012    &       284 &   253 &       0.99 &                 2.87 &               2.85 & $-$2.497&0.022 & $+$2.614&0.022 & 0.237&0.011  & 4240&$^{+200}_{-180}$   \\ % EDR3 orig
%O-012    &       206 &   194 &       1.15 &                 1.54 &               1.63 & $-$2.523&0.061 & $+$2.586&0.061 & 0.218&0.042  & 4850&$^{+1160}_{-800}$  \\ % DR2 results with spimax = 9.99
 O-012~N  &        63 &    61 &       0.87 &                 1.63 &               1.71 & $-$2.528&0.024 & $+$2.502&0.024 & 0.231&0.011  & 4350&$^{+230}_{-210}$   \\
%O-012~N  &        70 &    67 &       0.89 &                 1.62 &               1.77 & $-$2.525&0.024 & $+$2.505&0.024 & 0.231&0.011  & 4340&$^{+230}_{-200}$   \\ % EDR3 uncorrected PM + k6
%O-012~N  &        71 &    68 &       0.88 &                 1.58 &               1.74 & $-$2.525&0.024 & $+$2.509&0.024 & 0.232&0.012  & 4330&$^{+230}_{-200}$   \\ % EDR3 orig
%O-012~N  &        88 &    87 &       1.14 &                 1.41 &               1.18 & $-$2.544&0.067 & $+$2.487&0.067 & 0.204&0.044  & 5190&$^{+1380}_{-930}$  \\ % DR2 results with spimax = 9.99
 O-012~S  &       100 &    90 &       0.94 &                 2.98 &               2.80 & $-$2.506&0.023 & $+$2.667&0.023 & 0.236&0.011  & 4250&$^{+210}_{-190}$   \\
%O-012~S  &       111 &   101 &       0.97 &                 2.91 &               2.72 & $-$2.504&0.023 & $+$2.671&0.023 & 0.237&0.011  & 4240&$^{+210}_{-190}$   \\ % EDR3 uncorrected PM + k6
%O-012~S  &       112 &   103 &       0.99 &                 2.85 &               2.61 & $-$2.503&0.023 & $+$2.672&0.023 & 0.237&0.011  & 4240&$^{+210}_{-190}$   \\ % EDR3 orig
%O-012~S  &        67 &    62 &       1.10 &                 1.75 &               1.57 & $-$2.503&0.067 & $+$2.663&0.067 & 0.220&0.044  & 4830&$^{+1210}_{-830}$  \\ % DR2 results with spimax = 9.99
 O-013    &        11 &    11 &       0.86 &                 3.47 &               5.67 & $-$3.668&0.026 & $-$2.301&0.026 & 0.370&0.015  & 2710&$^{+110}_{-100}$   \\
%O-013    &        12 &    12 &       0.82 &                 3.37 &               5.43 & $-$3.670&0.026 & $-$2.306&0.026 & 0.370&0.015  & 2710&$^{+110}_{-100}$   \\ % EDR3 uncorrected PM + k6
%O-013    &        12 &    12 &       0.75 &                 3.57 &               5.22 & $-$3.695&0.026 & $-$2.318&0.026 & 0.370&0.015  & 2710&$^{+110}_{-100}$   \\ % EDR3 orig
%O-013    &        11 &    11 &       0.95 &                 1.96 &               3.95 & $-$3.708&0.068 & $-$2.214&0.068 & 0.356&0.044  & 2930&$^{+440}_{-340}$   \\ % DR2 results with spimax = 9.99
 O-014~SE &        13 &    13 &       1.09 &                 5.86 &               5.16 & $-$0.653&0.026 & $-$3.103&0.026 & 1.344&0.016  &  744&$^{+9}_{-9}$       \\ 
%O-014~SE &        14 &    13 &       1.09 &                 5.91 &               5.07 & $-$0.648&0.026 & $-$3.112&0.026 & 1.344&0.016  &  744&$^{+9}_{-9}$       \\ % EDR3 uncorrected PM + k6
%O-014~SE &        14 &    13 &       1.03 &                 5.78 &               4.93 & $-$0.636&0.026 & $-$3.100&0.026 & 1.344&0.016  &  744&$^{+9}_{-9}$       \\ % EDR3 orig
 O-014~NW &       248 &   144 &       1.01 &                 7.45 &               6.05 & $-$1.370&0.020 & $-$3.054&0.020 & 1.254&0.010  &  798&$^{+6}_{-6}$       \\ 
%O-014~NW &       264 &   156 &       1.01 &                 7.23 &               5.87 & $-$1.365&0.020 & $-$3.065&0.020 & 1.254&0.010  &  798&$^{+6}_{-6}$       \\ % EDR3 uncorrected PM  + k6
%O-014~NW &       199 &   102 &       1.00 &                 5.54 &               5.32 & $-$1.634&0.020 & $-$3.055&0.021 & 1.253&0.010  &  799&$^{+7}_{-7}$       \\ % EDR3 orig
%O-014    &        12 &    12 &       0.94 &                 2.98 &               2.63 & $-$0.092&0.055 & $-$4.176&0.055 & 1.404&0.036  &  714&$^{+19}_{-18}$     \\ % DR2 results with spimax = 9.99
 O-015    &       140 &   102 &       1.17 &                 4.60 &               3.70 & $-$2.570&0.022 & $-$6.370&0.022 & 1.000&0.011  & 1001&$^{+11}_{-11}$     \\
%O-015    &       146 &   105 &       1.16 &                 4.56 &               3.62 & $-$2.567&0.022 & $-$6.384&0.022 & 1.000&0.011  & 1001&$^{+11}_{-11}$     \\ % EDR3 uncorrected PM + k6
%O-015    &       147 &   106 &       1.16 &                 4.50 &               3.73 & $-$2.572&0.022 & $-$6.369&0.022 & 1.003&0.011  &  998&$^{+11}_{-10}$     \\ % EDR3 orig
%O-015    &        93 &    75 &       0.98 &                 3.20 &               2.82 & $-$2.605&0.048 & $-$6.390&0.048 & 0.997&0.035  & 1006&$^{+37}_{-34}$     \\ % DR2
 O-016    &       399 &   340 &       0.88 &                 8.04 &               5.27 & $-$1.852&0.020 & $-$3.688&0.020 & 1.423&0.010  &  703&$^{+5}_{-5}$       \\
%O-016    &       407 &   348 &       0.90 &                 8.03 &               5.27 & $-$1.851&0.020 & $-$3.682&0.020 & 1.423&0.010  &  703&$^{+5}_{-5}$       \\ % EDR3 uncorrected PM + k6
%O-016    &       398 &   342 &       0.94 &                 7.91 &               5.42 & $-$1.839&0.020 & $-$3.673&0.020 & 1.424&0.010  &  703&$^{+5}_{-5}$       \\ % EDR3 orig
%O-016    &       340 &   286 &       1.12 &                 5.15 &               3.56 & $-$1.885&0.041 & $-$3.716&0.041 & 1.394&0.031  &  719&$^{+16}_{-16}$     \\ % DR2
 O-016~N  &       123 &   119 &       0.87 &                 5.83 &               4.52 & $-$1.673&0.023 & $-$3.695&0.023 & 1.426&0.011  &  701&$^{+6}_{-5}$       \\
%O-016~N  &       129 &   125 &       0.90 &                 5.75 &               4.55 & $-$1.674&0.023 & $-$3.686&0.023 & 1.425&0.011  &  702&$^{+6}_{-5}$       \\ % EDR3 uncorrected PM + k6
%O-016~N  &       128 &   124 &       0.94 &                 5.79 &               4.93 & $-$1.682&0.023 & $-$3.672&0.023 & 1.425&0.011  &  702&$^{+6}_{-5}$       \\ % EDR3 orig
%O-016~N  &       102 &    99 &       1.04 &                 3.99 &               3.25 & $-$1.716&0.059 & $-$3.705&0.059 & 1.397&0.041  &  719&$^{+22}_{-21}$     \\ % DR2
 O-016~S  &       109 &   107 &       0.96 &                 8.12 &               4.94 & $-$2.026&0.023 & $-$3.759&0.023 & 1.416&0.011  &  706&$^{+6}_{-6}$       \\
%O-016~S  &       111 &   109 &       0.97 &                 8.11 &               4.89 & $-$2.030&0.023 & $-$3.755&0.023 & 1.416&0.011  &  707&$^{+6}_{-6}$       \\ % EDR3 uncorrected PM + k6
%O-016~S  &       105 &   104 &       0.96 &                 8.02 &               4.97 & $-$2.015&0.023 & $-$3.755&0.023 & 1.417&0.011  &  706&$^{+6}_{-6}$       \\ % EDR3 orig
%O-016~S  &        94 &    90 &       1.19 &                 5.15 &               3.43 & $-$2.077&0.057 & $-$3.788&0.057 & 1.390&0.041  &  722&$^{+22}_{-21}$     \\ % DR2
 O-017    &       480 &   343 &       1.22 &                 5.46 &               5.52 & $-$0.843&0.020 & $-$0.537&0.020 & 0.483&0.010  & 2075&$^{+44}_{-42}$     \\ % Heart nebula
%O-017    &       486 &   347 &       1.23 &                 5.51 &               5.54 & $-$0.856&0.020 & $-$0.548&0.020 & 0.483&0.010  & 2074&$^{+43}_{-42}$     \\ % EDR3 uncorrected PM + k6
%O-017    &       476 &   352 &       1.27 &                 5.58 &               5.59 & $-$0.862&0.020 & $-$0.556&0.020 & 0.483&0.010  & 2072&$^{+43}_{-42}$     \\ % EDR3 orig
 O-018    &       388 &   301 &       1.10 &                 8.24 &               8.53 & $+$1.385&0.022 & $-$2.024&0.022 & 0.811&0.010  & 1234&$^{+16}_{-16}$     \\ % Lagoon nebula
%O-018    &       545 &   409 &       1.11 &                 7.88 &               7.97 & $+$1.374&0.022 & $-$2.037&0.022 & 0.811&0.010  & 1233&$^{+16}_{-16}$     \\ % EDR3 uncorrected PM + k6
%O-018    &       532 &   404 &       1.13 &                 7.90 &               8.30 & $+$1.383&0.022 & $-$2.043&0.022 & 0.810&0.010  & 1235&$^{+16}_{-16}$     \\ % EDR3 orig           
 O-019    &       202 &   188 &       0.96 &                 5.26 &               7.16 & $+$0.200&0.022 & $-$1.627&0.022 & 0.590&0.011  & 1697&$^{+31}_{-30}$     \\ % Eagle nebula
%O-019    &       229 &   212 &       0.97 &                 5.10 &               7.07 & $+$0.192&0.022 & $-$1.642&0.022 & 0.590&0.011  & 1696&$^{+31}_{-30}$     \\ % EDR3 uncorrected PM + k6
%O-019    &       220 &   206 &       0.99 &                 5.17 &               7.35 & $+$0.193&0.022 & $-$1.652&0.022 & 0.590&0.011  & 1696&$^{+31}_{-30}$     \\ % EDR3 orig           
 O-020    &       268 &   223 &       1.07 &                 4.83 &               6.79 & $-$1.729&0.021 & $+$0.276&0.021 & 0.704&0.010  & 1421&$^{+21}_{-20}$     \\ % Rosette nebula
%O-020    &       283 &   236 &       1.06 &                 4.74 &               6.66 & $-$1.732&0.021 & $+$0.283&0.021 & 0.704&0.010  & 1421&$^{+21}_{-20}$     \\ % EDR3 uncorrected PM + k6
%O-020    &       276 &   234 &       1.05 &                 4.79 &               6.77 & $-$1.730&0.021 & $+$0.279&0.021 & 0.705&0.010  & 1419&$^{+21}_{-20}$     \\ % EDR3 orig           
 O-021    &       129 &   116 &       1.09 &                 5.15 &               3.51 & $-$2.776&0.023 & $+$2.990&0.023 & 0.815&0.011  & 1227&$^{+17}_{-16}$     \\ % tau CMa cluster
%O-021    &       130 &   119 &       1.10 &                 5.13 &               3.60 & $-$2.766&0.023 & $+$3.006&0.023 & 0.814&0.011  & 1228&$^{+17}_{-16}$     \\ % EDR3 uncorrected PM + k6
%O-021    &       142 &   121 &       1.08 &                 5.50 &               3.89 & $-$2.773&0.023 & $+$3.015&0.023 & 0.812&0.011  & 1232&$^{+17}_{-16}$     \\ % EDR3 orig           
 O-022    &       527 &   495 &       1.01 &                 5.08 &               5.72 & $-$0.544&0.022 & $-$2.210&0.022 & 0.645&0.010  & 1551&$^{+25}_{-24}$     \\ % NGC 6231
%O-022    &       530 &   497 &       1.01 &                 5.09 &               5.73 & $-$0.559&0.022 & $-$2.211&0.022 & 0.645&0.010  & 1551&$^{+25}_{-24}$     \\ % EDR3 uncorrected PM + k6
%O-022    &       530 &   494 &       1.08 &                 5.25 &               5.85 & $-$0.564&0.022 & $-$2.209&0.022 & 0.643&0.010  & 1556&$^{+25}_{-24}$     \\ % EDR3 orig           
 O-023    &        84 &    82 &       0.92 &                11.29 &              12.49 & $+$1.577&0.022 & $+$0.464&0.022 & 2.562&0.011  &  390&$^{+2}_{-2}$       \\ % Orion nebula
%O-023    &       116 &   114 &       0.99 &                10.83 &              12.18 & $+$1.549&0.022 & $+$0.495&0.022 & 2.563&0.011  &  390&$^{+2}_{-2}$       \\ % EDR3 uncorrected PM + k6
%O-023    &       112 &   108 &       0.97 &                10.92 &              11.85 & $+$1.518&0.022 & $+$0.453&0.022 & 2.570&0.011  &  389&$^{+2}_{-2}$       \\ % EDR3 orig           
 O-024    &       167 &   138 &       1.22 &                 7.03 &               6.44 & $-$6.560&0.020 & $+$9.802&0.020 & 2.974&0.010  &  336&$^{+1}_{-1}$       \\ % Pozzo 1
%O-024    &       173 &   144 &       1.26 &                 7.01 &               6.56 & $-$6.554&0.020 & $+$9.812&0.020 & 2.973&0.010  &  336&$^{+1}_{-1}$       \\ % EDR3 uncorrected PM + k6
%O-024    &       170 &   144 &       1.26 &                 6.81 &               6.36 & $-$6.588&0.020 & $+$9.833&0.020 & 2.973&0.010  &  336&$^{+1}_{-1}$       \\ % EDR3 orig
 O-025    &       177 &   175 &       0.90 &                 4.11 &               3.77 & $-$6.877&0.023 & $+$2.596&0.023 & 0.434&0.011  & 2311&$^{+58}_{-56}$     \\ % Trumpler 16 E
%O-025    &       203 &   198 &       0.96 &                 4.11 &               3.84 & $-$6.875&0.023 & $+$2.623&0.023 & 0.434&0.011  & 2309&$^{+58}_{-55}$     \\ % EDR3 uncorrected PM + k6
%O-025    &       205 &   199 &       0.95 &                 4.15 &               3.97 & $-$6.884&0.023 & $+$2.630&0.023 & 0.432&0.011  & 2318&$^{+58}_{-56}$     \\ % EDR3 orig     
 O-026    &        60 &    55 &       0.99 &                 7.16 &               6.81 & $+$1.350&0.023 & $-$0.568&0.023 & 2.520&0.012  &  397&$^{+2}_{-2}$       \\ % sigma Ori
%O-026    &        64 &    54 &       0.98 &                 7.19 &               6.76 & $+$1.346&0.023 & $-$0.566&0.023 & 2.520&0.012  &  397&$^{+2}_{-2}$       \\ % EDR3 uncorrected PM + k6
%O-026    &        67 &    59 &       1.00 &                 7.73 &               6.82 & $+$1.306&0.023 & $-$0.565&0.023 & 2.521&0.012  &  397&$^{+2}_{-2}$       \\ % EDR3 orig
\hline
\end{tabular}
\renewcommand{\arraystretch}{1.0}
}
\label{results}                  
\end{table*}

\begin{table*}
\caption{New GOSSS spectral classifications for \VO{002} to \VO{022}.}
\centerline{
\scriptsize
\addtolength{\tabcolsep}{-3pt}
\begin{tabular}{lcrrclllll}
\hline
Star name                 & GOS/GBS/             & \mci{{\it Gaia} EDR3}     & \mci{ALS}   & Group & \mci{Spect.} & \mci{Lum.} & \mci{Qual.} & \mci{Second.}     & Notes                                                              \\
                          & GAS/GWS ID           & \mci{ID}                  & \mci{ID}    & ID    & \mci{type}   & \mci{cl.}  &             & \mci{comp.}       &                                                                    \\
\hline
HD~\num{93129}~B          & 287.41$-$00.57$\_$02 & \num{5350363910256783744} & \num{19309} & O-002 & O3.5         & V          & ((fc))z     & ---               & ---                                                                \\
HD~\num{93128}~B          & 287.40$-$00.58$\_$02 & \num{5350363875897024256} & ---         & O-002 & B0.2         & V          & ---         & ---               & ---                                                                \\
ALS~\num{15219}           & 287.43$-$00.61$\_$01 & \num{5350362638946370688} & \num{15219} & O-002 & B0.7         & V          & ---         & B1 V              & ---                                                                \\
ALS~\num{15229}           & 287.41$-$00.58$\_$02 & \num{5350363807177545216} & \num{15229} & O-002 & B0           & V          & ---         & ---               & ---                                                                \\
ALS~\num{15234}           & 287.39$-$00.60$\_$01 & \num{5350363463580089856} & \num{15234} & O-002 & B1.5         & V          & ---         & ---               & ---                                                                \\
ALS~\num{15863}           & 287.41$-$00.59$\_$01 & \num{5350363807177527680} & \num{15863} & O-002 & B0.2         & V          & ---         & ---               & ---                                                                \\ % npi=$-$3.33
Trumpler~14~MJ~152        & 287.40$-$00.58$\_$04 & \num{5350363978961349120} & ---         & O-002 & B1.5:        & V          & (n)         & ---               & ---                                                                \\
ALS~\num{15203}           & 287.40$-$00.63$\_$01 & \num{5350363326141053312} & \num{15203} & O-002 & B0           & V          & ---         & B + B             & RUWE=1.69, \Cstar~=~0.67                                           \\
ALS~\num{15243}           & 287.52$-$00.71$\_$02 & \num{5350357549383373440} & \num{15243} & O-003 & B0.2         & V          & (n)         & ---               & ---                                                                \\
ALS~\num{19692}           & 353.11$+$00.65$\_$01 & \num{5976039249668077824} & \num{19692} & O-005 & O3:          & III        & (f)         & O7: III((f))      & RUWE=1.40, $r$=897\arcsec                                          \\
ALS~\num{18552}           & 289.76$-$01.28$\_$01 & \num{5337975437890183168} & \num{18552} & O-006 & O6.5         & IV         & ((f))       & B0: V             & $r$=226\arcsec                                                     \\
ALS~\num{15128}           & 080.14$+$00.73$\_$01 & \num{2067781871173898624} & \num{15128} & O-007 & O7.5         & V          & (n)z        & ---               & ---                                                                \\
Cyg~OB2-23                & 080.24$+$00.80$\_$02 & \num{2067785070923661440} & \num{15120} & O-008 & O9.5         & V          & ---         & ---               & ---                                                                \\
ALS~\num{15123}           & 080.22$+$00.76$\_$01 & \num{2067784173274044928} & \num{15123} & O-008 & O9.5         & V          & (n)         & ---               & ---                                                                \\
BD~$-$16~4818             & 015.07$-$00.71$\_$01 & \num{4097815171698857856} & \num{4933}  & O-009 & O9.5         & V          & ---         & ---               & RUWE=19.49                                                         \\ % npi=3.15
ALS~\num{19612}           & 015.01$-$00.70$\_$01 & \num{4097809055665428864} & \num{19612} & O-009 & O6.5         & V          & ((f))z      & ---               & ---                                                                \\
ALS~\num{19611}           & 015.08$-$00.66$\_$01 & \num{4098003188189682176} & \num{19611} & O-009 & O7:          & V          & (n)         & ---               & RUWE=8.38                                                          \\
ALS~\num{19608}           & 015.07$-$00.65$\_$01 & \num{4098003291268883584} & \num{19608} & O-009 & O9.5         & V          & ---         & ---               & \pic/\sigmae = 3.27                                                \\
HD~\num{150041}           & 336.65$-$01.50$\_$01 & \num{5940956238844663552} & \num{3695}  & O-010 & B0           & II         & ---         & ---               & ---                                                                \\
CPD~$-$48~8705            & 336.73$-$01.58$\_$01 & \num{5940954280339192064} & ---         & O-010 & B0           & V          & ---         & ---               & ---                                                                \\
CPD~$-$48~8704            & 336.73$-$01.57$\_$01 & \num{5940954280339207424} & \num{3702}  & O-010 & B1           & V          & ---         & ---               & ---                                                                \\
2MASS~J20350798$+$4649321 & 084.85$+$03.80$\_$01 & \num{2071521932995188352} & ---         & O-011 & O8           & IV         & ((f))e      & ---               & RUWE=6.75, \sigmae = 0.210~mas                                     \\
ALS~\num{11448}~B         & 084.88$+$03.80$\_$01 & \num{2071522345312075776} & ---         & O-011 & B4           & V          & ---         & ---               & ---                                                                \\
CPD~$-$26~2711            & 243.26$+$00.45$\_$01 & \num{5602022983468167680} & \num{839}   & O-012 & O7           & V          & z           & ---               & ---                                                                \\
ALS~\num{832}             & 243.18$+$00.45$\_$01 & \num{4098003188189682176} & \num{832}   & O-012 & B0.5         & V          & (n)         & ---               & RUWE=1.78                                                          \\
57~Cyg                    & 084.90$-$00.19$\_$01 & \num{2162953368584828032} & ---         & O-014 & B5           & V          & ---         & B5 V              & foreground star                                                    \\
Tyc~3179-01209-1          & 085.25$-$00.69$\_$01 & \num{2162998929612532608} & ---         & O-014 & A0           & III        & n           & ---               & $r$=2724\arcsec, runaway?                                          \\
HD~\num{195965}           & 085.71$+$05.00$\_$01 & \num{2083681294654108672} & ---         & O-014 & B0.2         & V          & n           & ---               & confirmed runaway                                                  \\
HD~\num{201795}           & 082.97$-$06.21$\_$01 & \num{1968091458287111296} & ---         & O-014 & B0.5         & V          & ---         & ---               & confirmed runaway                                                  \\
HDE~\num{228845}          & 078.31$+$02.85$\_$01 & \num{2068367812086551936} & ---         & O-015 & B1.5         & V          & (n)         & ---               & ---                                                                \\
HD~\num{47887}            & 203.34$+$02.04$\_$01 & \num{3326686263650895360} & \num{9091}  & O-016 & B0.7         & V          & ---         & ---               & \sigmae = 0.149~mas                                                \\
HD~\num{47961}            & 203.03$+$02.28$\_$01 & \num{3326737120359493504} & ---         & O-016 & B2           & V          & ---         & ---               & RUWE=1.62                                                          \\
HD~\num{47777}            & 203.12$+$02.03$\_$01 & \num{3326710010525962624} & \num{14664} & O-016 & B2           & IV         & ---         & ---               & \sigmae = 0.142~mas                                                \\
HD~\num{47732}            & 202.95$+$02.05$\_$01 & \num{3326716813754146176} & ---         & O-016 & B2           & V          & ---         & B3: V             & ---                                                                \\
HD~\num{47755}            & 202.99$+$02.08$\_$01 & \num{3326715714242517248} & \num{9088}  & O-016 & B2.5         & V          & ---         & B4: V             & ---                                                                \\ % npi=3.20
HD~\num{48055}            & 203.38$+$02.21$\_$01 & \num{3326644967541579776} & \num{9097}  & O-016 & B2.5         & V          & n           & ---               & ---                                                                \\
HDE~\num{261938}          & 202.96$+$02.20$\_$01 & \num{3326740693772848896} & ---         & O-016 & B5           & V          & (n)         & ---               & \sigmae = 0.110~mas                                                \\
HDE~\num{261810}          & 203.02$+$02.08$\_$01 & \num{3326715439364610816} & ---         & O-016 & B2.5         & V          & n           & ---               & ---                                                                \\
HDE~\num{261878}          & 202.95$+$02.16$\_$01 & \num{3326717260430731648} & ---         & O-016 & B5           & V          & ---         & ---               & ---                                                                \\
HDE~\num{261903}          & 203.34$+$02.01$\_$01 & \num{3326685924349755520} & ---         & O-016 & B7           & V          & nn          & ---               & ---                                                                \\
HDE~\num{261969}          & 202.97$+$02.24$\_$01 & \num{3326740006577519360} & ---         & O-016 & A0           & V          & ---         & ---               & ---                                                                \\
HDE~\num{261936}          & 202.74$+$02.35$\_$01 & \num{3326941865743978880} & ---         & O-016 & B9           & III        & ---         & ---               & ---                                                                \\
HDE~\num{261940}          & 203.25$+$02.06$\_$01 & \num{3326695854314127616} & ---         & O-016 & B8           & V          & (n)         & ---               & ---                                                                \\
BD~$+$60~493              & 134.62$+$00.58$\_$01 & \num{465503094013753728}  & \num{7249}  & O-017 & B0.5         & Ia         & ---         & ---               & ---                                                                \\
BD~$+$60~496              & 134.61$+$00.96$\_$01 & \num{465533158786645504}  & \num{7265}  & O-017 & B0.7         & V          & ---         & B1 V              & ---                                                                \\
KM~Cas                    & 134.36$+$00.82$\_$01 & \num{465523778576137600}  & \num{7225}  & O-017 & O9.2         & IV         & n           & ---               & ---                                                                \\
ALS~\num{7320}            & 134.91$+$00.95$\_$01 & \num{465487322893075840}  & \num{7320}  & O-017 & B0.7         & V          & ---         & B1.5: V           & ---                                                                \\
V1166~Cas                 & 134.59$+$01.07$\_$01 & \num{465551472526065152}  & \num{7268}  & O-017 & B1           & V          & ---         & ---               & ---                                                                \\
ALS~\num{7299}            & 134.78$+$00.95$\_$01 & \num{465533742902211072}  & \num{7299}  & O-017 & B0.7         & V          & ---         & ---               & RUWE=1.51                                                          \\
ALS~\num{7270}            & 134.70$+$00.81$\_$01 & \num{465526596075683072}  & \num{7270}  & O-017 & B            & ---        & e           & ---               & RUWE=3.15, $r_\mu$=0.83~$\mu$as/a, also Table~\ref{LiLi_spclas}    \\
ALS~\num{15320}           & 134.77$+$00.95$\_$01 & \num{465534498815432704}  & \num{15320} & O-017 & B2.5         & V          & nn          & ---               & ---                                                                \\
HD~\num{164906}           & 006.05$-$01.33$\_$01 & \num{4065974968548020352} & \num{4616}  & O-018 & B            & ---        & e           & ---               & ---                                                                \\
HDE~\num{315031}          & 006.07$-$01.32$\_$01 & \num{4065975002907765760} & \num{4618}  & O-018 & B1           & V          & ---         & ---               & ---                                                                \\
CPD~$-$24~6140            & 005.99$-$01.18$\_$01 & \num{4066021182400931072} & \num{4592}  & O-018 & B1.5         & V          & (n)         & ---               & ---                                                                \\
BD~$-$13~4930             & 016.94$+$00.77$\_$01 & \num{4146599476126383872} & \num{9505}  & O-019 & O9.7         & V          & ---         & ---               & ---                                                                \\
BD~$-$13~4929             & 016.98$+$00.87$\_$01 & \num{4146600781797073920} & \num{4913}  & O-019 & O8           & V          & ---         & B0.5: V + B0.5: V & RUWE=1.75, also Table~\ref{LiLi_spclas}                            \\
BD~$-$13~4921             & 016.90$+$00.85$\_$01 & \num{4146598995090108288} & \num{9491}  & O-019 & B0.7         & V          & ---         & ---               & ---                                                                \\
BD~$-$13~4928             & 016.97$+$00.82$\_$01 & \num{4146600678717290624} & \num{4911}  & O-019 & O9.7         & IV         & nn          & ---               & also Table~\ref{LiLi_spclas}                                       \\
ALS~\num{4903}            & 016.92$+$00.85$\_$01 & \num{4146599029451913984} & \num{4903}  & O-019 & O9           & V          & ---         & ---               & ---                                                                \\
ALS~\num{15352}           & 017.00$+$00.90$\_$01 & \num{4146613632338722432} & \num{15352} & O-019 & O8           & IV         & ---         & ---               & ---                                                                \\
ALS~\num{15369}           & 016.98$+$00.84$\_$01 & \num{4146612425449995648} & \num{15369} & O-019 & B0.5         & V          & ---         & ---               & ---                                                                \\ % npi=-3.17  
ALS~\num{15363}           & 016.99$+$00.86$\_$01 & \num{4146612803413634688} & \num{15363} & O-019 & B1.5         & V          & p           & ---               & ---                                                                \\
ALS~\num{15360}           & 017.00$+$00.87$\_$01 & \num{4146612872133107584} & \num{15360} & O-019 & O6.5         & V          & ((f))z      & ---               & ---                                                                \\
HDE~\num{259135}          & 206.37$-$02.08$\_$01 & \num{3131334452194616192} & \num{8986}  & O-020 & B0           & V          & ---         & B1: V             & also Table~\ref{LiLi_spclas}                                       \\
HDE~\num{259481}          & 206.43$-$01.84$\_$01 & \num{3131332940366123392} & \num{9001}  & O-020 & B1.5:        & V          & (n)e        & ---               & RUWE=1.66, $r_\mu$=1.02~$\mu$as/a                                  \\
HDE~\num{258691}          & 206.37$-$02.49$\_$01 & \num{3131301299346222080} & \num{8970}  & O-020 & O9.5         & V          & ---         & ---               & RUWE=6.97, $r$=1517\arcsec, runaway?, also Table~\ref{LiLi_spclas} \\ % also spi=0.247
HD~\num{46056}~B          & 206.33$-$02.25$\_$01 & \num{3131327653265431552} & ---         & O-020 & B2           & V          & ---         & ---               & ---                                                                \\ % npi=3.38   
WR~79                     & 343.49$+$01.16$\_$01 & \num{5966509541883497984} & \num{3810}  & O-022 & WC           & ---        & ---         & O4 III-I(f)       & also Table~\ref{LiLi_spclas}                                       \\
HD~\num{152234}           & 343.46$+$01.22$\_$01 & \num{5966510057279631488} & \num{15086} & O-022 & B0.5         & Ia         & ---         & ---               & RUWE=4.01, \sigmae = 1.397~mas, also Table~\ref{LiLi_spclas}       \\
HD~\num{152235}           & 343.31$+$01.10$\_$01 & \num{5966501501704611840} & \num{3807}  & O-022 & B0.7         & Ia         & ---         & ---               & \sigmae = 0.117~mas                                                \\
HD~\num{152076}           & 343.42$+$01.40$\_$01 & \num{5966520876291200256} & \num{3794}  & O-022 & B0           & III        & ---         & ---               & ---                                                                \\
CPD~$-41$~7723            & 343.45$+$01.19$\_$01 & \num{5966509816761424256} & \num{16063} & O-022 & B1.5         & V          & ---         & ---               & ---                                                                \\
CPD~$-41$~7724            & 343.46$+$01.19$\_$01 & \num{5966509846818671872} & \num{16493} & O-022 & B0.7         & V          & ---         & ---               & ---                                                                \\
HDE~\num{326330}          & 343.45$+$01.14$\_$01 & \num{5966508957767911040} & \num{15969} & O-022 & B1           & IV         & (n)         & ---               & ---                                                                \\
HDE~\num{326333}          & 343.53$+$01.10$\_$01 & \num{5966507823896515200} & \num{3822}  & O-022 & B0.5         & V          & ---         & ---               & ---                                                                \\
V945~Sco                  & 343.40$+$01.19$\_$01 & \num{5966504044325373440} & \num{16119} & O-022 & B1           & III        & n           & ---               & ---                                                                \\
HDE~\num{326320}          & 343.58$+$01.33$\_$01 & \num{5966523938614083712} & \num{16057} & O-022 & B1.5:        & V          & ---         & ---               & ---                                                                \\ % npi=3.26 
CPD~$-41$~7721~B          & 343.44$+$01.18$\_$01 & \num{5966503941246146176} & ---         & O-022 & B1.5         & V          & ---         & ---               & ---                                                                \\
ALS~\num{14761}           & 343.45$+$01.12$\_$01 & \num{5966508717249708672} & \num{14761} & O-022 & B2           & V          & ---         & ---               & ---                                                                \\
2MASS~J16540272$-$4150286 & 343.44$+$01.19$\_$01 & \num{5966503936943654912} & ---         & O-022 & B5:          & V          & ---         & ---               & ---                                                                \\
\hline
\end{tabular}
\addtolength{\tabcolsep}{3pt}
}
\label{GOSSS_spclas_01}
\end{table*}

\begin{table*}
\caption{New GOSSS spectral classifications for \VO{023} to \VO{026}.}
\centerline{
\scriptsize
\addtolength{\tabcolsep}{-3pt}
\begin{tabular}{lcrrclllll}
\hline
Star name                 & GOS/GBS/             & \mci{{\it Gaia} EDR3}     & \mci{ALS}   & Group & \mci{Spect.} & \mci{Lum.} & \mci{Qual.} & \mci{Second.}      & Notes                                                      \\
                          & GAS/GWS ID           & \mci{ID}                  & \mci{ID}    & ID    & \mci{type}   & \mci{cl.}  &             & \mci{comp.}        &                                                            \\
\hline
$\theta^2$~Ori~B          & 209.06$-$19.36$\_$01 & \num{3017360833515044480} & \num{16182} & O-023 & B0.7         & V          & ---         & ---                & \sigmae=0.133~mas, also Table~\ref{LiLi_spclas}            \\
$\theta^1$~Ori~D          & 209.01$-$19.38$\_$03 & \num{3017364063330465152} & \num{16710} & O-023 & B0           & V          & ---         & ---                & also Table~\ref{LiLi_spclas}                               \\ % npi=-6.53 
$\theta^1$~Ori~A          & 209.01$-$19.39$\_$01 & \num{3017364132050194688} & \num{14651} & O-023 & B0.2         & V          & ---         & ---                & RUWE=1.49, \sigmae=0.279~mas, also Table~\ref{LiLi_spclas} \\
NU~Ori                    & 208.92$-$19.27$\_$01 & \num{3017367396223983616} & \num{16712} & O-023 & B0           & V          & (n)         & ---                & RUWE=2.41, \sigmae=0.226~mas, also Table~\ref{LiLi_spclas} \\
HD~\num{36981}            & 208.82$-$19.34$\_$01 & \num{3209522411971821824} & ---         & O-023 & B5           & V          & ---         & ---                & ---                                                        \\
$\theta^1$~Ori~Ba,Bb      & 209.01$-$19.38$\_$02 & \num{3017364132049943680} & \num{14652} & O-023 & B3           & V          & n           & ---                & RUWE=1.46, \sigmae=0.146, also Table~\ref{LiLi_spclas}     \\
$\theta^2$~Ori~C          & 209.07$-$19.34$\_$01 & \num{3017360799155290368} & ---         & O-023 & B4           & V          & ---         & ---                & $r_\mu$=3.39~$\mu$as/a, also Table~\ref{LiLi_spclas}       \\
HD~\num{36982}            & 209.07$-$19.44$\_$01 & \num{3017360348171372672} & \num{16708} & O-023 & B2           & V          & ---         & ---                & also Table~\ref{LiLi_spclas}                               \\ % npi=-3.33              
HD~\num{36939}            & 209.08$-$19.52$\_$01 & \num{3017265961968363904} & ---         & O-023 & B8           & V          & (n)         & ---                &                                                            \\ % npi=-3.69         
$\gamma^1$~Vel~A,B        & 262.81$-$07.70$\_$01 & \num{5519266900766220800} & \num{14916} & O-024 & B1.5         & V          & ---         & ---                & \sigmae=0.338~mas                                          \\
$\gamma^2$~Vel            & 262.80$-$07.69$\_$01 & ---                       & \num{980}   & O-024 & WC           & ---        & ---         & O7/8.5 III-II((f)) & not in {\it Gaia} EDR3, also Table~\ref{LiLi_spclas}       \\
$\eta$~Car                & 287.60$-$00.63$\_$01 & \num{5350358584482202880} & \num{1868}  & O-025 & LBV          & ---        & ---         & ---                & no parallax in {\it Gaia} EDR3                             \\
CPD~$-$59~2598            & 287.56$-$00.66$\_$01 & \num{5350358859375074816} & \num{1854}  & O-025 & B1.5         & V          & ---         & ---                & ---                                                        \\
CPD~$-$59~2640            & 287.61$-$00.60$\_$01 & \num{5350381914760913024} & \num{15223} & O-025 & B1.5         & V          & (n)         & ---                & ---                                                        \\
CPD~$-$59~2616            & 287.61$-$00.67$\_$01 & \num{5350357931646616448} & \num{19745} & O-025 & B2           & V          & (n)         & ---                & ---                                                        \\
2MASS~J10452265$-$5942596 & 287.65$-$00.64$\_$01 & \num{5350311374212229120} & ---         & O-025 & B7           & V          & (n)e        & ---                & RUWE=1.92                                                  \\
$\sigma$~Ori~D            & 206.82$-$17.33$\_$01 & \num{3216486478101982592} & \num{8474}  & O-026 & B2           & V          & (n)         & ---                & \sigmae=0.160~mas, also Table~\ref{LiLi_spclas}            \\
$\sigma$~Ori~E            & 206.82$-$17.32$\_$01 & \num{3216486478101981056} & \num{8475}  & O-026 & B2           & V          & p           & ---                & \sigmae=0.168~mas, also Table~\ref{LiLi_spclas}            \\
HDE~\num{294271}          & 206.76$-$17.34$\_$01 & \num{3216489776637339136} & ---         & O-026 & B5           & V          & (n)         & ---                & RUWE=1.44                                                  \\
\hline
\end{tabular}
\addtolength{\tabcolsep}{3pt}
}
\label{GOSSS_spclas_02}
\end{table*}

\begin{table*}
\caption{New \lili\ spectral classifications.}
\centerline{
\scriptsize
\addtolength{\tabcolsep}{-3pt}
\begin{tabular}{lcrrclllll}
\hline
Star name                 & GOS/GBS/             & \mci{{\it Gaia} EDR3}     & \mci{ALS}   & Group & \mci{Spect.} & \mci{Lum.} & \mci{Qual.} & \mci{Second.}      & Notes                                                                  \\
                          & GAS/GWS ID           & \mci{ID}                  & \mci{ID}    & ID    & \mci{type}   & \mci{cl.}  &             & \mci{comp.}        &                                                                        \\
\hline
NGC~3603~HST-5            & 291.62$-$00.52$\_$36 & \num{5337418397799185536} & ---         & O-001 & B1.5         & Ia         & ---         & ---                & ---                                                                    \\
HD~\num{93161}~A          & 287.44$-$00.59$\_$02 & \num{5350362982543827456} & \num{1832}  & O-002 & O7           & V          & ((f))       & O9 IV              & also \citet{Sotaetal14}                                                \\
ALS~\num{7270}            & 134.70$+$00.81$\_$01 & \num{465526596075683072}  & \num{7270}  & O-017 & B            & ---        & e           & ---                & RUWE=3.15, $r_\mu$=0.83~$\mu$as/a, also Table~\ref{GOSSS_spclas_01}    \\
HD~\num{164816}           & 006.06$-$01.20$\_$01 & \num{4066023415781257984} & \num{4597}  & O-018 & O9.5         & V          & ---         & O9.7 V             & also \citet{Sotaetal14}                                                \\
HD~\num{165052}           & 006.12$-$01.48$\_$01 & \num{4066064956700837248} & \num{4635}  & O-018 & O6.5         & V          & ((f))z      & O6.5 V((f))z       & also \citet{Maizetal16}                                                \\
BD~$-$13~4930             & 016.94$+$00.77$\_$01 & \num{4146599476126383872} & \num{9505}  & O-019 & O9.7         & V          & ---         & ---                & also Table~\ref{GOSSS_spclas_01}                                       \\
BD~$-$13~4929             & 016.98$+$00.87$\_$01 & \num{4146600781797073920} & \num{4913}  & O-019 & O8           & V          & ---         & B0.5: V + B0.5: V  & RUWE=1.75, also Table~\ref{GOSSS_spclas_01}                            \\
BD~$-$13~4928             & 016.97$+$00.82$\_$01 & \num{4146600678717290624} & \num{4911}  & O-019 & O9.7         & IV         & nn          & ---                & also Table~\ref{GOSSS_spclas_01}                                       \\
HD~\num{46106}            & 206.20$-$02.09$\_$01 & \num{3131384484272514688} & \num{8981}  & O-020 & O9.7         & V          & ---         & B2: Vn             & also \citet{Sotaetal14}                                                \\
HDE~\num{259135}          & 206.37$-$02.08$\_$01 & \num{3131334452194616192} & \num{8986}  & O-020 & B0           & V          & ---         & B1: V              & also Table~\ref{GOSSS_spclas_01}                                       \\
HDE~\num{258691}          & 206.37$-$02.49$\_$01 & \num{3131301299346222080} & \num{8970}  & O-020 & O9.5         & V          & ---         & ---                & RUWE=6.97, $r$=1517\arcsec, runaway?, also Table~\ref{GOSSS_spclas_01} \\ % also spi=0.247
HD~\num{152234}           & 343.46$+$01.22$\_$01 & \num{5966510057279631488} & \num{15086} & O-022 & B0.5         & Ia         & ---         & ---                & RUWE=4.01, \sigmae = 1.397~mas, also Table~\ref{GOSSS_spclas_01}       \\
HD~\num{152248}~Aa,Ab     & 343.46$+$01.18$\_$01 & \num{5966509885480902656} & \num{15087} & O-022 & O7           & Ib         & (f)         & O7 Ib(f)           & also \citet{Sotaetal14}                                                \\
HD~\num{152218}           & 343.53$+$01.28$\_$01 & \num{5966522667303699072} & \num{15085} & O-022 & O8.5         & IV         & ---         & B0 IV              & also \citet{Sotaetal14}                                                \\
WR~79                     & 343.49$+$01.16$\_$01 & \num{5966509541883497984} & \num{3810}  & O-022 & WC7          & ---        & ---         & O4 III-I(f)        & also Table~\ref{GOSSS_spclas_01}                                       \\
CPD~$-$41~7733            & 343.46$+$01.17$\_$01 & \num{5966509022176616448} & \num{15756} & O-022 & O9           & IV         & ---         & B                  & also \citet{Sotaetal14}                                                \\
V1034~Sco                 & 343.48$+$01.15$\_$01 & \num{5966509095207261056} & \num{15757} & O-022 & O9.5         & IV         & ---         & B1: V              & also \citet{Maizetal16}                                                \\
$\theta^2$~Ori~B          & 209.06$-$19.36$\_$01 & \num{3017360833515044480} & \num{16182} & O-023 & B0.7         & V          & ---         & ---                & \sigmae=0.133~mas, also Table~\ref{GOSSS_spclas_02}                    \\
$\theta^1$~Ori~D          & 209.01$-$19.38$\_$03 & \num{3017364063330465152} & \num{16710} & O-023 & B0           & V          & ---         & ---                & also Table~\ref{GOSSS_spclas_02}                                       \\ % npi=-6.53
$\theta^1$~Ori~A          & 209.01$-$19.39$\_$01 & \num{3017364132050194688} & \num{14651} & O-023 & B0.2         & V          & ---         & ---                & RUWE=1.49, \sigmae=0.279~mas, also Table~\ref{GOSSS_spclas_02}         \\
NU~Ori                    & 208.92$-$19.27$\_$01 & \num{3017367396223983616} & \num{16712} & O-023 & B0           & V          & (n)         & ---                & RUWE=2.41, \sigmae=0.226~mas, also Table~\ref{GOSSS_spclas_02}         \\
$\theta^1$~Ori~Ba,Bb      & 209.01$-$19.38$\_$02 & \num{3017364132049943680} & \num{14652} & O-023 & B3           & V          & n           & ---                & RUWE=1.46, \sigmae=0.146~mas, also Table~\ref{GOSSS_spclas_02}         \\
$\theta^2$~Ori~C          & 209.07$-$19.34$\_$01 & \num{3017360799155290368} & ---         & O-023 & B4           & V          & ---         & ---                & $r_\mu$=3.39~$\mu$as/a, also Table~\ref{GOSSS_spclas_02}               \\
HD~\num{36982}            & 209.07$-$19.44$\_$01 & \num{3017360348171372672} & \num{16708} & O-023 & B2           & V          & ---         & ---                & also Table~\ref{GOSSS_spclas_02}                                       \\ % npi=-3.33       
$\gamma^2$~Vel            & 262.80$-$07.69$\_$01 & ---                       & \num{980}   & O-024 & WC8          & ---        & ---         & O7/8.5 III-II((f)) & not in {\it Gaia} EDR3, also Table~\ref{GOSSS_spclas_02}               \\
V572~Car                  & 287.59$-$00.69$\_$01 & \num{5350358069101053184} & \num{1861}  & O-025 & O6.5         & V          & z           & B0 V + B0.2 V      & also \citet{Maizetal16}                                                \\
CPD~$-$59~2635            & 287.64$-$00.68$\_$02 & \num{5350358069101053184} & \num{1872}  & O-025 & O8           & V          & z           & O9.2 V             & also \citet{Sotaetal14}                                                \\ % npi=4.22   
HD~\num{93343}            & 287.64$-$00.68$\_$01 & \num{5350310824456389504} & \num{16717} & O-025 & O7.5         & V          & z           & O7.5: V(n)         & also \citet{Maizetal16}                                                \\
V573~Car                  & 287.60$-$00.62$\_$01 & \num{5350358481418098944} & \num{1871}  & O-025 & O9.5         & IV         & ---         & B0.5 V             & also \citet{Sotaetal14}                                                \\
$\sigma$~Ori~D            & 206.82$-$17.33$\_$01 & \num{3216486478101982592} & \num{8474}  & O-026 & B2           & V          & (n)         & ---                & \sigmae=0.160~mas, also Table~\ref{GOSSS_spclas_02}                    \\
$\sigma$~Ori~E            & 206.82$-$17.32$\_$01 & \num{3216486478101981056} & \num{8475}  & O-026 & B2           & V          & p           & ---                & \sigmae=0.168~mas, also Table~\ref{GOSSS_spclas_02}                    \\
\hline
\end{tabular}
\addtolength{\tabcolsep}{3pt}
}
\label{LiLi_spclas}
\end{table*}

\end{appendix}

\end{document}